\documentclass[%
 aip,
 amsmath,amssymb,
preprint,
]{revtex4-1}

\usepackage{dcolumn}
\usepackage[mathlines]{lineno}

\usepackage[utf8]{inputenc}
\usepackage[T1]{fontenc}
\usepackage{mathptmx}

\usepackage{hyperref}
\usepackage{latexsym}
\usepackage{mathrsfs}
\usepackage{graphicx}
\usepackage{subfig}
\usepackage{bm} 
\usepackage[dvipsnames]{xcolor}
\usepackage[section]{placeins}
\usepackage{amsmath}
\usepackage[normalem]{ulem}

\usepackage[titletoc,title]{appendix}

\newcommand{\aeq}{\begin{equation}}
\newcommand{\eeq}{\end{equation}}
\newcommand{\aeqn}{\begin{eqnarray}}
\newcommand{\eeqn}{\end{eqnarray}}
\newcommand{\aeqns}{\begin{eqnarray*}}
\newcommand{\eeqns}{\end{eqnarray*}}
\newcommand{\yb}[1]{\mathbf{#1}}
\newcommand{\ybs}[1]{\boldsymbol{#1}}

\newcommand*\diff{\mathop{}\!\mathrm{d}}

\newcommand{\ySubFigSL}[2]{
	\subfloat{\includegraphics[width=0.5\textwidth]{./#1}\label{#2}}
}

\newcommand{\yFigOne}[3]{
	\begin{figure}[!t]
		\includegraphics[width=\textwidth]{./#1}
		\caption{#2 \label{#3}}
	\end{figure}
}

\newcommand{\yFigTwo}[6]{
	\begin{figure}[!t]
		\ySubFigSL{#1}{#2}
		\ySubFigSL{#3}{#4}
		\caption{#5 \label{#6}}
	\end{figure}
}
\newcommand{\yFigTwoV}[6]{
	\begin{figure}[!t]
		\centering
		\subfloat{\includegraphics[width=0.5\textwidth]{./#1}\label{#2}}\\
		\subfloat{\includegraphics[width=0.5\textwidth]{./#3}\label{#4}}
		\caption{#5 \label{#6}}
	\end{figure}
}

\newcommand{\yFigThreeV}[8]{
	\begin{figure}[!t]
		\subfloat{\includegraphics[width=0.5\textwidth]{./#1}\label{#2}}\\
		\subfloat{\includegraphics[width=0.5\textwidth]{./#3}\label{#4}}\\
		\subfloat{\includegraphics[width=0.5\textwidth]{./#5}\label{#6}}
		\caption{#7 \label{#8}}
	\end{figure}
}

\begin{document}

\preprint{AIP/123-QED}

\title[]{Nonlinear dynamics of energetic-particle driven geodesic acoustic modes in ASDEX Upgrade.}

\author{I. Novikau}
\email{ivan.novikau@ipp.mpg.de}
\author{A. Biancalani}
\author{A. Bottino}
\author{Ph. Lauber}
\author{E. Poli}
\author{P. Manz}
\author{G. D. Conway}
\author{A. Di Siena}
\affiliation{Max-Planck-Institut f\"ur Plasmaphysik, 85748 Garching, Germany}

\author{N. Ohana}
\author{E. Lanti}
\author{L. Villard}
\affiliation{{\'E}cole Polytechnique F\'ed\'erale de Lausanne, Swiss Plasma Center, CH-1015 Lausanne, Switzerland}

\author{ASDEX Upgrade Team\footnote{See \href{https://doi.org/10.1088/1741-4326/ab18b8}{H. Meyer \textit{et al 2009 Nucl. Fusion} \textbf{59} 112014} for the ASDEX Upgrade Team.}}
\affiliation{Max-Planck-Institut f\"ur Plasmaphysik, 85748 Garching, Germany}

\date{\today}

\begin{abstract}
Turbulence in tokamaks generates radially sheared zonal
flows. Their oscillatory counterparts, geodesic
acoustic modes (GAMs), appear due to the action of the
magnetic field curvature. The GAMs can be driven unstable by an
anisotropic energetic particle (EP) population leading to
the formation of global radial structures, called EGAMs.
The EGAMs can redistribute EP energy to the bulk plasma
through collisionless wave-particle interaction. 
In such a way, the EGAMs might contribute to the plasma heating. 
Thus, investigation of EGAM properties, especially in the velocity space, is
necessary for precise understanding of the transport
phenomena in tokamak plasmas.

In this work, the nonlinear dynamics of EGAMs without considering the mode interaction with the turbulence is investigated
with the help of a Mode-Particle-Resonance (MPR)
diagnostic implemented in the global
gyrokinetic particle-in-cell code ORB5.
An ASDEX Upgrade discharge is chosen as a reference
case for this investigation due to its rich EP nonlinear dynamics.
An experimentally relevant magnetic field configuration, thermal species profiles and an EP density profile are taken for EGAM chirping modelling and its comparison with available empirical data. 
The same magnetic configuration is used to explore energy transfer by the mode from the energetic particles to the thermal plasma including kinetic electron effects.
For a given EGAM level the plasma heating by the mode can be significantly enhanced by varying the EP parameters. 
Electron dynamics decreases the EGAM saturation amplitude and consequently reduces the plasma heating,
even though the mode transfers its energy to thermal ions much more than to electrons.


\end{abstract}

\maketitle

\section{Introduction}

An energetic particle (EP) beam, injected in tokamak plasma, can excite a variety of modes. 
One of these modes, called energetic-particle-driven geodesic acoustic mode (EGAM)~\cite{Boswell06, Fu08}, can have a significant influence on the EP dynamics and on the plasma confinement. First of all, being excited, this mode provides an additional mechanism of the energy exchange between the energetic particles and the thermal plasma~\cite{Sasaki11, Osakabe14, Toi19, Wang19} due to the wave-particle interaction.
Moreover, because of its interaction with the turbulence\cite{Zarzoso17, MSasaki17}, this mode might be used as an additional knob in the turbulence regulation. Significant progress has been made in the last decade in building a theoretical model which can explain the main nonlinear physics of the EGAM\cite{Qiu17, Qiu18, Chen18}. Yet, when one wants to quantitatively compare with experimental measurements, some of the approximations in the previous models can be limiting. 
For example, due to the importance of the wave-particle resonances with the thermal and energetic species, a kinetic treatment should be used as in Ref.~\onlinecite{Zhang10, Merlo18, Novikau19}. 
Apart from that, effect of the magnetic surface shape\cite{Siena18} can be considered by employing an experimental magnetic equilibrium. 
In this work, the importance of these effects on an experimental ASDEX-Upgrade (AUG) case by means of numerical simulations with the nonlinear gyrokinetic code ORB5~\cite{Jolliet07, Bottino15, Lanti19} is investigated. 
ORB5 has been previously used for nonlinear studies of EGAMs in simplified configurations\cite{BiancalaniJPP17, Biancalani18}, and it is used here to compare the nonlinear EGAM dynamics with experimental measurements.

In this work, only the EGAM dynamics, leaving aside its interaction with the turbulence is going to be considered. 
The nonlinear behaviour of this mode, 
such as the mode chirping\cite{Berk06, Berk10}, plasma heating by the EGAMs, and its dependence on the plasma parameters are of interest here.
It should be noted here that in contrast to finite-$n$ modes (where $n$ is a toroidal number), which also can transfer energy from the EPs to the thermal plasma due to the wave-particle interaction, the EGAM dynamics is practically not associated with additional particle loss, at least if one does not consider the mode propagation and topological orbital changes\cite{Zarzoso18}.

\begin{figure}[!ht]
\subfloat{
\includegraphics[width=0.52\textwidth]
{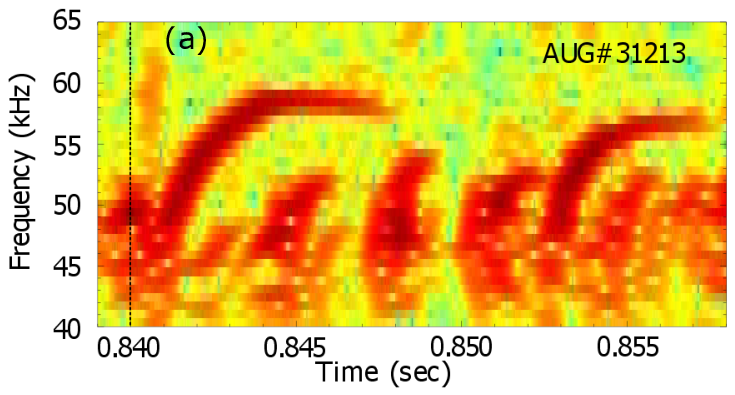}
\label{fig:expEGAM-a}
}
\subfloat{
\includegraphics[width=0.48\textwidth]
{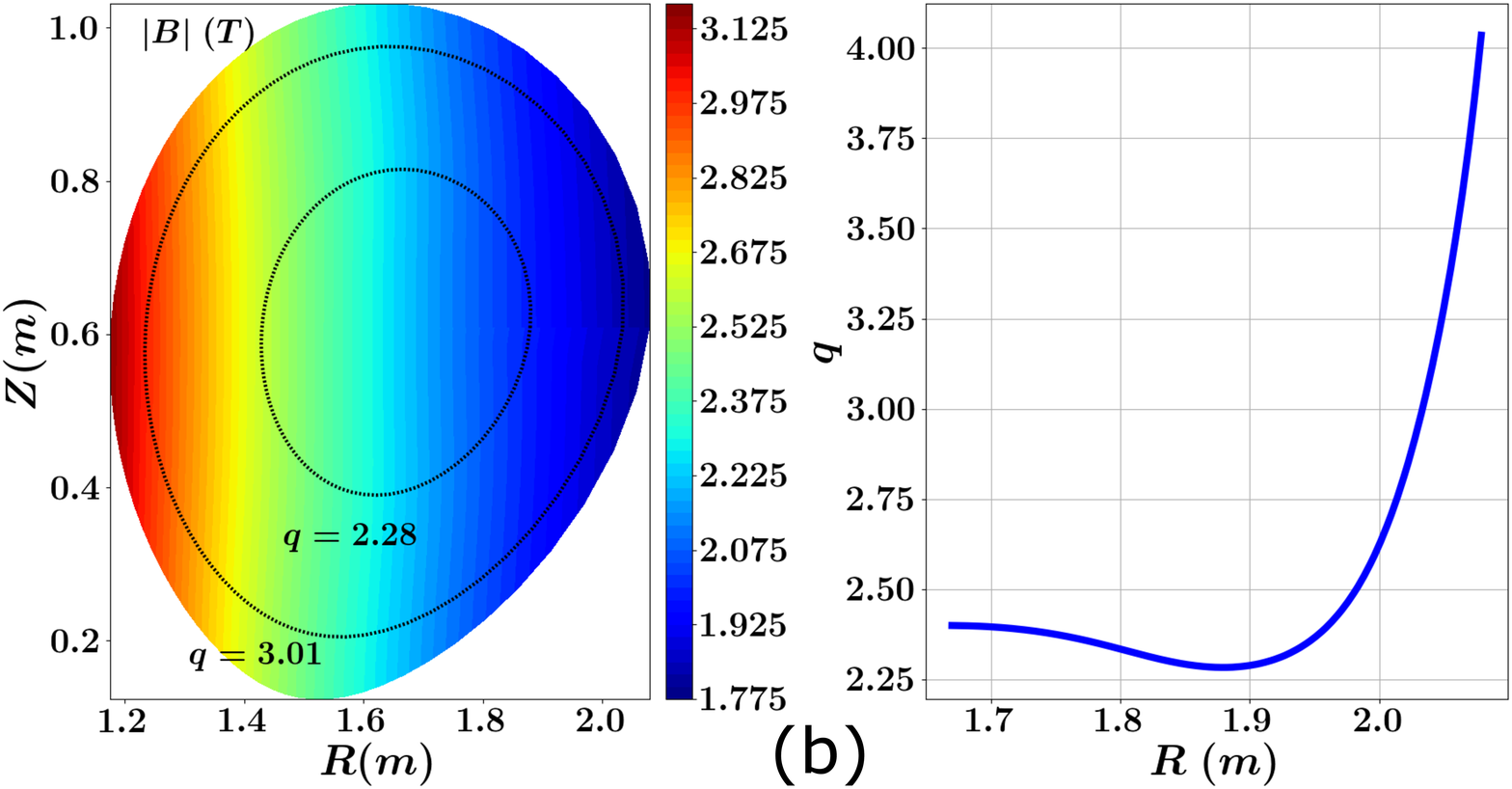}
\label{fig:expEGAM-b}
}
\caption{Experimental EGAM spectrum\cite{Horvath16, LauberSitee} in the NLED AUG discharge \#31213@0.84 s , obtained from Mirnov coils (Fig.~\ref{fig:expEGAM-a}). Fig.~\ref{fig:expEGAM-b} shows the magnetic field configuration and the safety factor profile. \label{fig:expEGAM}}
\end{figure}

The EGAM dynamics is going to be investigated in the plasma configuration of the ASDEX Upgrade (AUG) discharge \#31213@0.84 s (referred to as the NLED-AUG case)\cite{LauberSitee, Lauber18}, where a rich EP nonlinear dynamics is present\cite{Horvath16}. 
In this discharge, various types of EP-driven instabilities were identified, among which there are Alfv\'en instabilities (see, for example, Ref.~\onlinecite{Vannini19}) and EGAMs. In particular, a clear EGAM chirping was observed. 
More precisely, at the experimental EGAM frequency spectrogram 
(Fig.~\ref{fig:expEGAM-a}) one can see at least two different branches of the EGAM evolution in time. A "short" one is with a $\sim 2.5$ ms period with the frequency rise from $\sim 45$ kHz till $\sim 55$ kHz. A "long" branch is characterised by an ascending frequency from $\sim 45$ kHz till $\sim 58$ kHz and frequency saturation with a period of change around $13$ ms. 


The paper structure is as follows. The ASDEX Upgrade configuration is presented in Section~\ref{sec:conf}. 
The Mode-Particle-Resonance (MPR) diagnostic\cite{Novikau19}, implemented in ORB5 to study field-particle interaction in velocity space, is briefly described in Section~\ref{sec:mpr}.
The mode chirping and corresponding comparison with the AUG experiment is presented in Section ~\ref{sec:chirp}. Using two shifted Maxwellians as an EP distribution function, a nonlinear electrostatic (ES) gyrokinetic simulation with the code ORB5 produces a relative EGAM up-chirping close to the experimental one.
In this simulation, EP parameters have been taken consistent with previous works (Refs.~\onlinecite{Siena18, Novikau19}), where in the same AUG discharge the ordering of the EGAM frequency was reproduced in linear electrostatic and electromagnetic (EM) simulations, performed with the codes GENE\cite{Jenko00} and ORB5.
In Section~\ref{sec:distr} the EGAM linear behaviour is investigated by varying EP parameters. 
This study is used later for the analysis of the plasma heating by the EPs through the EGAMs in Section~\ref{sec:heating}. 
Existence of an energy exchange between EPs and thermal particles are emphasized, and dominant resonant thermal species are indicated using the MPR diagnostic. Such kind of plasma heating has been recently demonstrated in Ref.~\onlinecite{Wang19} in a realistic 3D equilibrium of the LHD stellarator, using the hybrid code MEGA with kinetic treatment of both thermal and energetic ions.
 
By comparing ES simulations with adiabatic electrons and electromagnetic (EM) simulations with drift-kinetic electrons, it is shown how the inclusion of the electron dynamics affects the EGAM formation and the plasma heating by this mode in Section~\ref{sec:kin}. 
Although even in the presence of drift-kinetic electrons the EGAM transfers most of its energy to the thermal ions, electron dynamics still clearly decreases the mode level in the considered AUG discharge, and in such a way reduces the energy flow from the EPs to the thermal ions.

\section{Numerical AUG configuration}
\label{sec:conf}
\yFigThreeV
{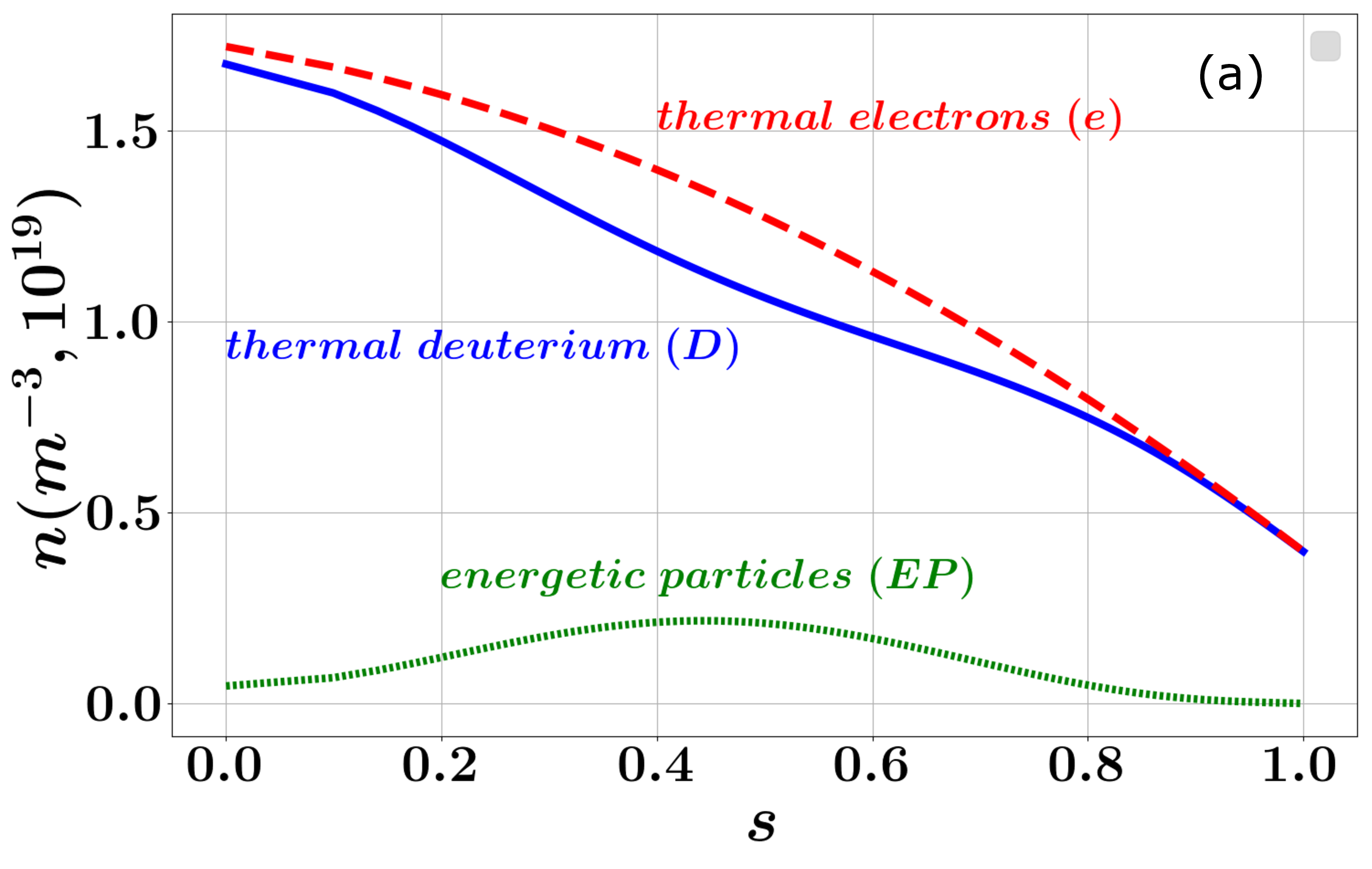}{fig:nT-species-a}
{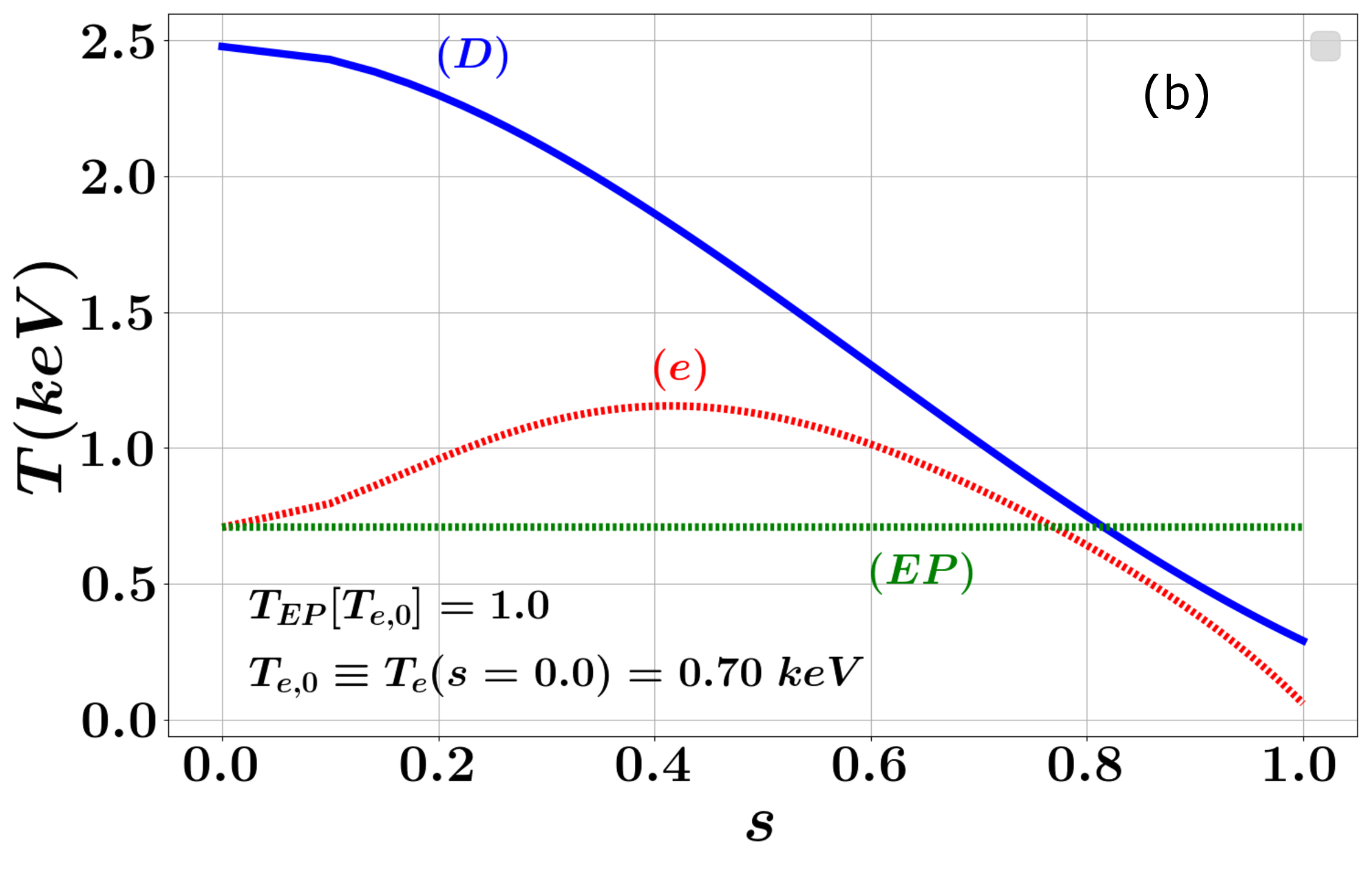}{fig:nT-species-b}
{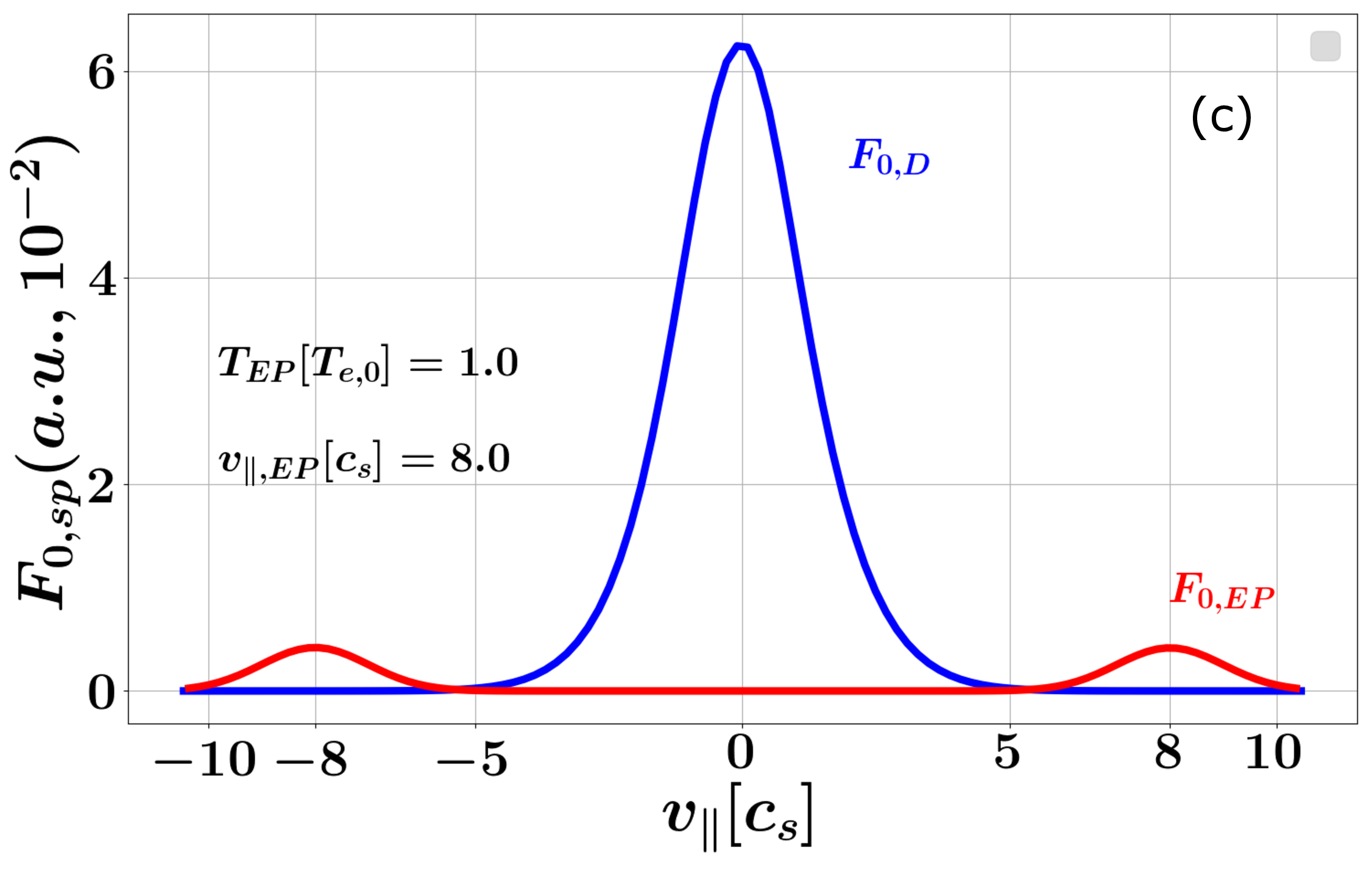}{fig:nT-species-c}
{Thermal species density (Fig.~\ref{fig:nT-species-a}) and temperature 
(Fig.~\ref{fig:nT-species-b}) profiles (blue and red lines), and corresponding typical EP profiles (green lines). Distribution functions of the thermal and energetic ions in velocity space are shown in Fig.~\ref{fig:nT-species-c}.}
{fig:nT-species}
The AUG plasma is simulated in the gyrokinetic (GK) particle-in-cell (PIC) global code ORB5\cite{Jolliet07, Bottino15, Lanti19}, where a realistic magnetic configuration is reconstructed from experimental data using the Grad-Shafranov solver code CHEASE \cite{Lutjens96},  without including the magnetic separatrix as last closed magnetic flux surface 
(Fig.~\ref{fig:expEGAM-b}).
As it is described farther in this section, the radial domain of the plasma configuration is slightly reduced for numerical reasons.
The magnetic field at the magnetic axis is $B_0 = 2.2$ T.
The major radius at the axis is $R_0 = 1.67$ m, the minor radius is $a_0 = 0.482$ m. 
Realistic temperature and density profiles of the thermal species (deuterium and electrons) are used as well (Figs.~\ref{fig:nT-species-a}-\ref{fig:nT-species-b}). 
Under "energetic particles" we understand here a deuterium beam, which actually drives EGAMs in a plasma system. This species is described by a two-bumps-on-tail distribution function (also called as two shifted Maxwellians) in velocity space (Fig.~\ref{fig:nT-species-c})
\begin{align}
F_{0,EP}(p_{z},\mu) = &A_{EP}(\psi)\exp\left[
		- \frac{m_{EP}}{T_{EP}} 
		\left( \frac{1}{2}\left(\frac{p_{z}}{m_{EP}}\right)^2 + 
			\frac{\mu B}{m_{EP}} \right) - 
		\frac{v_{\parallel, EP}^2}{2 T_{EP}}
	\right]\notag\\
	&\cosh\left(
		\frac{p_{z}}{m_{EP}}\frac{v_{\parallel, EP}}{T_{EP}} 
	\right),\label{eq:two-bumps}\\
A_{EP}(\psi) =& \frac{n_{EP}(\psi)}{(2\pi)^{3/2} T_{EP}^{3/2}},
\end{align}
which leads to the EGAM formation through the wave-particle interaction (due to the mechanism of the inverse Landau damping).
Here, $p_z$ is the canonical parallel momentum, $\mu$ is the magnetic moment, $B$ is the background magnetic field, while $v_{\parallel, EP}$, $T_{EP}$ being constant input parameters, which specify a shift and width of the bumps respectively.
It means, that the EPs are described by a flat temperature profile and a constant in space parallel velocity shift.
From Fig.~\ref{fig:nT-species-b} one can see that the EPs have relatively low temperature, which just indicates the fact that the width of the EP bumps is smaller in comparison to that of the thermal ion Maxwellian. 
A typical shift of the EP bumps in the parallel velocity is $v_{\parallel,EP}[c_s] = 8.0$, while typical EP temperature is $T_{EP}[T_e(s = 0)] = 1.0$ or $T_{EP}[keV] = 0.70$. Here, $c_s = \sqrt{T_{e, max}/m_{d}}$ is the sound speed with $m_d$ being the deuterium mass, $T_{e, max} = 1.15$ keV. 
Such kind of normalization for $v_{\parallel, EP}$ and $T_{EP}$ is used further in this work.
The EPs are modelled by means of the two shifted Maxwellians to qualitatively investigate the EGAM behaviour in the AUG plasma configuration.
The presence of two bumps in the velocity space is explained by the necessity to avoid the input of an additional momentum into the plasma system, which might have some effect on the EGAM dynamics.

In the nonlinear (NL) electrostatic simulation, performed in Section~\ref{sec:chirp} to study the EGAM up-chirping, the EPs have a realistic density profile, shown in Fig.~\ref{fig:nT-species-a}. In other sections, the EPs are described by an axisymmetric Gaussian distribution function:
\aeqn
F_{0,EP}(s) \sim \exp\left(- \frac{(s - s_{EP})^2}{2\sigma_{EP}^2}\right),
\label{eq:gaussian}
\eeqn
and are localised at a radial point $s_{EP} = 0.50$, with $s = \sqrt{\psi/\psi_{edge}}$ being a radial coordinate, where $\psi$ is the poloidal flux coordinate. Such a simplification has been done to have more freedom in the variation of the EP parameters.
The radial width of the EP Gaussian $\sigma_{EP}$, used in this work, is equal to $0.10$ to have a localised EP beam. 
An EP concentration is defined as $n_{EP}/n_e$, where $n_{EP}$, $n_e$ are the EP and electron densities, averaged in volume, respectively.   

In the simulations, presented in this work, three species are taken into account: gyro-kinetic thermal deuterium, gyro-kinetic energetic deuterium (EP), and electrons, either adiabatic (AE) or drift-kinetic (KE). 
Simulations with AE have been performed electrostatically, while the ones with KE have been made electromagnetically. The experimental pressure ratio at the reference magnetic surface has been adopted:
\aeqn
\beta_e = \frac{n_e T_e(s = 0)}{B_0^2/(2\mu_0)} = 2.7\cdot 10^{-4}.
\eeqn
Electromagnetic simulations have the advantage of the absence of the unstable $\omega_H$-mode\cite{Lee87}, observed in ES simulations with KEs. The KEs have a realistic deuterium/electron mass ratio $m_D/m_e = 3672$.
The pullback scheme\cite{Mishchenko17} has been used for the mitigation of the cancellation problem\cite{Hatzky07}. 
Real space in ES simulations has been discretized using the following grid parameters:
$n_s = 128, n_{\chi} = 64, n_{\phi} = 32$, - for the radial, poloidal and toroidal directions respectively.
Only $n = 0$ toroidal and $|m| = [0,..,3]$ poloidal mode numbers have been taken into account, since generally the geodesic acoustic modes have mainly low $m$ numbers.
The ES cases have been simulated in a radial domain $s = [0.0, 0.95]$ with a time step $dt[\omega_{ci}^{-1}] = 20$ (where $\omega_{ci} = eB_0/m_d$ and $e$ being the absolute value of the electron charge). The number of numerical markers for the thermal deuterium ($N_d$) and for the EPs ($N_{EP}$) in these cases are $N_d = N_{EP} = 6\cdot 10^7$. 
The EM cases have been simulated with $n_s = 256,\ dt[\omega_{ci}^{-1}] = 5$ in a radial domain $s = [0.0, 0.90]$, taking $N_e = 1.2\cdot 10^8$ with the same $N_d$ and $N_{EP}$.
The radial domain has been reduced to make the EM simulations more stable by avoiding the abrupt increase of the safety factor at the edge, reducing in such a way the restriction on the numerical time step. 
Since EGAMs are localised mainly in the core, this reduction of the radial domain has a negligible influence on the EGAM dynamics.

\subsection{Mode-Particle-Resonance diagnostic}
\label{sec:mpr}
To localise velocity domains of the most intensive EGAM-particle interaction,  the power balance diagnostic, implemented in ORB5\cite{Tronko16}, has been recently extended\cite{Novikau19} by keeping information about the wave-plasma interaction in velocity space. Using this diagnostic, a damping/growth rate $\gamma$ of an ES wave can be calculated as\cite{Tronko16, Novikau19}
\aeqn
\gamma &=& - \frac{1}{2} \frac{1}{n_{w}T_{w}}
	\int^{n_{w}T_{w}}_0\frac{\mathcal{P}}{\mathcal{E}_f}\diff t,
	\label{eq:mpr-gamma}\\
\mathcal{P}   &=& \sum_{sp} \mathcal{P}_{sp} = 
	- \sum_{sp} Z_{sp}e \int\diff V \diff W_{sp} f_{sp} 
	\yb{\dot{R}_{0,sp}}\cdot\ybs{\nabla}(J_{0,sp}\Phi),
	\label{eq:P}\\
\mathcal{E}_f &=& \sum_{sp}\frac{1}{2}Z_{sp} e\int\diff V\diff W_{sp} 
	 f_{sp} \Phi.\label{eq:Ef}
\eeqn
where $\mathcal{P}$ is the energy transfer signal (heating rate), $\mathcal{E}_f$ is the field energy, $f_{sp}(\yb{r}, p_z, \mu, t) = F_{0,sp}(\yb{r}, p_z, \mu) + \delta f_{sp}(\yb{r}, p_z, \mu, t)$ is the species distribution function, $\Phi$, $J_{0,sp}\Phi$ are the ES potential perturbation and the gyroaveraged one, 
$\yb{\dot{R}_{0,sp}}$ is the species equations of motion, unperturbed by the field perturbations, $Z_{sp}e$ is the species charge. 
The time integration in Eq.~\ref{eq:mpr-gamma} is performed on several wave periods $n_{w}T_{w}$. Finally, the integration in Eq.~\ref{eq:P} and Eq.~\ref{eq:Ef} is performed over the real $V$ and species velocity $W_{sp}$ domains.
For the equilibrium distribution functions $F_{0,sp}$, considered in this work, the part of $\mathcal{P}$ related to $F_{0,sp}$ is negligible. 
The energy transfer is normalized in the following way:
\aeqn
\mathcal{P} = \frac{\mathcal{P}[W]}{T_e(s=0)[J]\omega_{ci}[s^{-1}]}.
\label{eq:P-norm}
\eeqn

\section{GK simulation of EGAM chirping in the AUG discharge}
\label{sec:chirp}
\yFigTwoV
{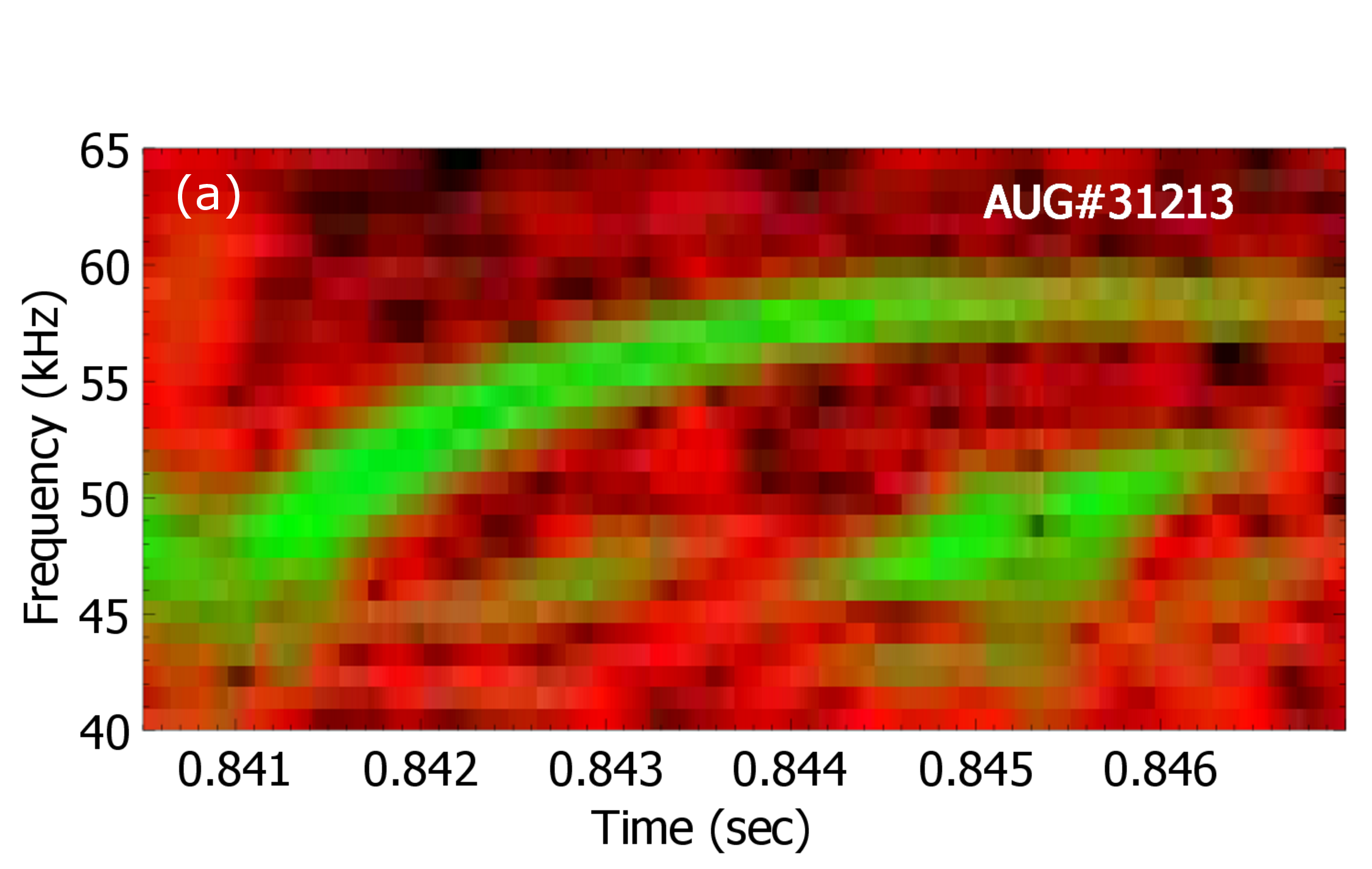}{fig:egam-chirp-a}
{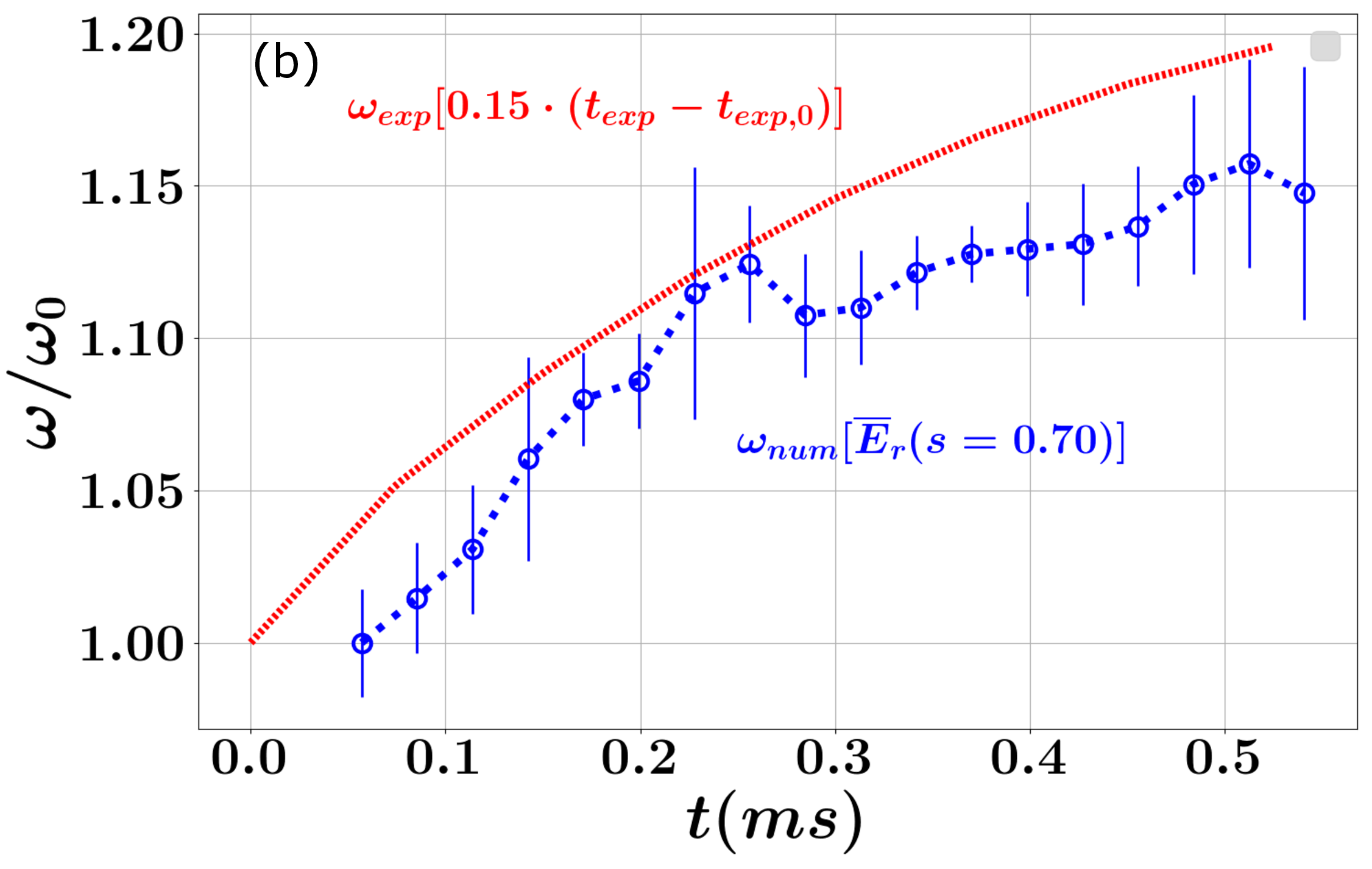}{fig:egam-chirp-b}
{A zoom of a 'long' branch of the EGAM up-chirping from the experimental AUG spectrogram\cite{Horvath16} is shown in Fig.~\ref{fig:egam-chirp-a}.
Numerical and experimental EGAM spectrograms, normalized to corresponding initial EGAM frequencies ($\sim 39$ kHz and $\sim 47$ kHz respectively), are indicated in Fig.~\ref{fig:egam-chirp-b}. 
The numerical chirping is calculated by nonlinear fitting of zonal electric field $\overline{E}$ at a radial point of the EGAM localisation (Appendix~\ref{app:conv-tests}).}
{fig:egam-chirp}

To model the nonlinear evolution of the EGAM frequency, the EP parameters are taken consistent with previous works (Refs.~\onlinecite{Siena18, Novikau19}), where the order of the mode frequency has been reproduced in linear GK simulations. 
The following EP velocity $v_{\parallel, EP} = 8.0$, temperature $T_{EP} = 1.0$ and concentration $n_{EP}/n_e = 0.095$ are applied.
In Fig.~\ref{fig:egam-chirp-a}, one can see a zoom of the 'long' branch, described previously, of the experimental EGAM up-chirping.
In Fig.~\ref{fig:egam-chirp-b}, experimental and numerical spectrograms normalized to an initial EGAM frequency are indicated.
The blue curve shows the mode up-chirping, obtained from a nonlinear ES simulation with adiabatic electrons. 
The numerical spectrogram has been measured by nonlinear fitting of zonal electric field in different time windows at the radial point of the mode localisation $s = 0.70$. In the experimental discharge, the mode was found in a radial domain $s \sim [0.50, 0.67]$ according to soft X-ray emission data\cite{Horvath16}.
The red curve corresponds to the experimental EGAM spectrogram, but with a squeezing time scale $0.15$, since the characteristic time of the numerical frequency change is shorter than the experimental one.
As one can see from  Fig.~\ref{fig:egam-chirp-b}, despite the two shifted Maxwellians, which are used as a distribution function of the EPs, the GK model is able to simulate the EGAM relative up-chirping in this AUG discharge.
In Appendix~\ref{app:conv-tests}, the chirping is calculated at other radial positions, and some convergence tests of this computation are presented.
On the other hand, to perform quantitative comparison with the experiment of absolute chirping and of the time scales, one should use an EP distribution function closer to the experimental one such as a slow-down distribution function with a pitch-angle dependence. Since the EGAM is a strongly driven mode, it significantly depends on EP parameters.
Moreover, the considered simulation is performed electrostatically without taking into account the dynamics of the drift-kinetic electrons. As it is shown in Section~\ref{sec:kin}, electrons can significantly modify the EGAM behaviour and, as a result, might have some effect on the mode chirping as well.

\section{Influence of EP parameters on the EGAM dynamics}
\label{sec:distr}

\yFigTwo
{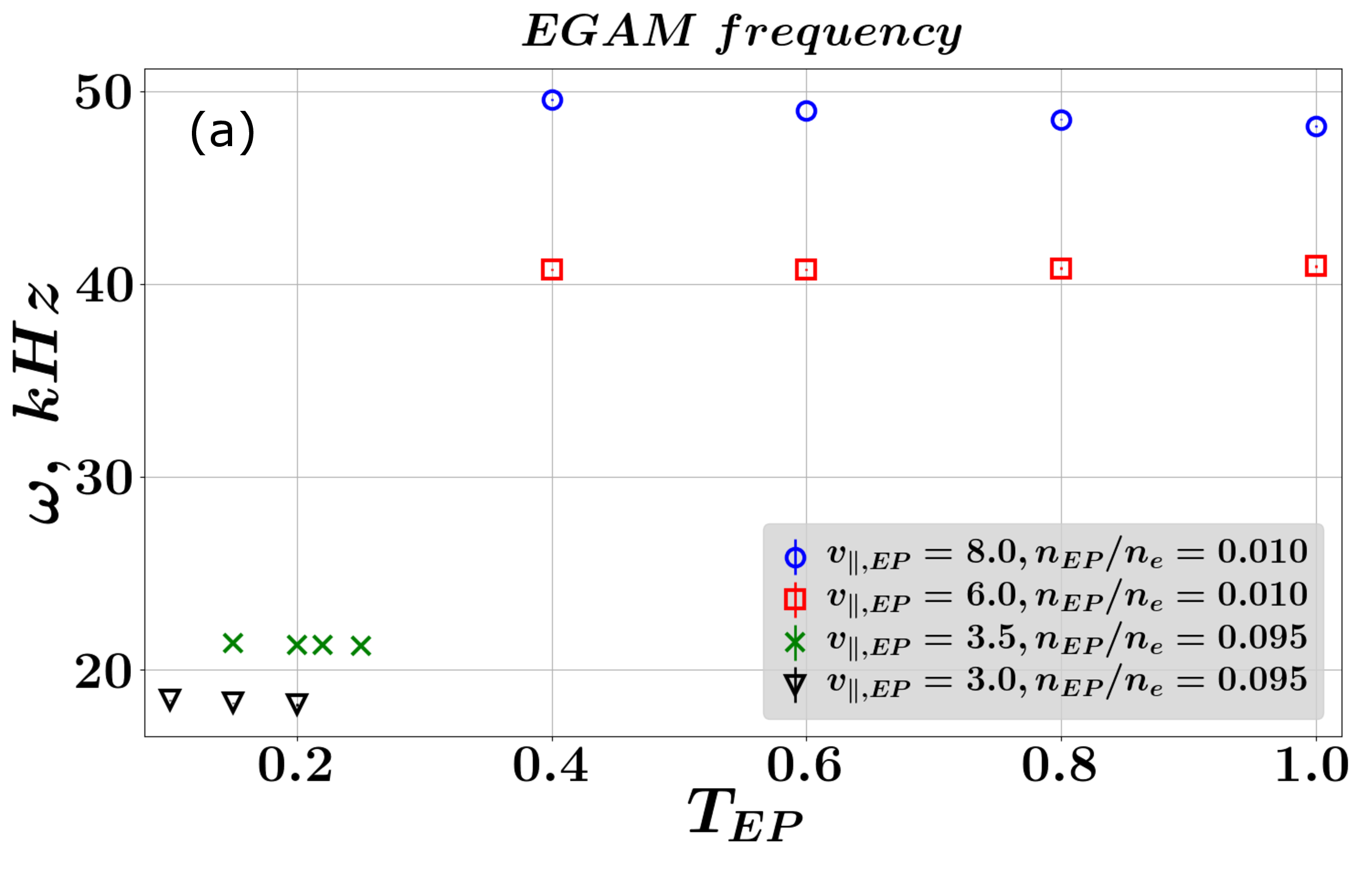}{fig:lin-scan-Tv-a}
{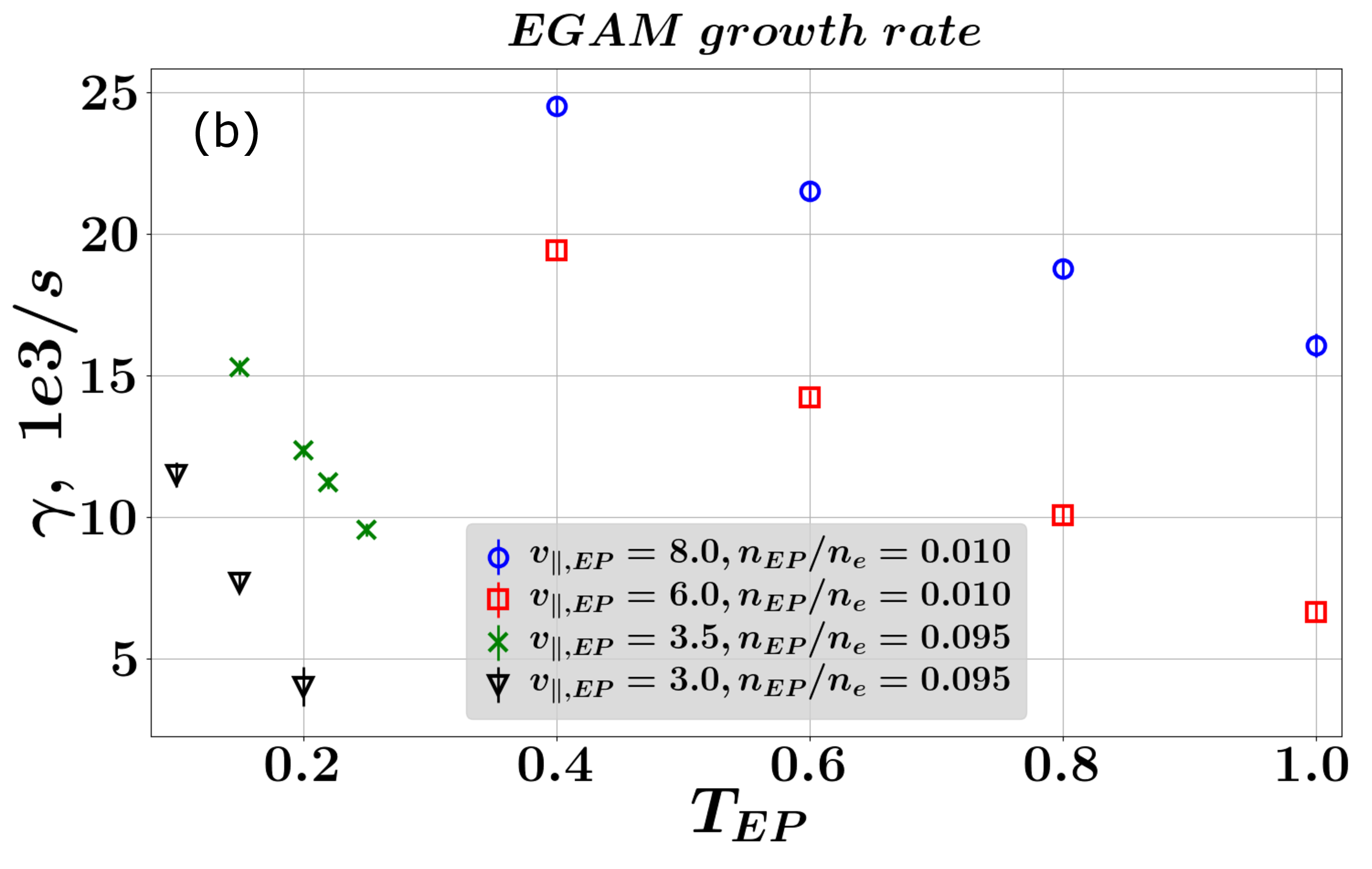}{fig:lin-scan-Tv-b}
{Dependence of the EGAM frequency (Fig.~\ref{fig:lin-scan-Tv-a}) and linear growth rate (Fig.~\ref{fig:lin-scan-Tv-b}) on the temperature and velocity shift of the EP beam.}
{fig:lin-scan-Tv}

To investigate the plasma heating by the EGAMs, we start from the study of the mode behaviour dependence on EP parameters. 
First of all, the linear growth rate is strongly modified by EP temperature variations, as one can see in Fig.~\ref{fig:lin-scan-Tv-b}, while the EGAM frequency remains practically the same and does not depend on $T_{EP}$. 
As it was already shown in different works (Refs.~\onlinecite{Girardo14, Zarzoso14}), the EPs can excite an EGAM corresponding to a damped mode in the absence of the EPs. 
By reducing the EP parallel velocity by a factor around of two, an EGAM with half the frequency of the original is excited (compare blue points and green crosses in Fig.~\ref{fig:lin-scan-Tv-a}).   
It should be noted that since the mode excited by a smaller EP velocity is less unstable, it was necessary to increase the EP concentration from $n_{EP}/n_e \approx 0.01$ to $n_{EP}/n_e \approx 0.09$.
\yFigTwoV
{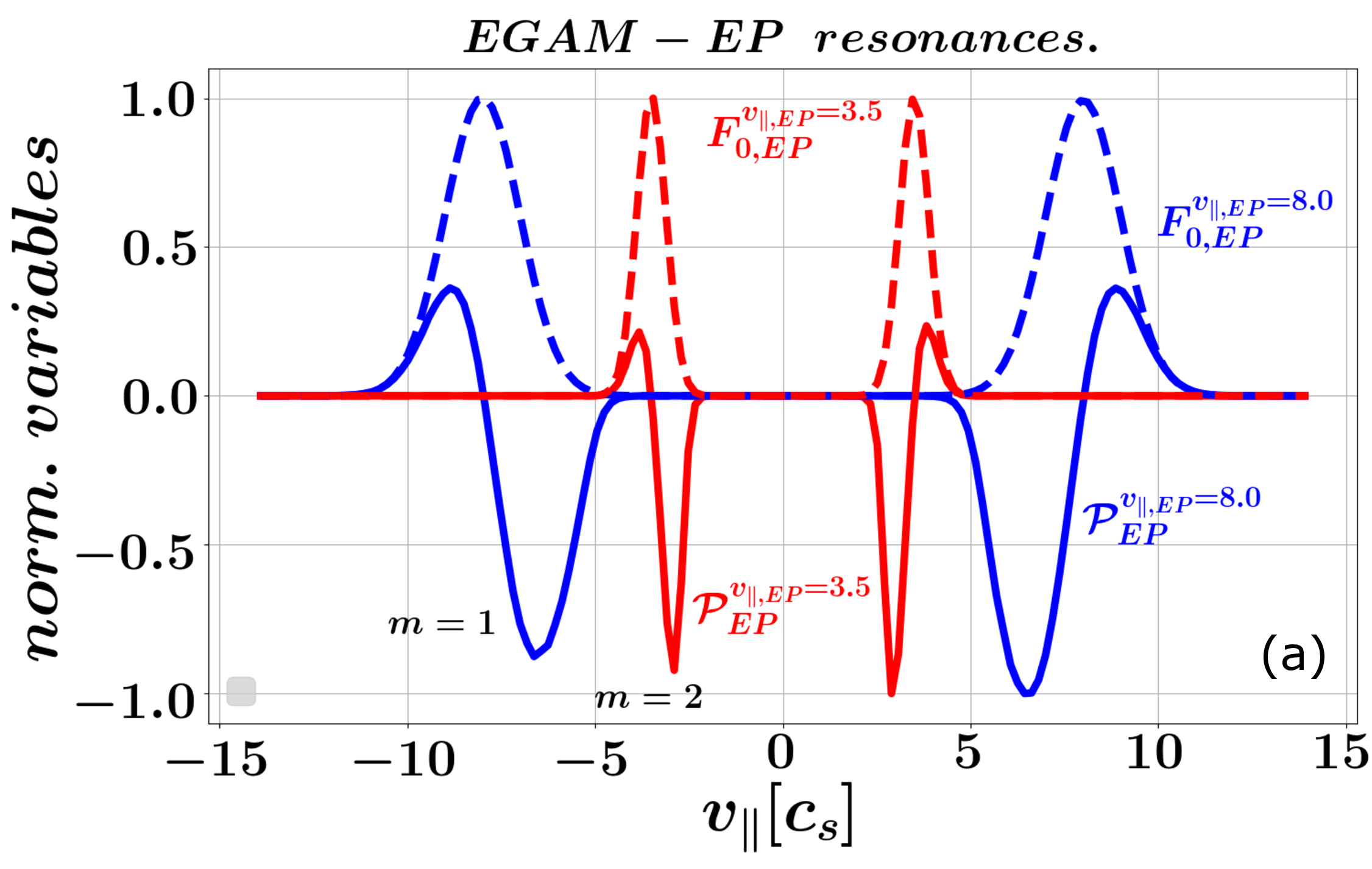}{fig:EGAM-EP-thD-res-a}
{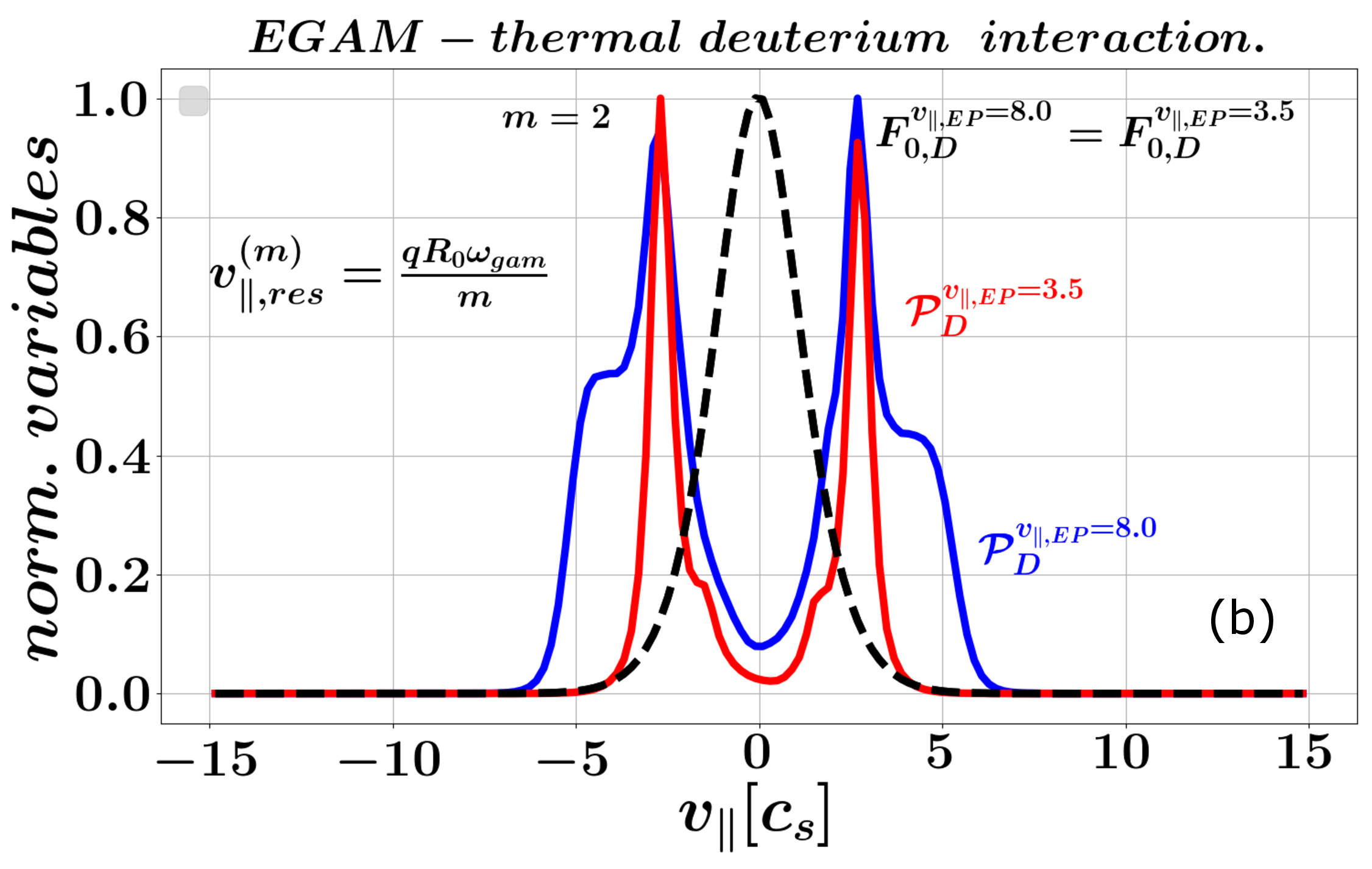}{fig:EGAM-EP-thD-res-b}
{Localisation of the EGAM interaction with the EPs (Fig.~\ref{fig:EGAM-EP-thD-res-a}) and with thermal deuterium (Fig.~\ref{fig:EGAM-EP-thD-res-b}). 
Here, two simulations are considered: with $v_{\parallel, EP} = 8.0, T_{EP} = 1.0$ (blue lines), and with $v_{\parallel, EP} = 3.5, T_{EP} = 0.25$ (red ones). The solid lines indicate the mode-species energy transfer $\mathcal{P}_{sp}$, summed on the perpendicular velocity and averaged on several EGAM periods in time. 
The dashed lines depict the localisation of the species initial distribution functions.}
{fig:EGAM-EP-thD-res}

\yFigTwo
{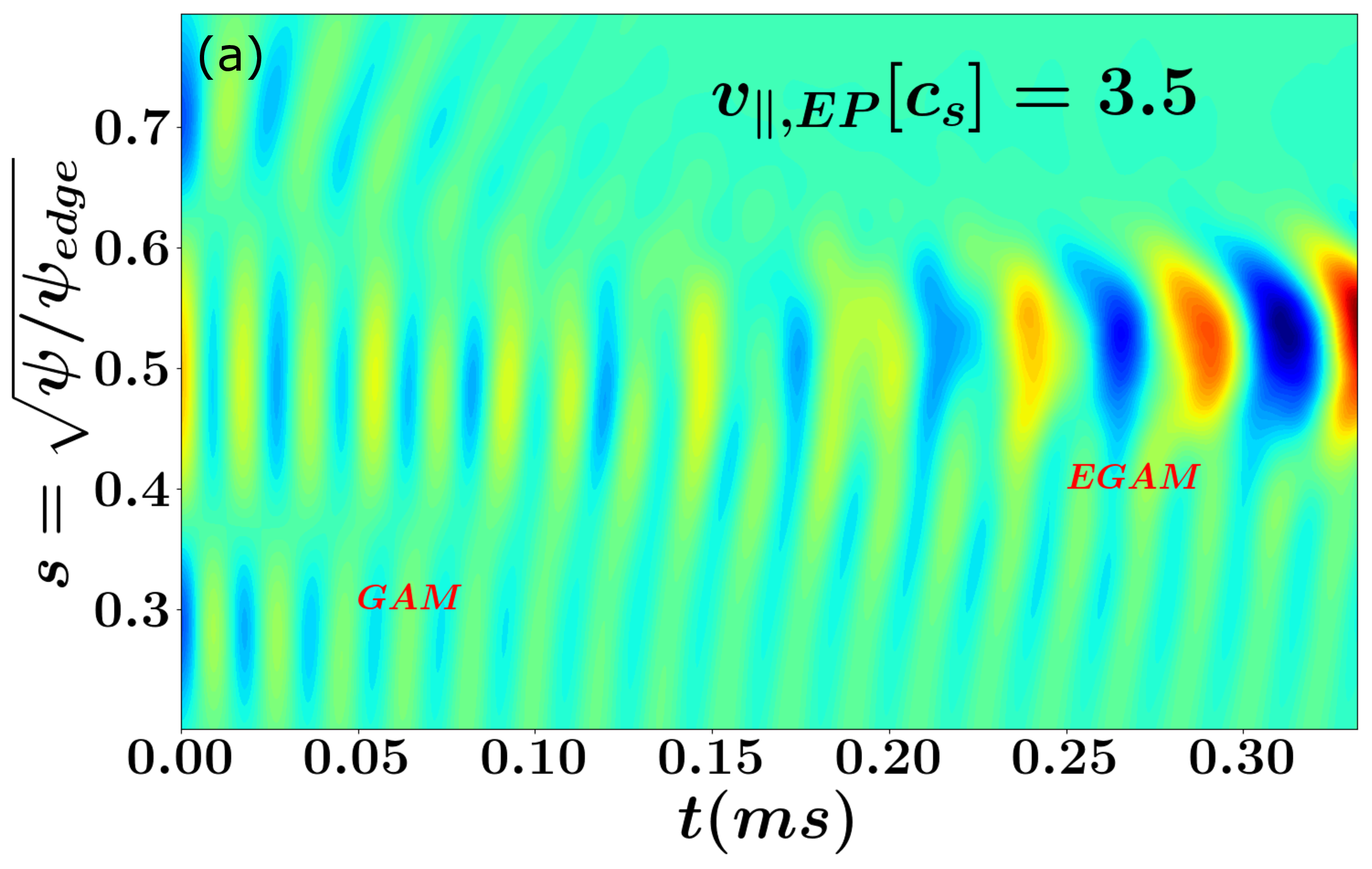}{fig:lin-erbar-st-a}
{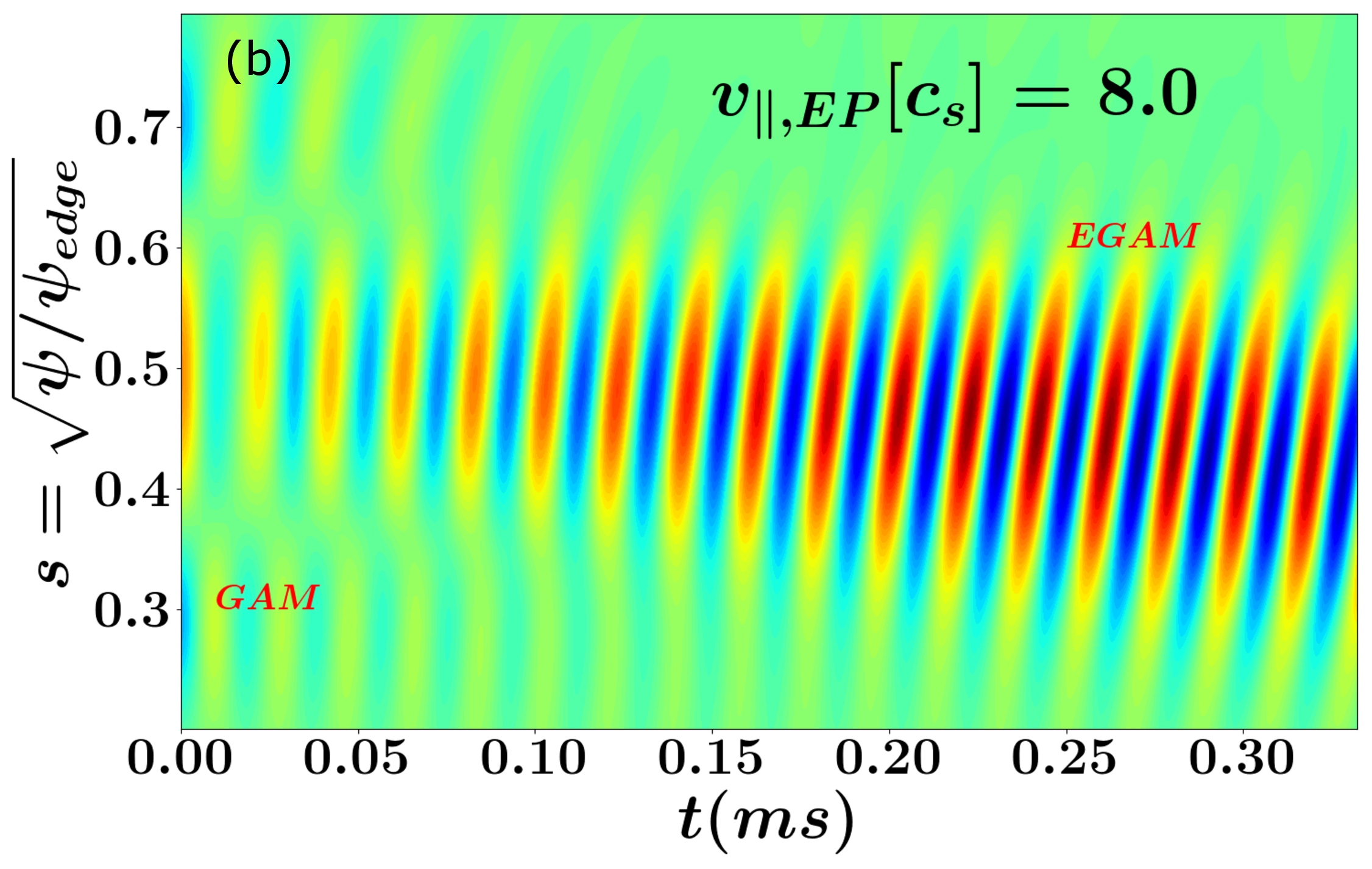}{fig:lin-erbar-st-b}
{Time evolution of zonal electric field $\overline{E}$ in two cases with low  (Fig.~\ref{fig:lin-erbar-st-a}) and high (Fig.~\ref{fig:lin-erbar-st-b}) EP velocities.}
{fig:lin-erbar-st}

The localisation of the GAM-particle resonances of order $m$ can be analytically estimated from the equation
\aeqn
v^{(m)}_{\parallel,res} = \frac{q R_0 \omega_{GAM}}{m}.\label{eq:vresm}
\eeqn
By reducing the EP velocity $v_{\parallel, EP}$ one can excite an EGAM which is driven by higher order GAM-EP resonances. 
In Fig.~\ref{fig:EGAM-EP-thD-res-a}, one can see that by injecting an EP beam with $v_{\parallel, EP} = 8.0$ (blue dashed line), one obtains an EGAM driven by the low order resonance $m = 1$ (blue solid line). 
On the other hand, injection of an EP beam with $v_{\parallel, EP} = 3.5$ (red dashed line) leads to the EGAM drive on the $m = 2$ resonance (red solid line). 
The time evolution of the zonal electric field in these two cases with low and high EP velocities is shown in Fig.~\ref{fig:lin-erbar-st}. 
The high velocity energetic beam (Fig.~\ref{fig:lin-erbar-st-b}) drives an EGAM through a low order $m=1$ resonance, and the resulting mode has a frequency close to the original GAM frequency.
The low velocity energetic beam (Fig.~\ref{fig:lin-erbar-st-a}), which interacts with the geodesic mode through a high order $m = 2$ resonance, drives an EGAM with a frequency close to half the GAM frequency.
Indeed, according to Fig.~\ref{fig:EGAM-EP-thD-res}, the mode-species resonances are localised at 
\aeqn
v_{\parallel, res, num}^{m=1} \approx 6.5,\ \ \ 
v_{\parallel, res, num}^{m=2} \approx 2.7.
\eeqn
A GAM frequency can be computed from a simulation without EPs:
\aeqn
\omega_{gam}(s = 0.50) = 49\ kHz,
\eeqn
to compare with corresponding EGAM frequencies
\aeqn
\omega^{EP8.0}_{egam}(s = 0.50) = 40\ kHz,\ \ \ 
\omega^{EP3.5}_{egam}(s = 0.50) = 21\ kHz
\eeqn
Using these frequencies and a safety factor value $q (s = 0.50) = 2.3$, (E)GAM-plasma resonance positions can be analytically estimated from Eq.~\ref{eq:vresm}:
\aeqn
v^{EP8.0}_{\parallel, res, theor} = 
	q R_0 \omega^{EP8.0}_{egam} &&\approx 5.2,\ \ \ 
v^{EP3.5}_{\parallel, res, theor} = 
	q R_0 \omega^{EP3.5}_{egam} \approx 2.8,\\
v^{gam, m = 1}_{\parallel, res, theor} &&\approx 6.4,\ \ \ 
v^{gam, m = 2}_{\parallel, res, theor} \approx 3.2.
\eeqn
The difference between $v^{EP8.0}_{\parallel, res, theor}$ and 
$v_{\parallel, res, num}^{m=1}$ might be explained by the fact that $v_{\parallel, res, num}^{m=1}$ is estimated from a signal averaged in a whole space domain while $v^{EP8.0}_{\parallel, res, theor}$ is calculated by using local plasma parameters. Moreover, the first order resonance $v_{\parallel, res, num}^{m=1}$ has a significant width along parallel velocity.
Since $m = 1$ and $m=2$ GAM resonance velocities are close to $v_{\parallel, res, num}^{m=1}$ and $v_{\parallel, res, num}^{m=2}$ respectively, it is reasonable to consider an EGAM as a mode that is driven by the first order resonance in case with $v_{\parallel, EP} = 8.0$, and by the second order resonance in case with $v_{\parallel, EP} = 3.5$.

As it was mentioned in Refs.~\onlinecite{Zarzoso18, Wang19}, interactions between the EPs and thermal species significantly benefit from the existence of high order resonances, which is confirmed here as well. 
The EGAM-thermal ion energy exchange occurs at the higher order ($m=2$) resonance in both cases, independently of the EP parallel velocity (Fig.~\ref{fig:EGAM-EP-thD-res-b}).
In other words, even when the EGAM is driven by an EP beam with a high parallel speed, the EGAM is still damped by the thermal deuterium through resonances at higher order. 
This can be simply explained by the fact that the position of these resonances are closer to the bulk of the thermal ion distribution function.
Since the thermal ion energy transfer signal (Fig.~\ref{fig:EGAM-EP-thD-res-b}) is positive, and the EP energy transfer signal (Fig.~\ref{fig:EGAM-EP-thD-res-a}) is mostly negative, there is an energy flow from the EPs to the thermal ions establishing bulk plasma heating.

\section{Plasma heating by EGAMs}
\label{sec:heating}

As it has been already discussed previously in Refs.~\onlinecite{Sasaki11, Osakabe14, Toi19, Wang19}, the EGAMs might provide an additional route for plasma heating by EPs. 
The thermal species can obtain energy from the EGAM due to the Landau damping of the mode, which in turn is driven by the energetic particles through the inverse Landau damping. 
Note that this is also true for Alfv\'en modes, with the substantial difference that the EP radial redistribution due to EGAMs is negligible with respect to Alfv\'en modes. Therefore, EGAMs represent a privileged mode for this plasma heating mechanism.
This process is investigated here by varying the EP parameters in the AUG plasma configuration by performing nonlinear GK simulations firstly in the ES limit with adiabatic electrons.

\yFigOne
{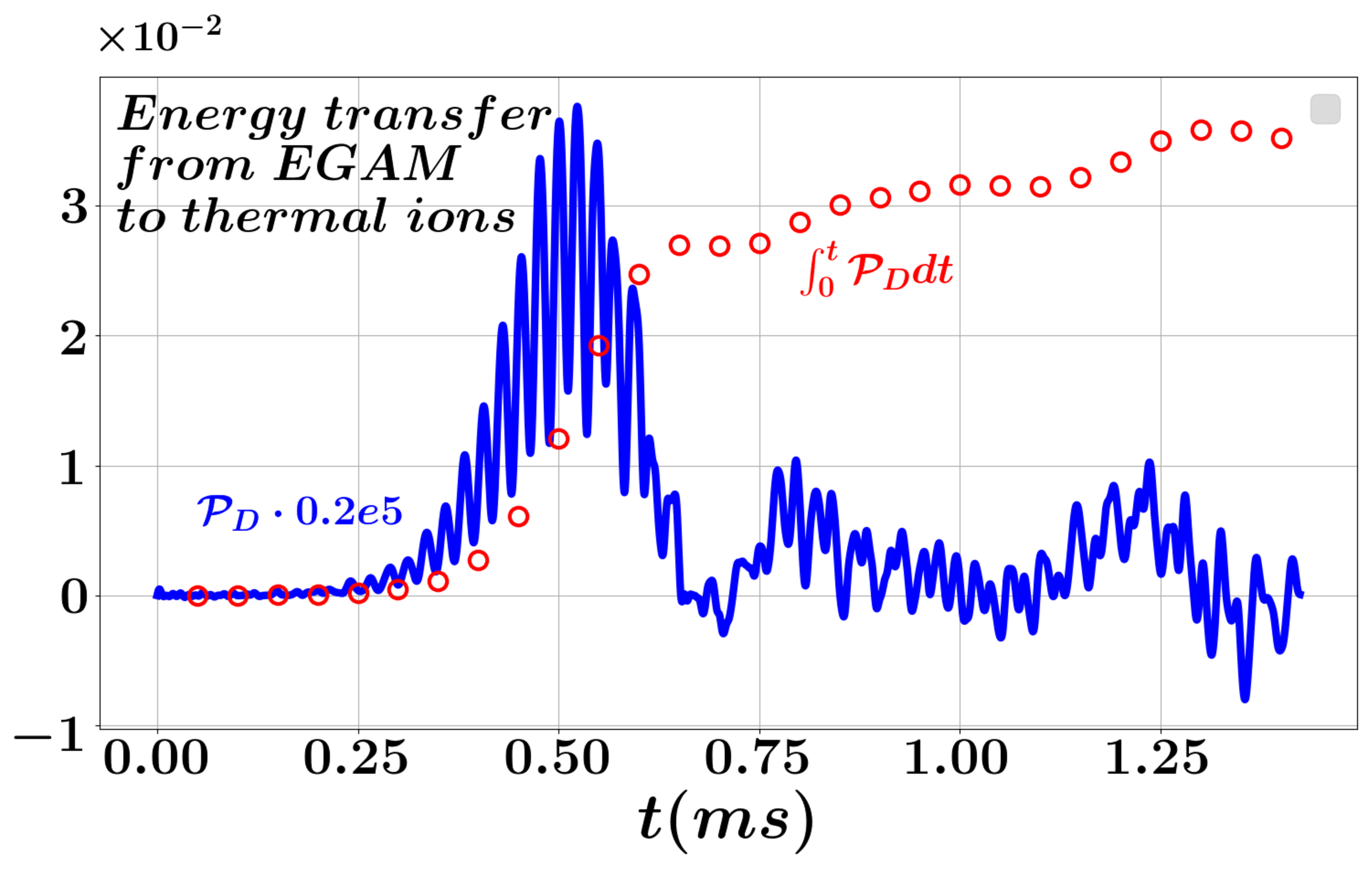}
{Time evolution of the heating rate (solid blue line), and the amount of energy transferred from the EGAM to the bulk deuterium plasma (red points). 
A case with $n_{EP}/n_e = 0.09, v_{\parallel, EP} = 3.5, T_{EP} = 0.25$ is considered here. The heating rate $\mathcal{P}_D$ is normalized according to Eq.~\ref{eq:P-norm}.}
{fig:NL-P-E}

\yFigTwo
{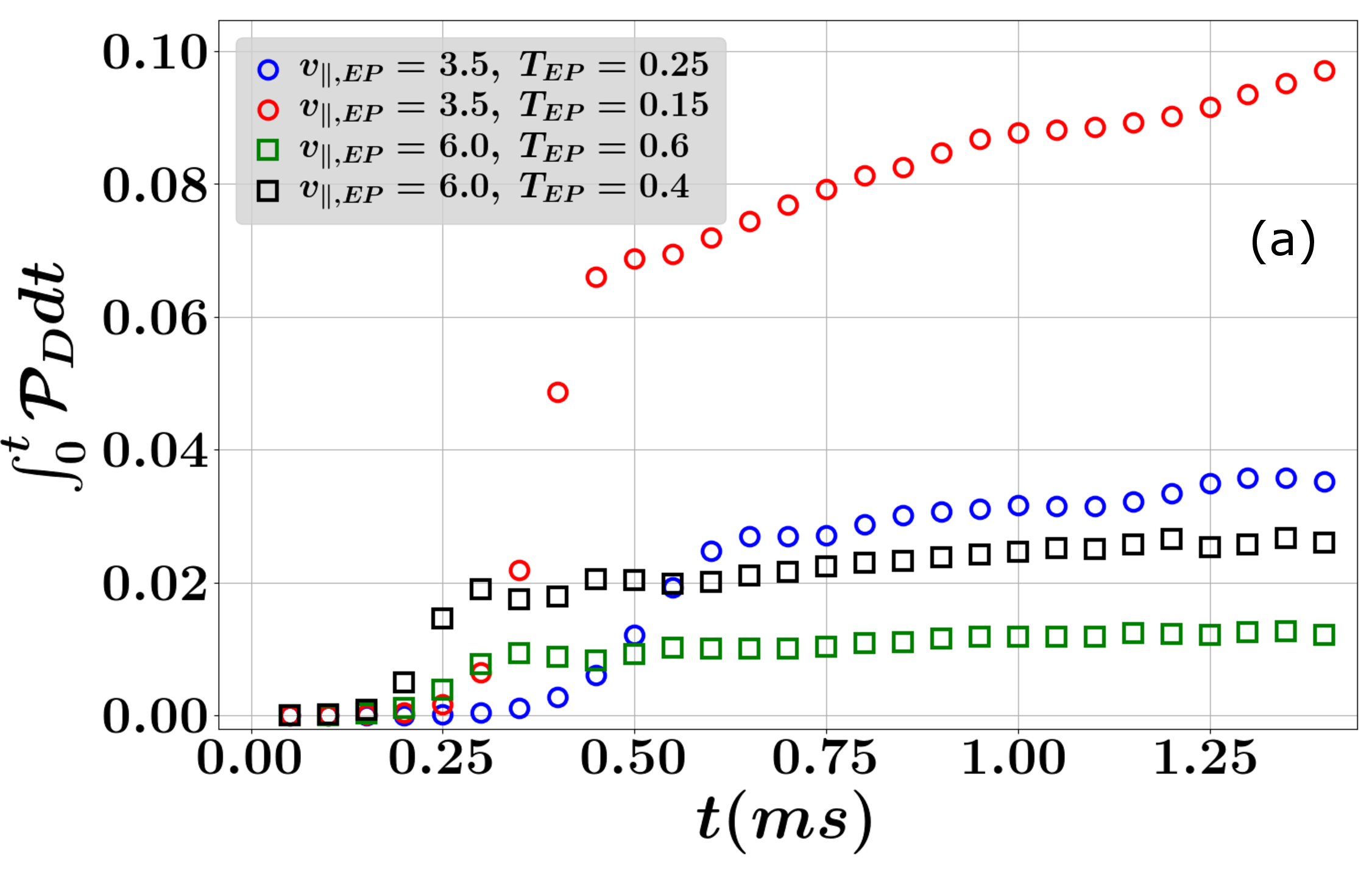}{fig:NL-T-energy-a}
{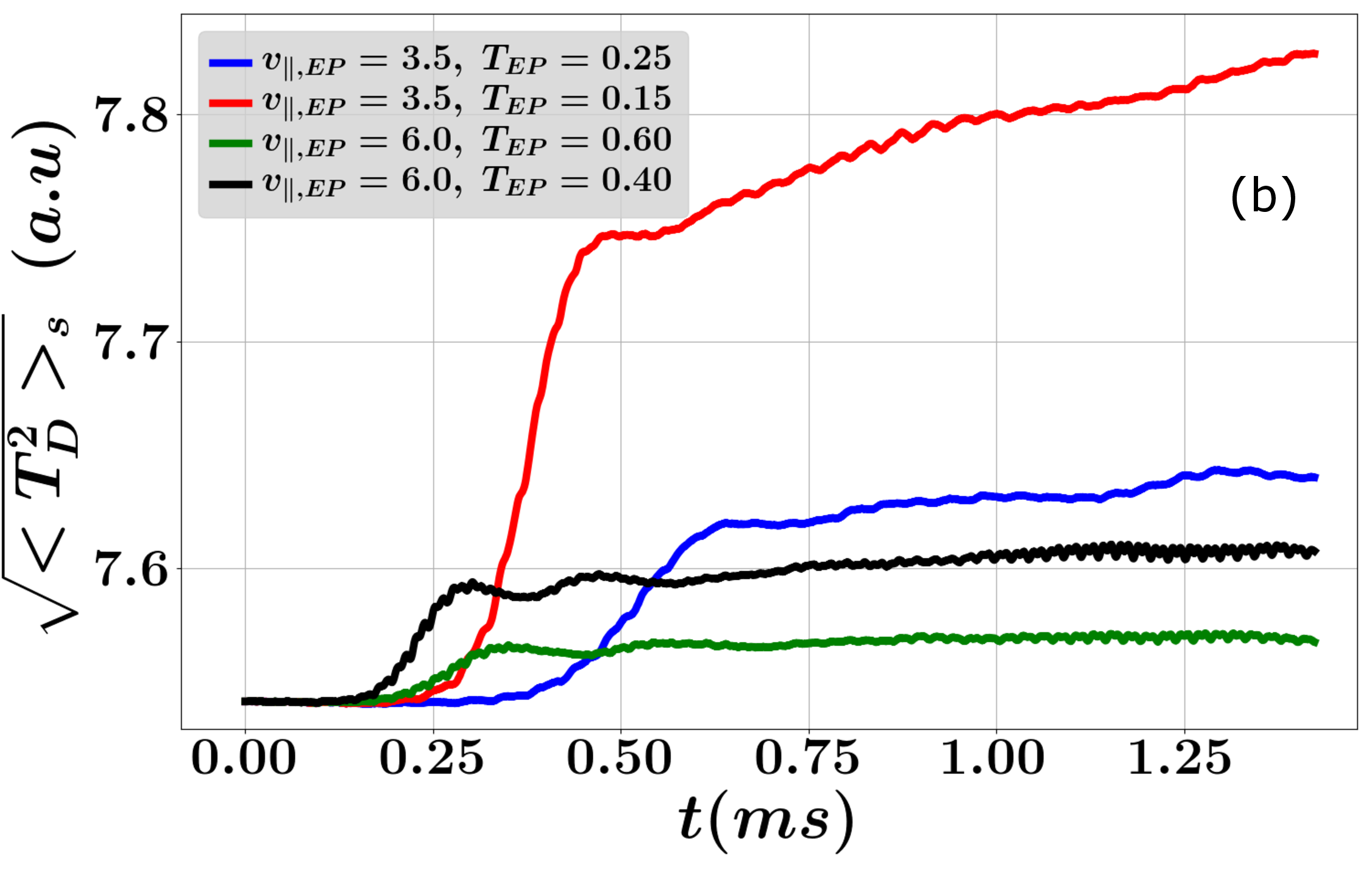}{fig:NL-T-energy-b}
{Energy transferred from the EGAM to the bulk ions for different EP temperatures and parallel velocities (Fig.~\ref{fig:NL-T-energy-a}).
Evolution of the thermal ion temperature in the same cases (Fig.~\ref{fig:NL-T-energy-b}).}
{fig:NL-T-energy}

To calculate the amount of energy transferred from the mode to the thermal ions, the EGAM-species energy exchange signal $\mathcal{P}_{sp}$ is integrated in time.
As it is shown in Fig.~\ref{fig:NL-P-E}, at the beginning, a rise of the heating rate due to the growth of the mode is observed. 
After a while, the mode saturates, the heating rate's level decreases due to the relaxation of the EP distribution function, and in the deep saturated domain the energy transfer from the mode to the thermal ions is practically suppressed. 
It means that the total energy transferred to the bulk plasma (red points in Fig.~\ref{fig:NL-P-E}) remains practically unchanged after the mode saturation.
The amount of this energy depends on the EP parameters (Fig.~\ref{fig:NL-T-energy-a}), which is directly reflected in the bulk ion temperature evolution (Fig.~\ref{fig:NL-T-energy-b}).


\yFigOne
{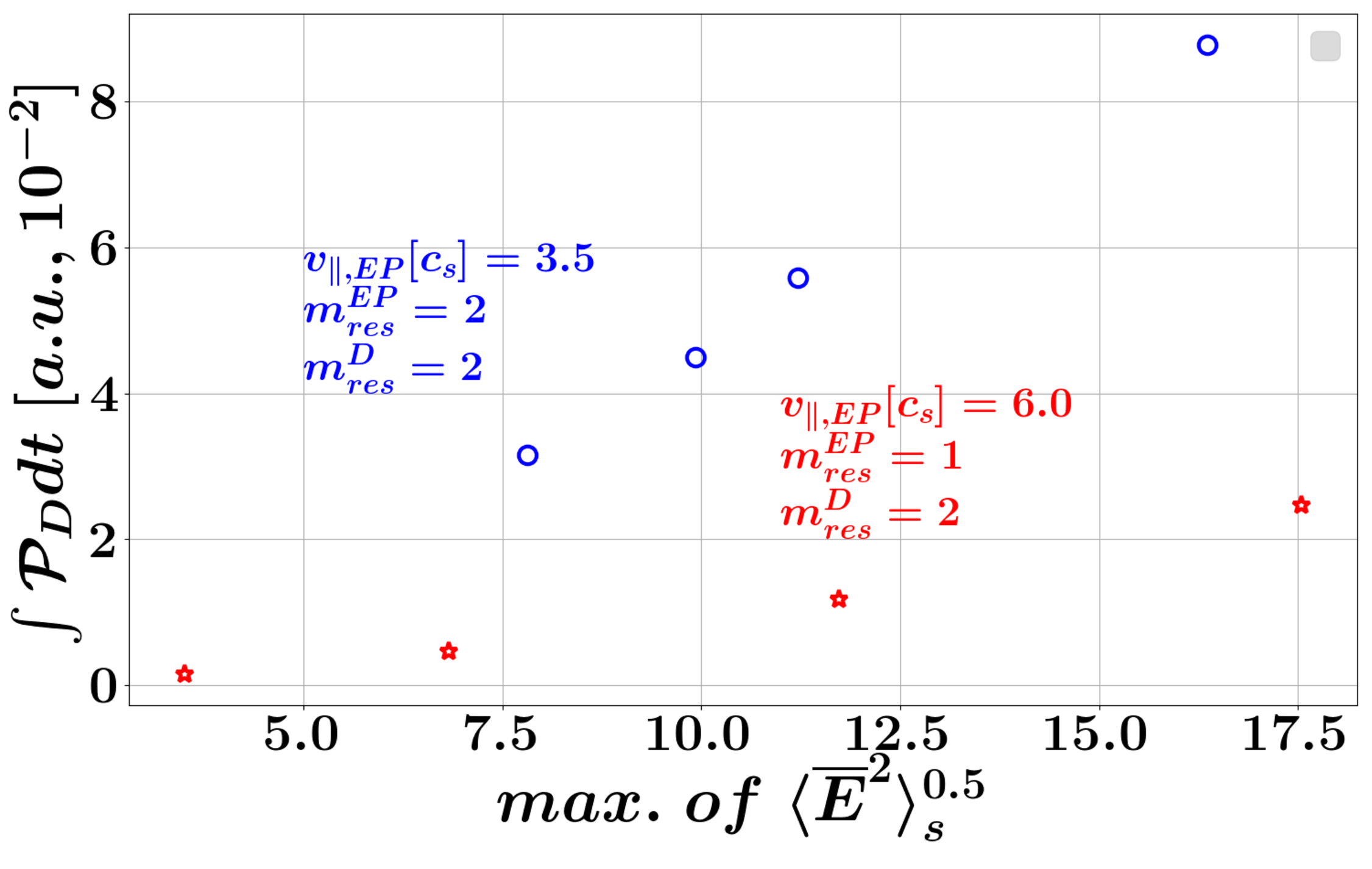}
{Dependence of the EGAM-thermal deuterium energy exchange on the EGAM saturation level in nonlinear ES simulations with adiabatic electrons. 
The saturation levels are calculated as a maximum value in time of r.m.s in space of the zonal electric field $\overline{E}$. The electric field is normalized to $T_e(s = 0.0)[eV]\omega_{ci}[s^{-1}]/c_s[m/s]$. Here, $m_{res}^{sp}$ indicates an order of a resonance, where the mode-species energy exchange occurs. The blue points correspond to the cases with $v_{\parallel, EP} = 3.5$, $n_{EP}/n_e = 0.09$, $T_{EP} = [0.25, 0.22, 0.20, 0.15]$, while the red stars correspond to the cases with $v_{\parallel, EP} = 6.0$, $n_{EP}/n_e = 0.01$, $T_{EP} = [1.0, 0.8, 0.6, 0.4]$.}
{fig:NL-scan-saturation}

By varying the EP parameters in NL electrostatic simulations, one can observe a general correlation between the EGAM saturation levels and the amount of energy transferred from the mode to the bulk deuterium plasma (Fig.~\ref{fig:NL-scan-saturation}). 
However, for example, the case with 
$v_{\parallel, EP} = 6.0,\ T_{EP} = 0.4,\  n_{EP}/n_e = 0.01$ 
(the most right red star) has practically the same saturation level as the case with 
$v_{\parallel, EP} = 3.5,\ T_{EP} = 0.15,\  n_{EP}/n_e = 0.09$ 
(the most right blue point). 
Having said that, we should notice that the corresponding EGAM-bulk ion energy exchange for these cases is significantly different.
This means that having the same mode level, one can achieve higher plasma heating by varying EP parameters.
In Appendix~\ref{app:details-plasma-heating}, these two cases are analysed in detail. 
Here, in particular, increase of the EP density $n_{EP}/n_e$ and lowering of the EP temperature $T_{EP}$ has led to a high energy transfer from the EPs to the mode.
On the other hand, decrease of the EP velocity has made the mode transfer higher part of its energy to the thermal plasma enhancing in such a way the energy exchange between energetic and thermal ions, and keeping the EGAM amplitude on the same level.

\section{Influence of electron dynamics}
\label{sec:kin}

\yFigTwo
{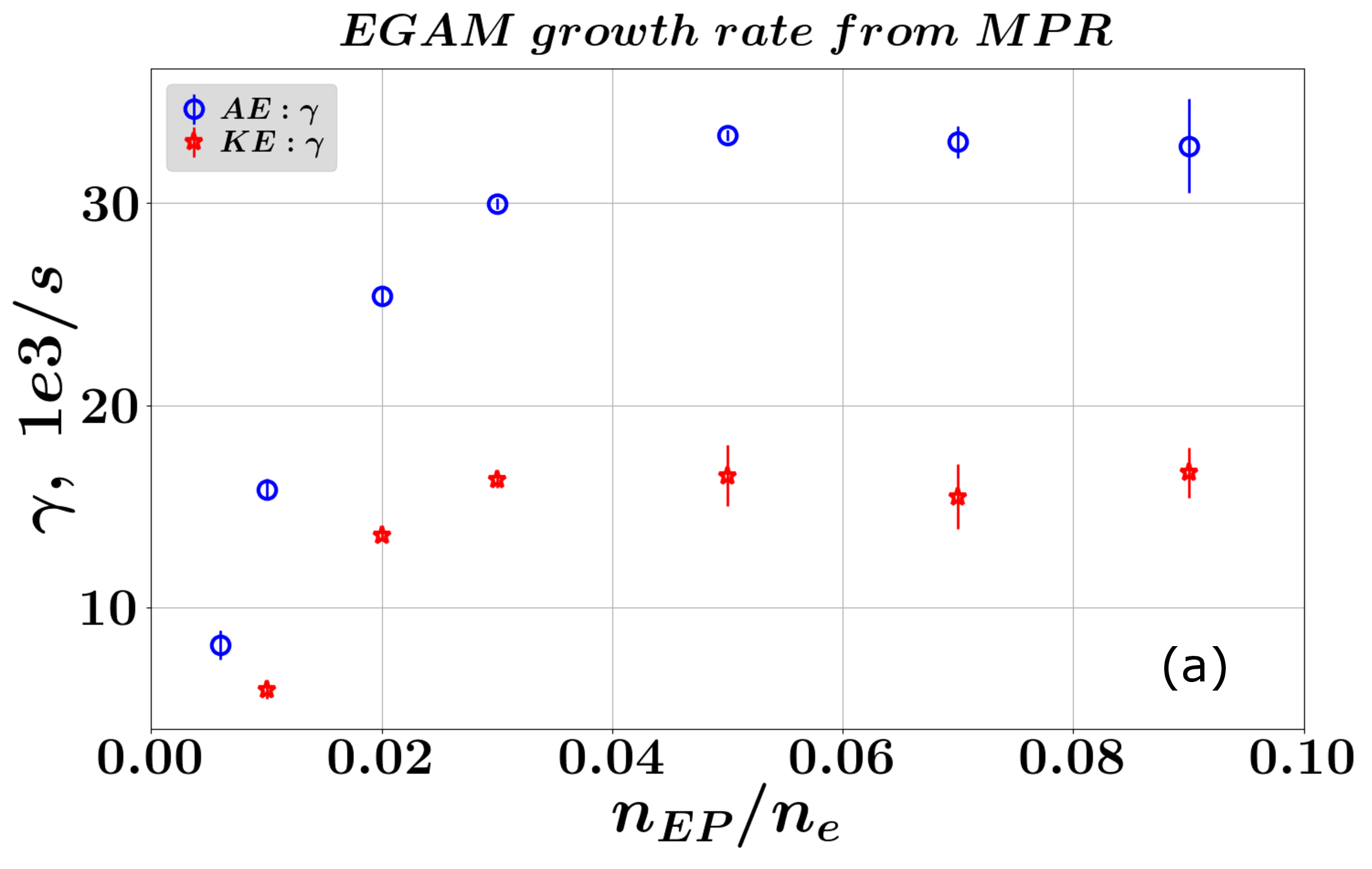}{fig:lin-AE-KE-a}
{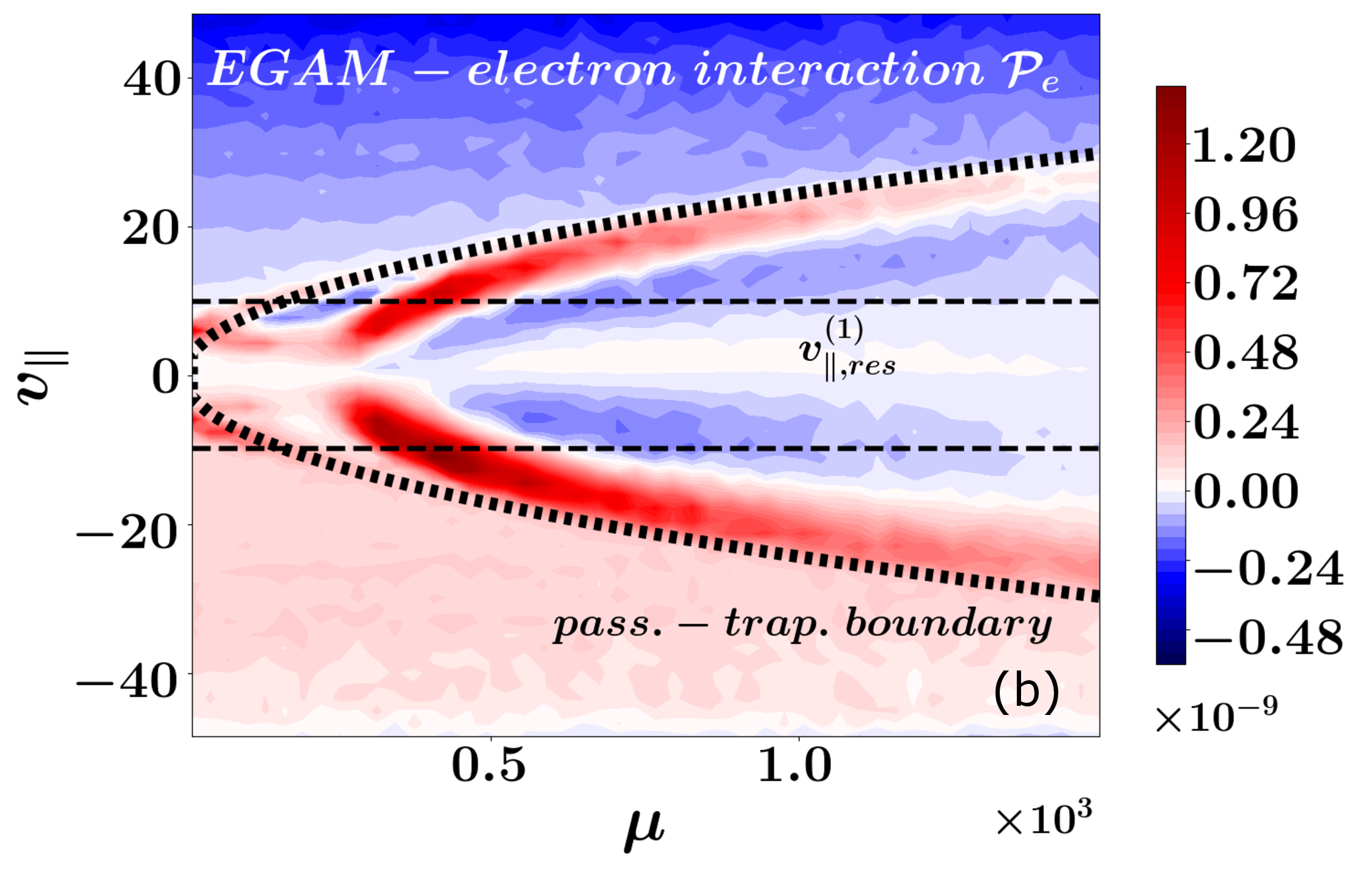}{fig:lin-AE-KE-b}
{Influence of electron dynamics on the EGAM linear growth rate for different EP concentrations (Fig.~\ref{fig:lin-AE-KE-a}). "AE" indicates ES simulations with adiabatic electrons, "KE" indicates EM simulations with drift-kinetic electrons.
Localisation of the EGAM interaction with the electrons (case with $n_{EP}/n_e = 0.01$, Fig.~\ref{fig:lin-AE-KE-b}).
The black cone indicates the approximate position of the boundary between the passing and trapped electrons of Eq.~\ref{eq:p-tr-boundary}. 
The horizontal lines are the analytical estimation of the EGAM-plasma resonances, calculated in Eq.~\ref{eq:vresm} (for $m = 1$). 
Here, all simulations have been performed linearly with $v_{\parallel, EP} = 8.0, T_{EP} = 1.0$.}
{fig:lin-AE-KE}

In this section, the effect of the drift-kinetic electrons on the EGAM dynamics is examined, by considering electrons with a realistic mass $m_d/m_e = 3672$.
In Fig.~\ref{fig:lin-AE-KE-a}, one can observe the EGAM growth rate significantly decreases when drift-kinetic electrons are included.  
The interaction between the EGAMs and the electrons is investigated by the wave-particle energy transfer signal in the velocity space. 
In Fig.~\ref{fig:lin-AE-KE-b}, one can see the EGAM-electron resonances, where the horizontal dashed lines indicate the analytical estimation of the resonance positions (Eq.~\ref{eq:vresm}). 
To guide the eye, the passing-trapped boundary for electrons is shown as well:
\aeqn
v^{p-tr}_\parallel = \sqrt{2\epsilon \mu},\label{eq:p-tr-boundary}
\eeqn
where $\epsilon$ is an inverse aspect ratio, the parallel velocity $v_{\parallel}$ is normalized to the sound speed $c_s$, and the magnetic moment $\mu$ is normalised to $m_d c_s^2/(2 B_0)$.
The energy exchange between the EGAM and trapped electrons occurs inside of this cone. 
The fact that the energy transfer resonances are close to the passing-trapped boundary, means that the mode interacts mainly with the barely trapped electrons with a bounce frequency close to that of the mode.

\yFigTwo
{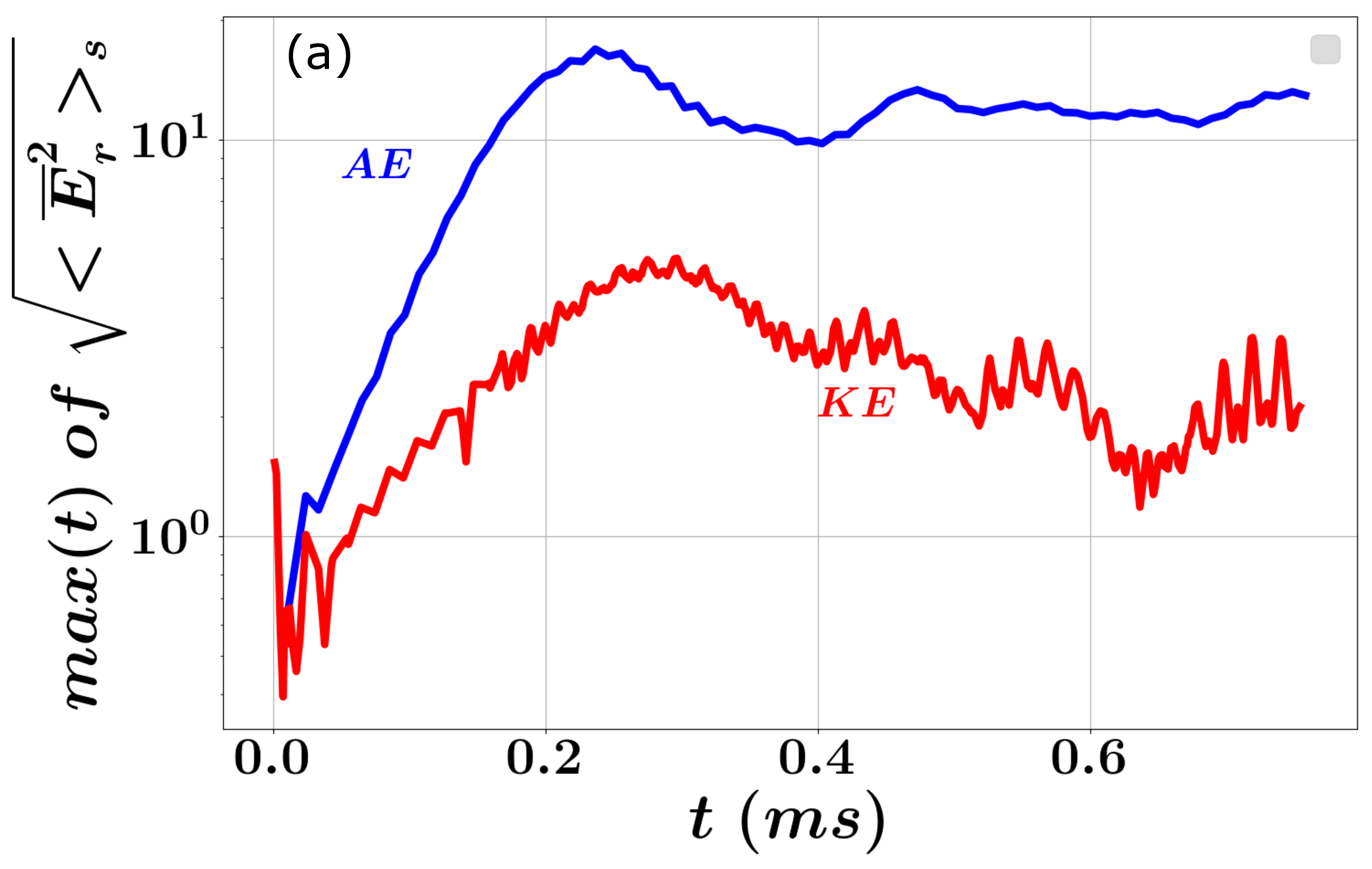}{fig:NL-KE-sat-T-a}
{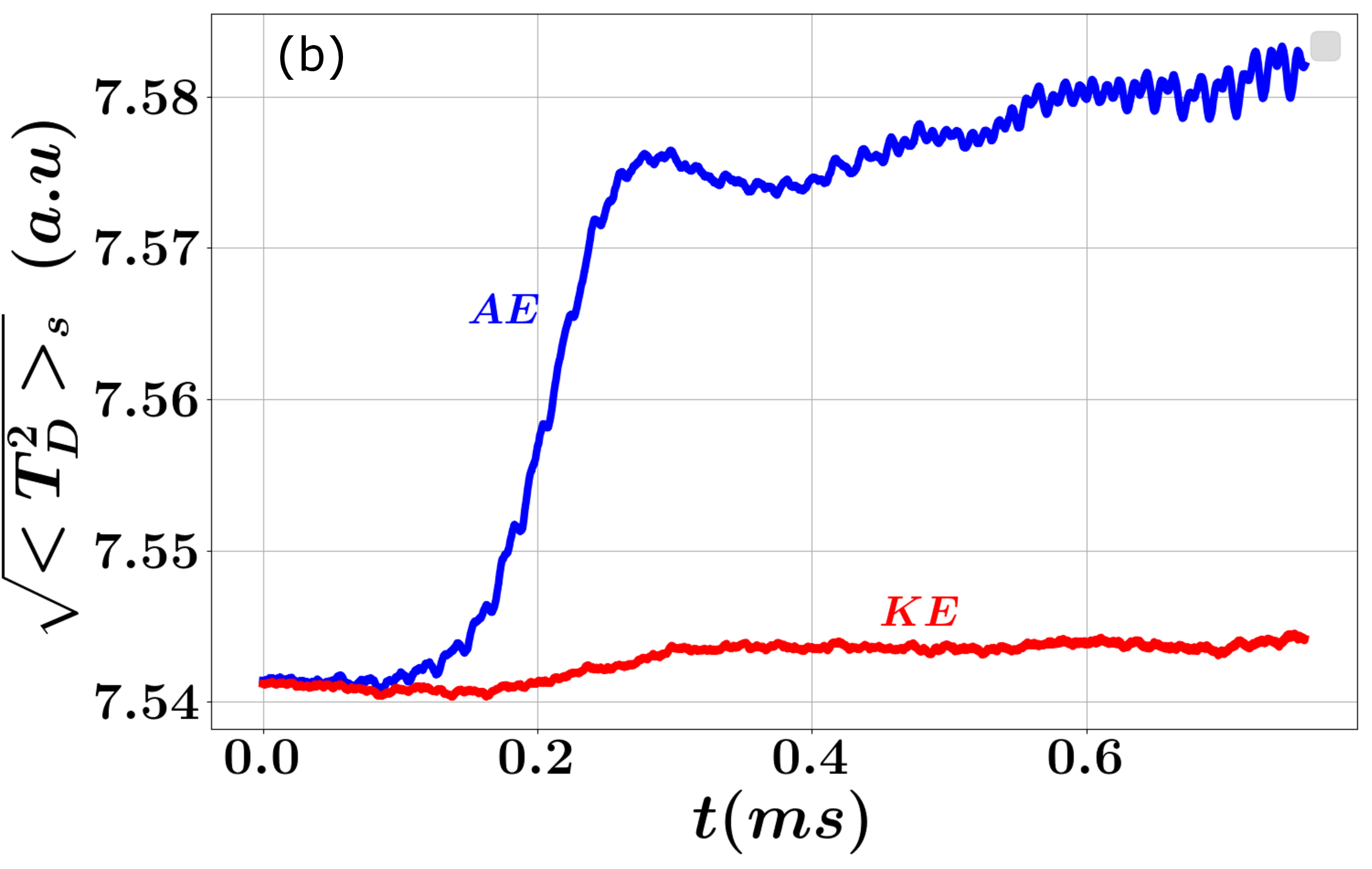}{fig:NL-KE-sat-T-b}
{Comparison between an electrostatic NL simulation assuming adiabatic electrons and an electromagnetic NL simulation with the drift-kinetic electrons. In both cases $n_{EP}/n_e = 0.01, v_{\parallel, EP} = 8.0, T_{EP} = 1.0$ are taken.
EGAM saturation levels (Fig.~\ref{fig:NL-KE-sat-T-a}).
Evolution of the thermal deuterium temperature (Fig.~\ref{fig:NL-KE-sat-T-b}).}
{fig:NL-KE-sat-T}

\yFigTwo
{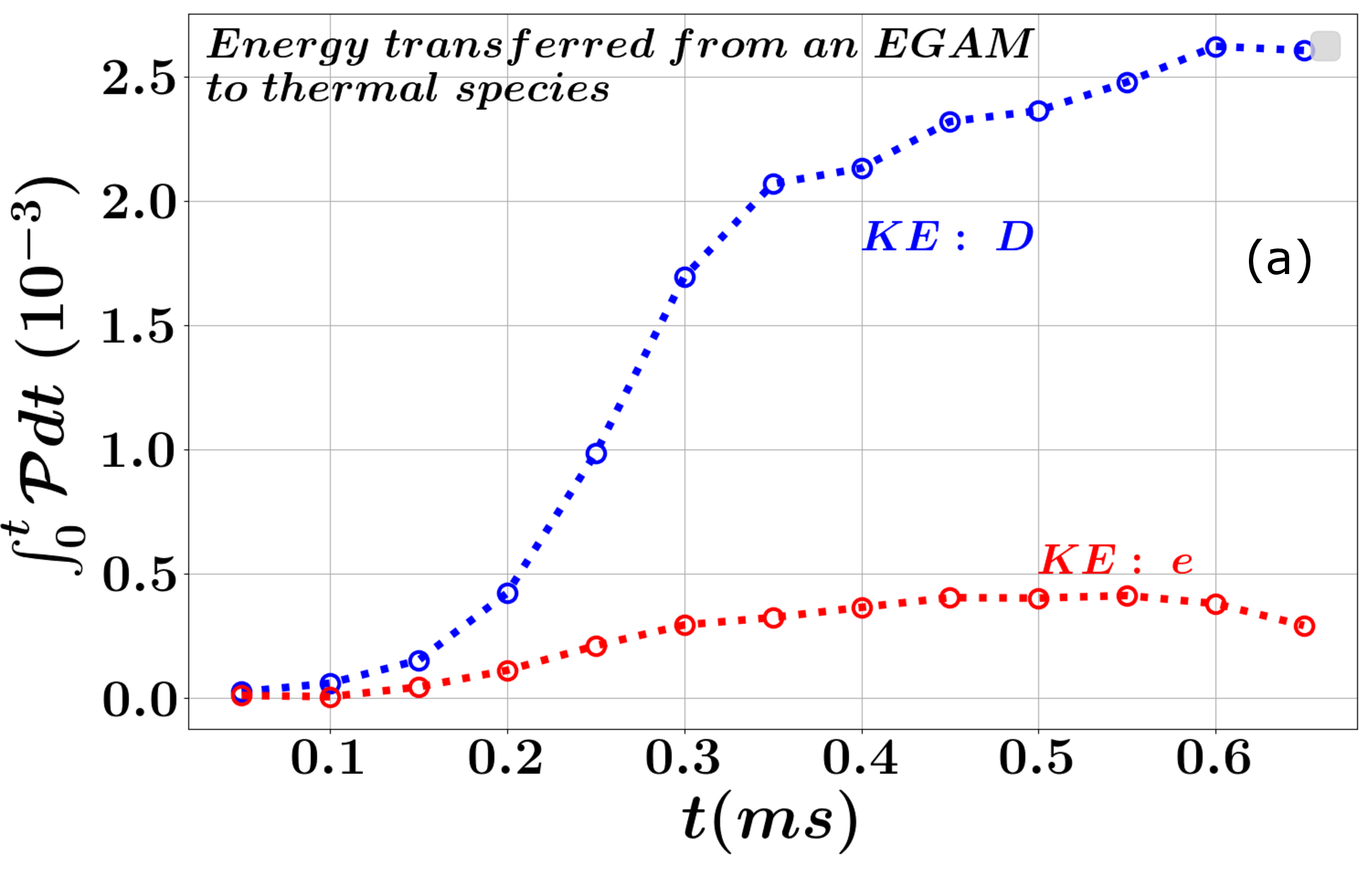}{fig:NL-energy-transfer-KE-a}
{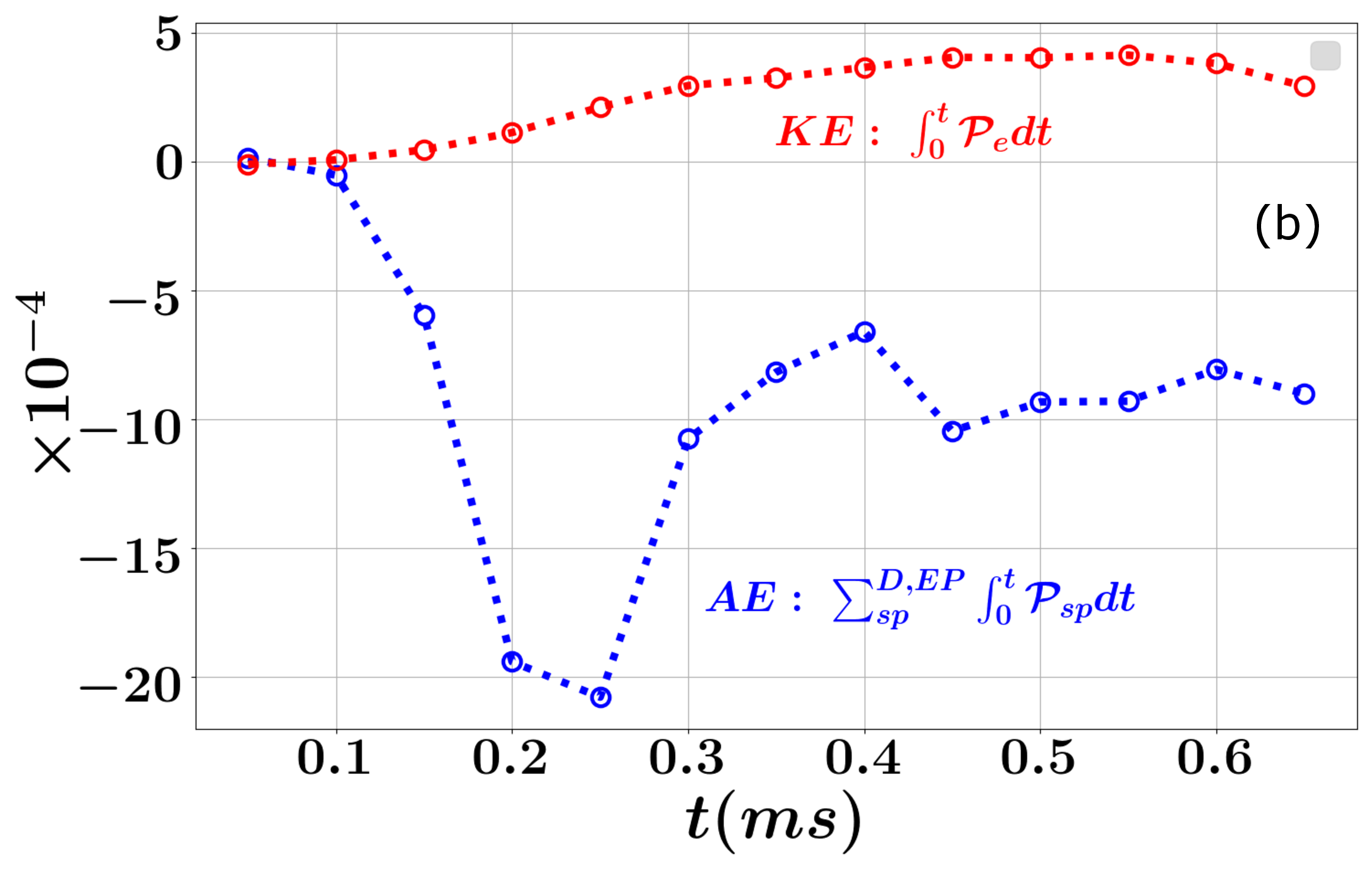}{fig:NL-energy-transfer-KE-b}
{Nonlinear ES and EM cases with $n_{EP}/n_e = 0.01, v_{\parallel, EP} = 8.0, T_{EP} = 1.0$ are considered.
Energy transferred from the mode to the thermal species (Fig.~\ref{fig:NL-energy-transfer-KE-a}, blue line - thermal ions, red line - electrons) in the nonlinear EM case with drift-kinetic electrons. 
Total energy balance between the EGAM and all kinetic species in the ES case (Fig.~\ref{fig:NL-energy-transfer-KE-b}, blue line), and the electron contribution in the EM simulation (red line).}
{fig:NL-energy-transfer-KE}

The inclusion of the electron dynamics influences the nonlinear EGAM behaviour as well. 
Consistently with the decrease of the EGAM linear growth rate, the mode saturation levels are also reduced in NL simulations 
(Fig.~\ref{fig:NL-KE-sat-T-a}) that leads to the lowering of the bulk plasma heating by the mode. This reduction is clearly observed in the time evolution of the thermal ion temperature (Fig.~\ref{fig:NL-energy-transfer-KE-b}). 
However, even in the case with the drift-kinetic electrons, the EGAM transfers most of its energy to the thermal ions, and not to the electrons, that can be seen in Fig.~\ref{fig:NL-energy-transfer-KE-a}.
In other words, it is the ion plasma heating, which mainly benefits due to the presence of the EGAM dynamics.
Having said that, we should emphasize, that the electron contribution is still not negligible in comparison to the total energy, obtained by the EGAM due to the wave-particle interaction with all kinetic species in the considered plasma system.
More precisely, in Fig.~\ref{fig:NL-energy-transfer-KE-b}, one can see that the amount of energy that the mode stores in the ES simulation (blue line) and the energy that the mode transfers to the electrons (red line) in the EM simulation are of the same order. 
Negative blue line indicates the fact that the EGAM interaction with the EPs and the thermal ions result in the increase of the mode energy.
The red curve is positive that demonstrates the energy flow from the mode to the electrons.
Although this flow is significantly smaller than that to the thermal ions, it is still comparable to the total mode energy. 
Therefore, the electron dynamics can noticeably decrease the amount of energy, stored in the mode, and in such a way reduce the plasma heating by the EGAM.

\section{Conclusions}
Energetic-particle-driven geodesic acoustic modes (EGAMs) have a significant influence on the EP dynamics and on the plasma confinement.
These modes provide an additional mechanism of the energy exchange between the energetic particles and the thermal plasma, enhancing direct heating of the bulk ions.
The significant progress done in the last decade in studying the EGAM nonlinear physics has been expanded here, by including kinetic electrons and the geometrical effects of a realistic magnetic shape in the GK simulations. 

In this work, linear and nonlinear global GK simulations of the EGAMs in an ASDEX Upgrade discharge have been presented. 
While a corresponding tokamak configuration and thermal species profiles have been used here, the EPs have been represented by two shifted Maxwellians in velocity space.
Using such a model, an EGAM relative up-chirping close to the experimental one has been obtained. It has demonstrated that the GK code is able to handle zonal structures and anisotropic distribution functions properly in experimentally relevant cases which is a crucial ingredient before going onto more complex plasma systems.
On the other hand, to compare absolute mode chirping and characteristic time scales of the mode frequency evolution, it is necessary to use an EP distribution function as close as possible to the experimental one.

By varying the EP parameters, general correlation between the mode level and the EP-thermal plasma energy flow has been revealed. 
Apart from that, it has been emphasized that the plasma heating by EGAMs benefits from the presence of high order mode-particle resonances which are often responsible for the EGAM-thermal species interaction.
More precisely, it has been shown that having the same mode levels, 
one can achieve a higher energy exchange between energetic and thermal species through an EGAM by adjusting EP parameters.
Finally, it was indicated that although the EGAM transfers most of its energy to the thermal ions, and not to the electrons, the electron dynamics might significantly reduce the plasma heating by EGAMs by lowering the mode amplitude.

Since the nonlinear GK simulations with drift-kinetic electrons have shown a significant influence of the electron dynamics on the EGAM behaviour, it would be reasonable to study effect of the kinetic electrons on the mode chirping.
The results obtained here will be useful in a future work to perform investigation of the EGAM influence on the turbulent dynamics in the ASDEX Upgrade, that will be a next step in the code validation and an interesting example of a complex plasma system.

\begin{acknowledgments}
The authors would like to acknowledge stimulating discussions with the Multi-scale Energetic particle Transport in fusion devices (MET) team, and especially with Prof. Fulvio Zonca.
This work has been carried out within the framework of the EUROfusion Consortium and has received funding from the Euratom research and training program 2014-2018 and 2019-2020 under grant agreement N$^o$ 633053. 
The views and opinions expressed herein do not necessarily reflect those of the European Commission.
Simulations, presented in this work, have been performed on the CINECA Marconi supercomputer within the framework of the OrbZONE and ORBFAST projects.
\end{acknowledgments}

\begin{appendices}
\section{Convergence tests}
\label{app:conv-tests}
\numberwithin{equation}{section}
\setcounter{equation}{0}

\begin{figure}[!b]
\subfloat{\includegraphics
	[width=0.50\textwidth]
	{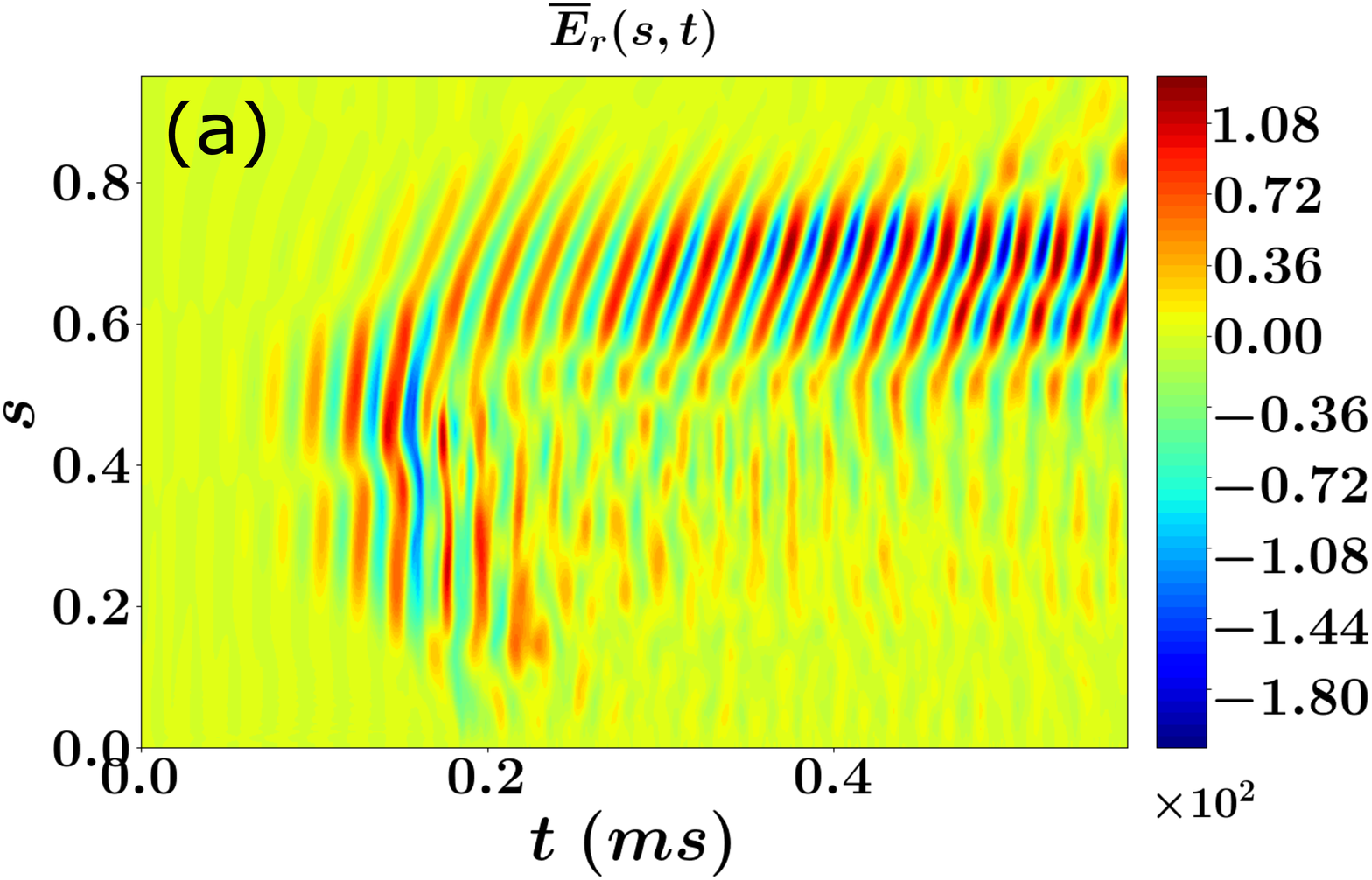}
	\label{fig:conv-chirp-st-a}
}
\subfloat{\includegraphics
	[width=0.50\textwidth]
	{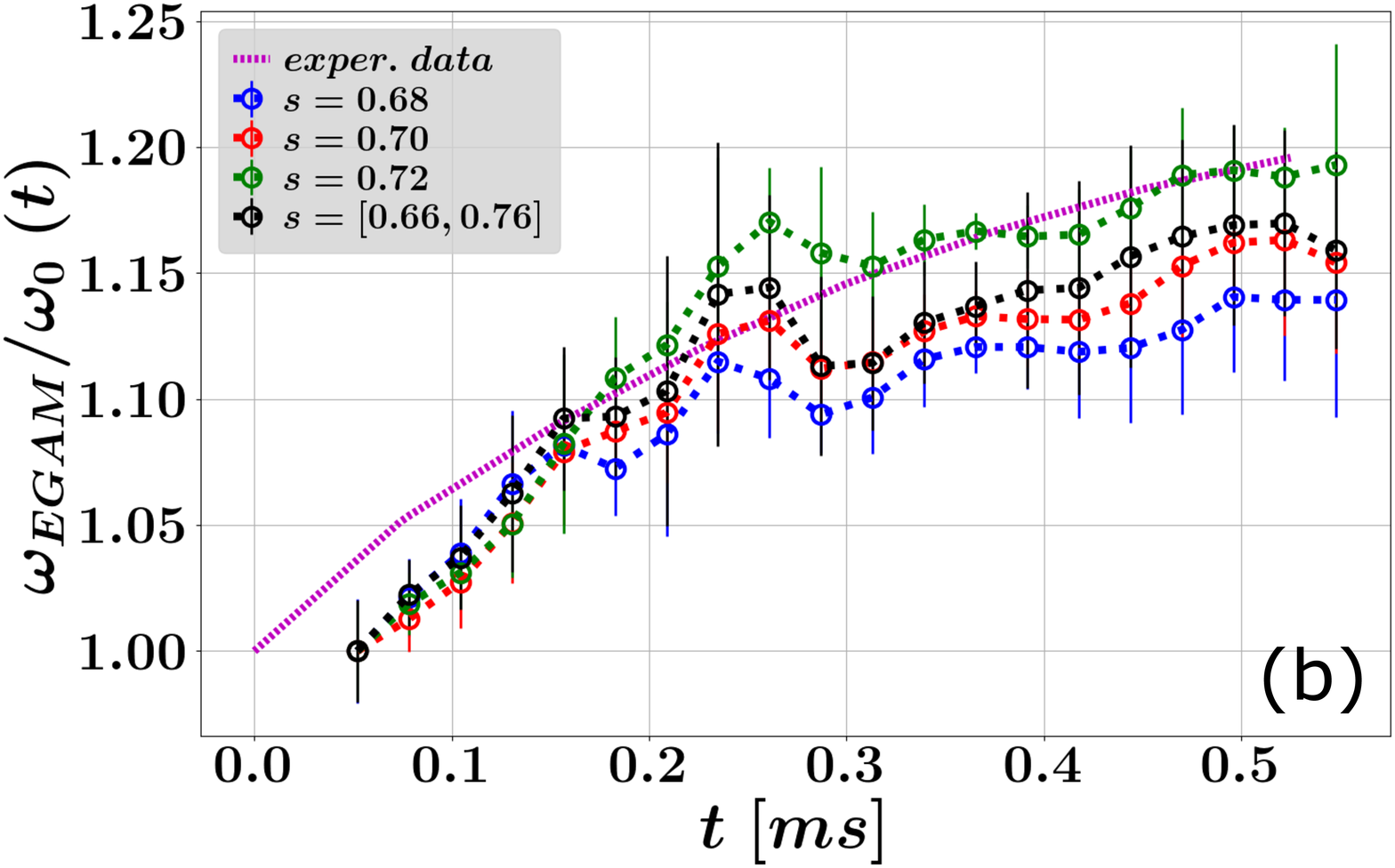}
	\label{fig:conv-chirp-st-b}
}\\
\subfloat{\includegraphics
	[width=0.50\textwidth]
	{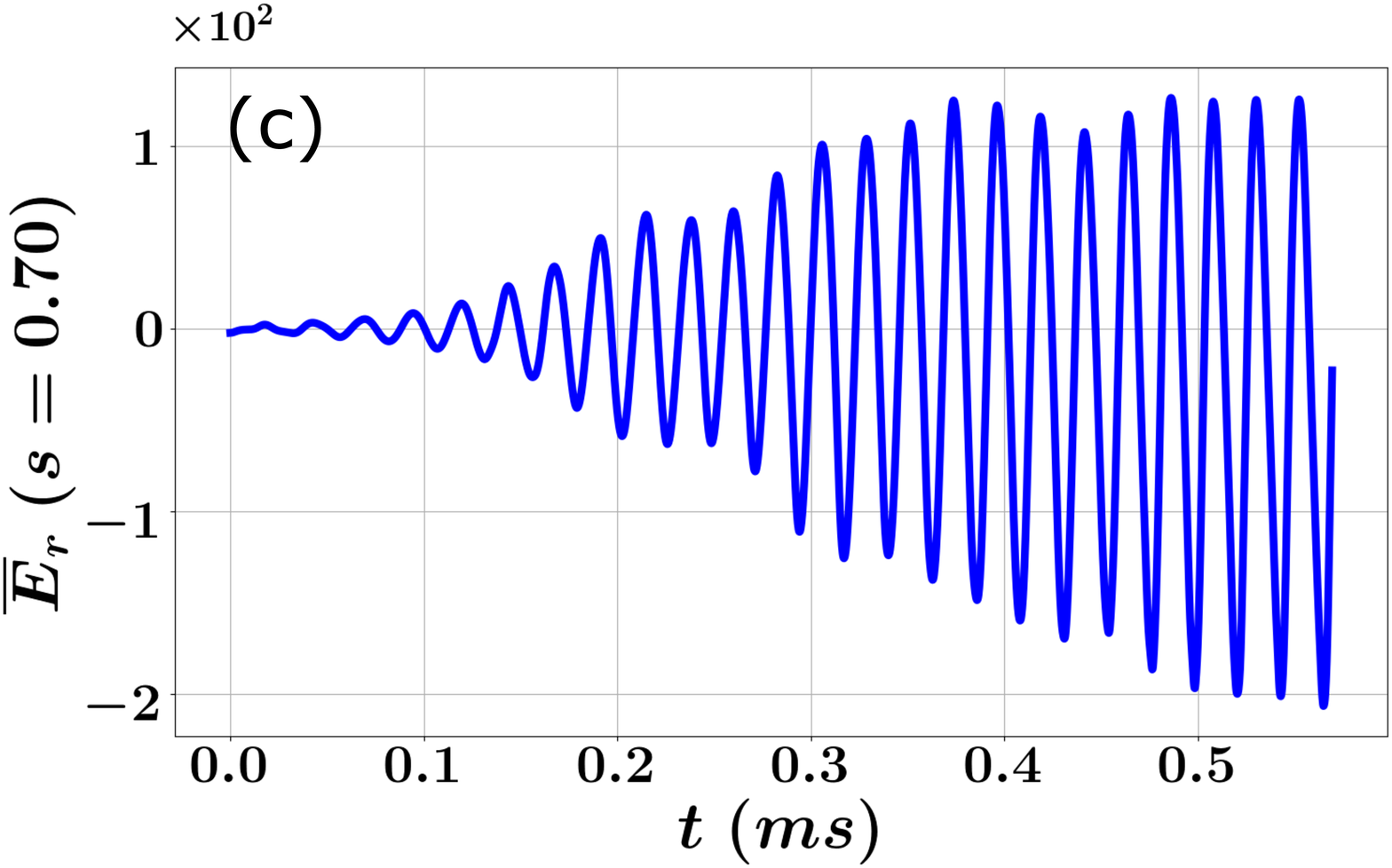}
	\label{fig:conv-chirp-st-c}
}
\subfloat{\includegraphics
	[width=0.50\textwidth]
	{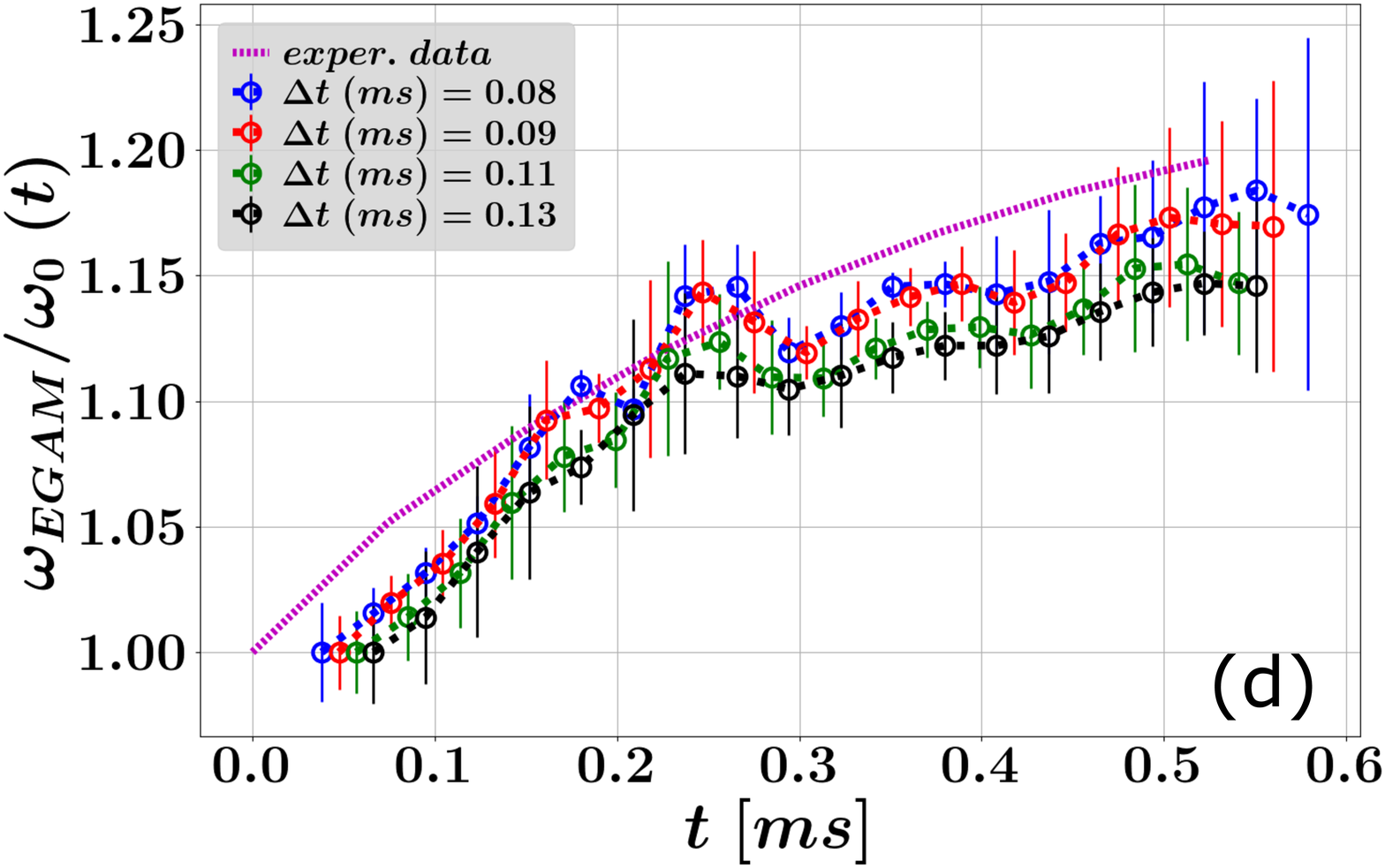}
	\label{fig:conv-chirp-st-d}
}
\caption{
\label{fig:conv-chirp-st}
EGAM radial structure and its time evolution at $s = 0.70$ (Figs.~\ref{fig:conv-chirp-st-a} and ~\ref{fig:conv-chirp-st-c} respectively).
Fig.~\ref{fig:conv-chirp-st-b}: EGAM chirping at different radial points, where every curve is normalized to its own initial frequency. The black markers indicate frequency evolution of an averaged zonal electric field in $s=[0.66, 0.76]$ interval. Here, $\Delta t = 0.11$.
Fig.~\ref{fig:conv-chirp-st-d}: frequency evolution at $s = 0.70$ with different time interval lengths $\Delta t$ for the nonlinear fitting.
}
\end{figure}

Convergence tests of some simulations discussed above are presented here. 
The main attention is dedicated to the mode saturation levels and the EGAM chirping.
First of all, the mode frequency evolution considered in Sec.~\ref{sec:chirp} is analysed. 
That computation has been done with $dt = 20, n_s = 256, N_d = N_{EP} = 1.2\cdot 10^8$, which has a finer space grid (and, respectively, a higher number of markers) than a typical electrostatic simulation described in Sec.~\ref{sec:conf}. 
The radial structure of the mode (of the zonal electric field) is shown in Fig.~\ref{fig:conv-chirp-st-a}, where one can see that the mode is localised around a radial point $s = 0.70$.
In Fig.~\ref{fig:conv-chirp-st-b}, the EGAM chirping is investigated at different radial positions near the mode localisation. 
There is some dependence of the frequency evolution on the radial position. 
The point $s = 0.70$ is chosen since the chirping there is closer to the frequency evolution of a signal averaged around the mode localisation (black markers in Fig.~\ref{fig:conv-chirp-st-b}). 
The mode frequency has been calculated by using a nonlinear fitting of the zonal electric field (Fig.~\ref{fig:conv-chirp-st-c}) to a test function
\aeqn
\sim \cos(\omega t) \exp(\gamma t)\label{eq:nonlinear-fitting}
\eeqn
Time evolution of the signal is split on a set of time intervals of a length $\Delta t$ with several EGAM oscillations.
By performing the nonlinear fitting in every time domain, one can find a mode frequency there from Eq.~\ref{eq:nonlinear-fitting}.
In Fig.~\ref{fig:conv-chirp-st-d}, one can see how the mode chirping depends on the choice of a time interval length $\Delta t$. The case with $\Delta t = 0.11$ is chosen since 
the corresponding chirping has one of the smallest errorbars.

\yFigTwo
{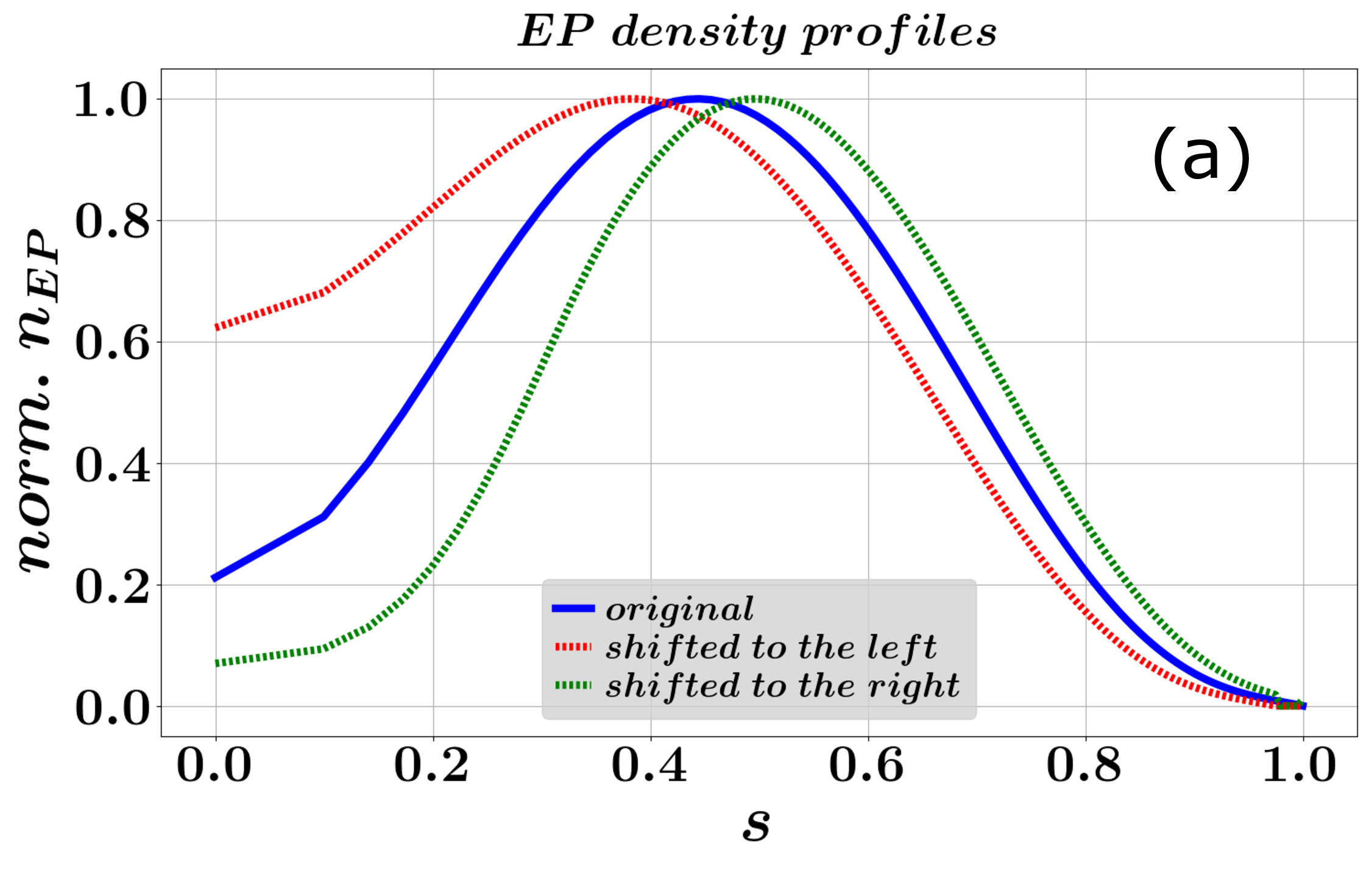}{fig:conv-chirp-profiles-a}
{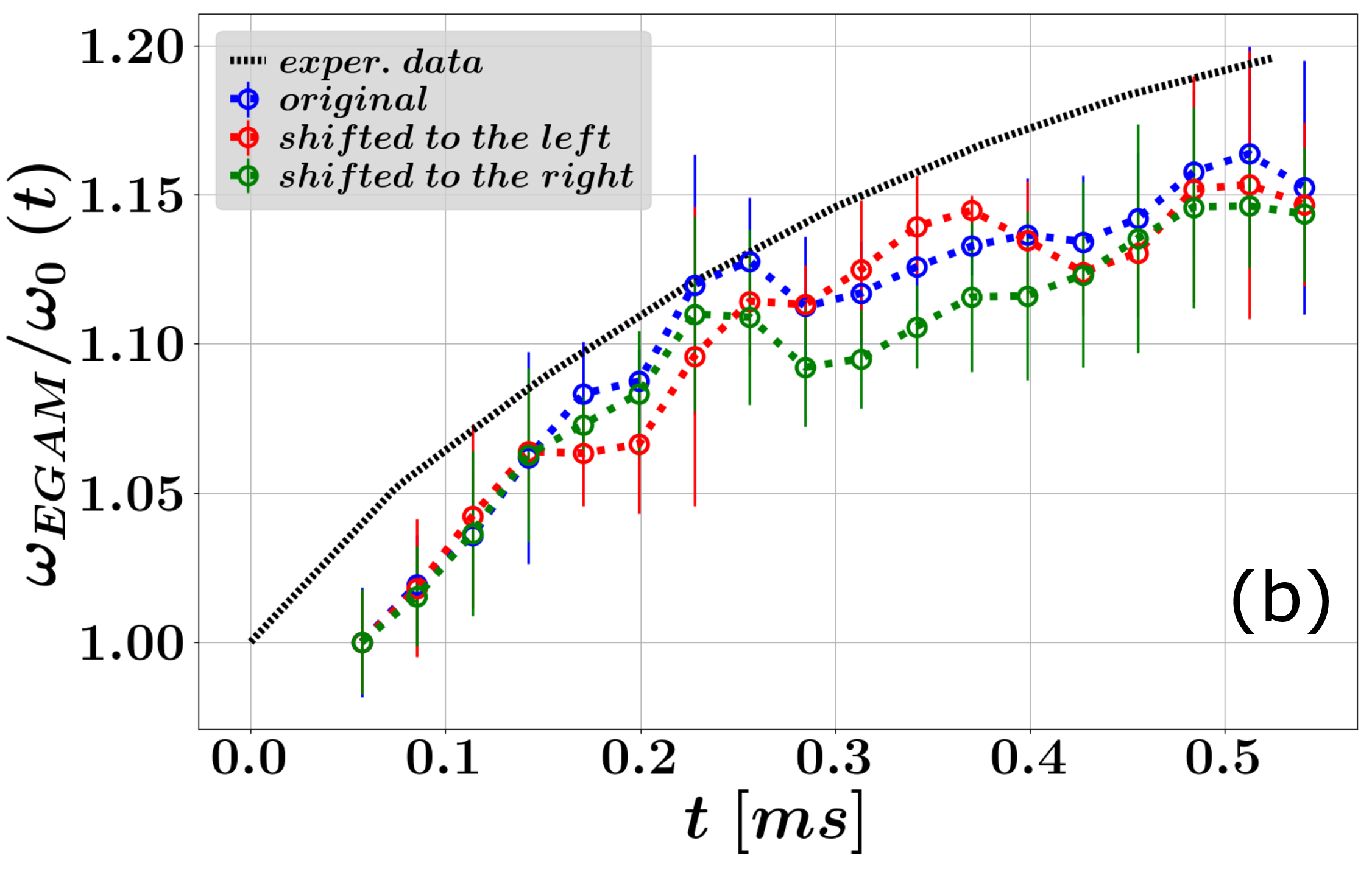}{fig:conv-chirp-profiles-b}
{Variation of the EP density profiles (Fig.~\ref{fig:conv-chirp-profiles-a}). 
Fig.~\ref{fig:conv-chirp-profiles-b}: corresponding mode chirping calculated at $s = 0.70$.
}
{fig:conv-chirp-profiles}

By slightly varying the localisation of the energetic particles (Fig.~\ref{fig:conv-chirp-profiles-a}), we analyse how the mode chirping changes due to the uncertainty in the definition of the EP density profile. 
In Fig.~\ref{fig:conv-chirp-profiles-b}, one can see that the chirping does not disappear and remains more less stable with respect to the indicated density variation.

\begin{figure}[!t]
\subfloat{\includegraphics
	[width=0.50\textwidth]
	{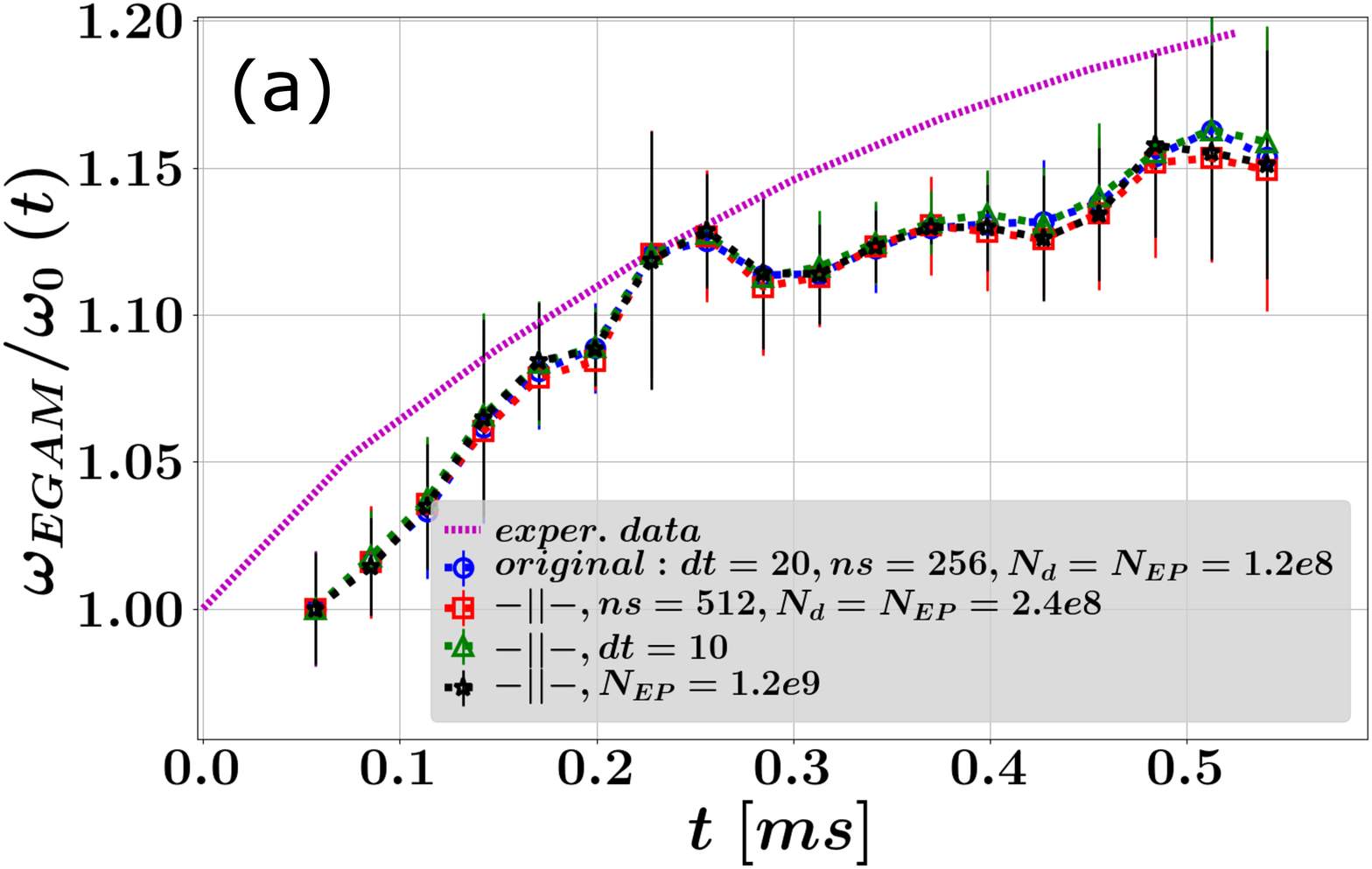}
	\label{fig:es-conv-a}
}\\
\subfloat{\includegraphics
	[width=0.50\textwidth]
	{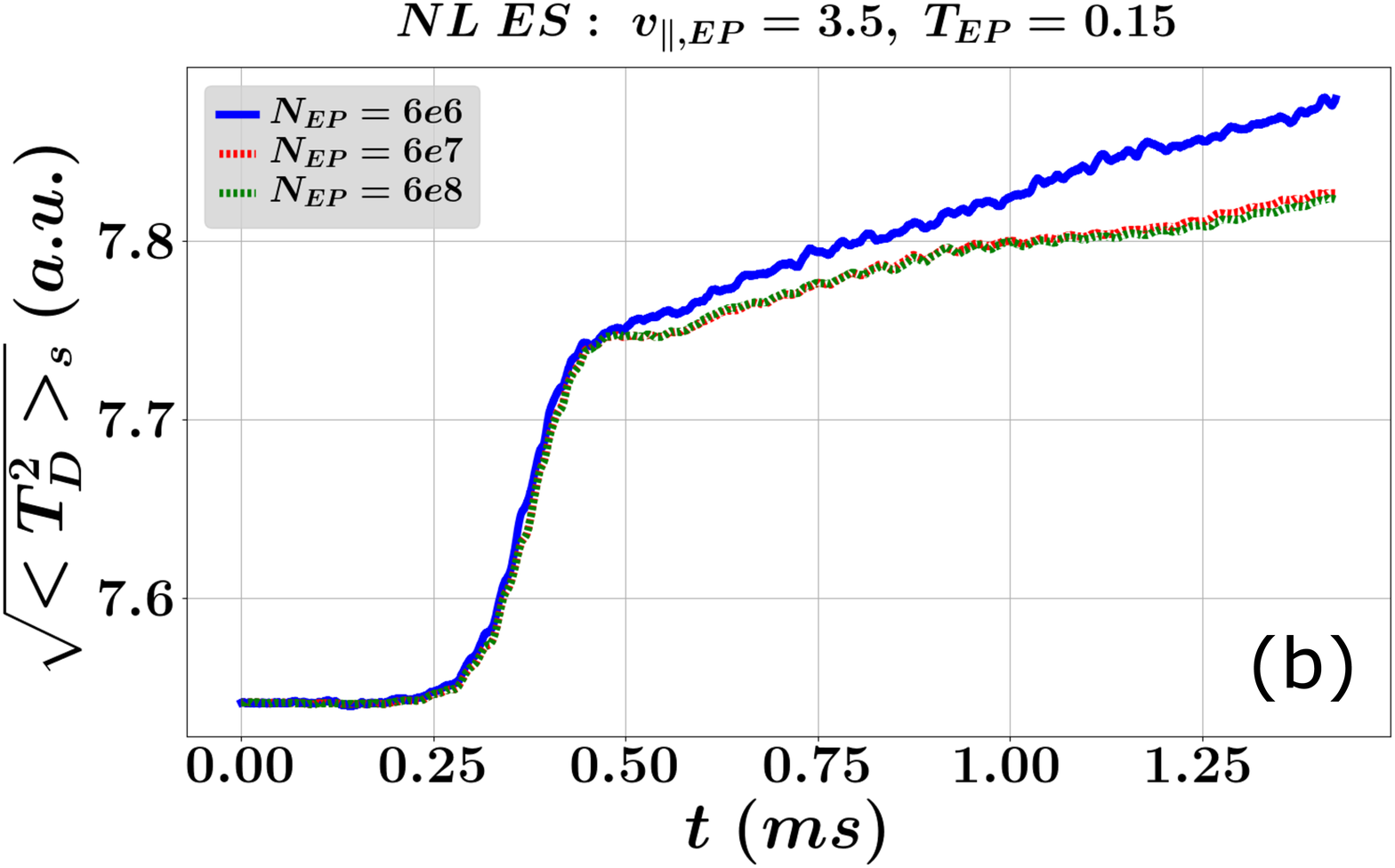}
	\label{fig:es-conv-b}
}\\
\subfloat{\includegraphics
	[width=0.50\textwidth]
	{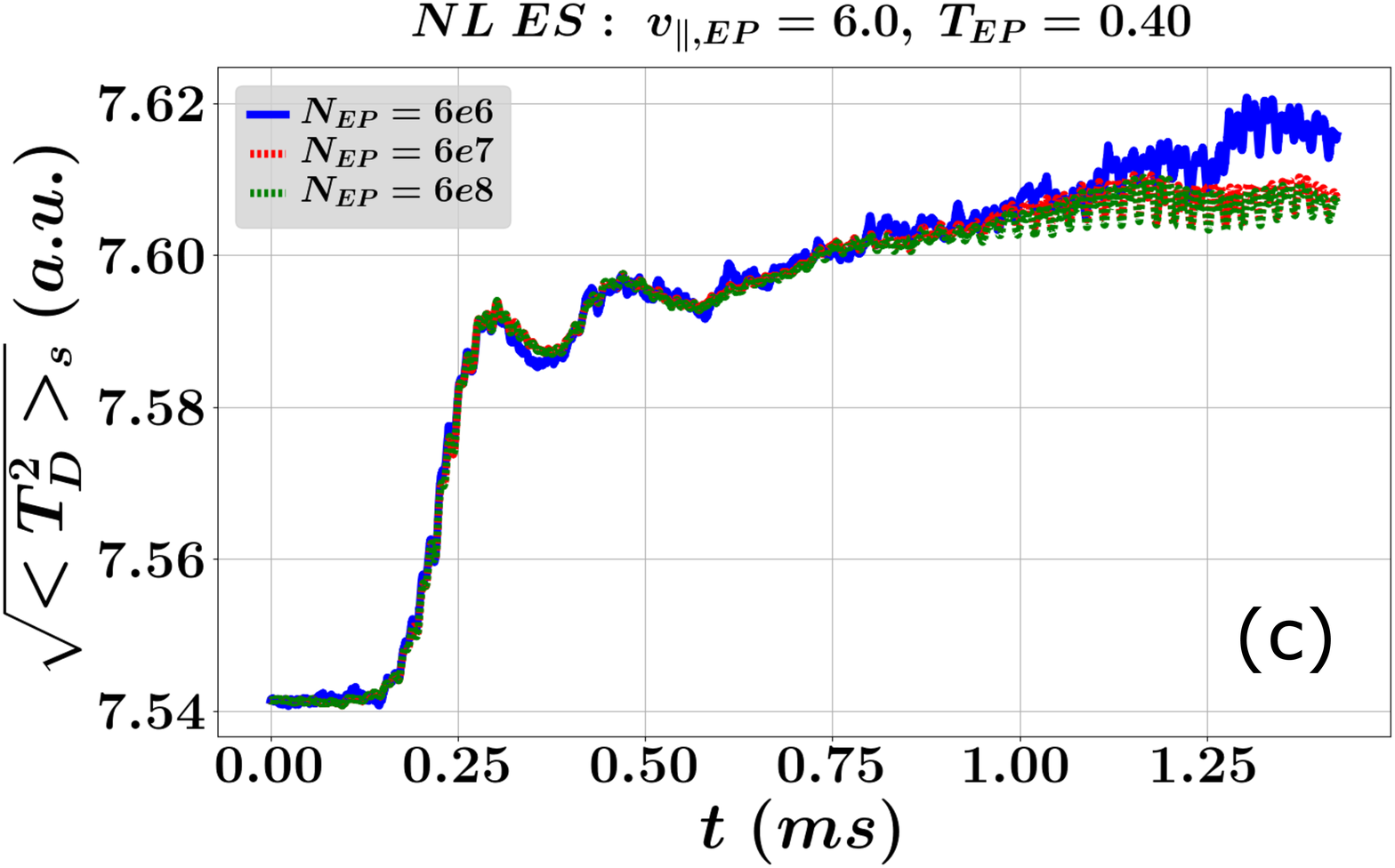}
	\label{fig:es-conv-c}
}
\caption{
EGAM chirping in simulations with different numerical parameters is shown in Fig.~\ref{fig:es-conv-a}.
Thermal plasma heating with different number of EP markers for two cases:
case $v_{\parallel, EP} = 3.5, T_{EP} = 0.15$ (Fig.~\ref{fig:es-conv-b}),
case $v_{\parallel, EP} = 6.0, T_{EP} = 0.40$ (Fig.~\ref{fig:es-conv-c}). 
\label{fig:es-conv}
}
\end{figure}

Convergence tests of some nonlinear ES simulations are shown in 
Fig.~\ref{fig:es-conv}.
Computation of the EGAM chirping with different numerical parameters are presented in Fig.~\ref{fig:es-conv-a}, where one can see that the simulation is converged.
As an additional test, we present here a convergence test (with respect to a number of EP markers $N_{EP}$) of ES nonlinear simulations with different EP velocities (Fig.~\ref{fig:es-conv-b} and Fig.~\ref{fig:es-conv-c}).
In Sec.~\ref{sec:heating}, all simulations have $N_{EP} = 6e7$. 
Increase of $N_{EP}$ by a factor of $10$ does not change $T_D$ growth rate and $T_D$ saturation level.

\yFigTwo
{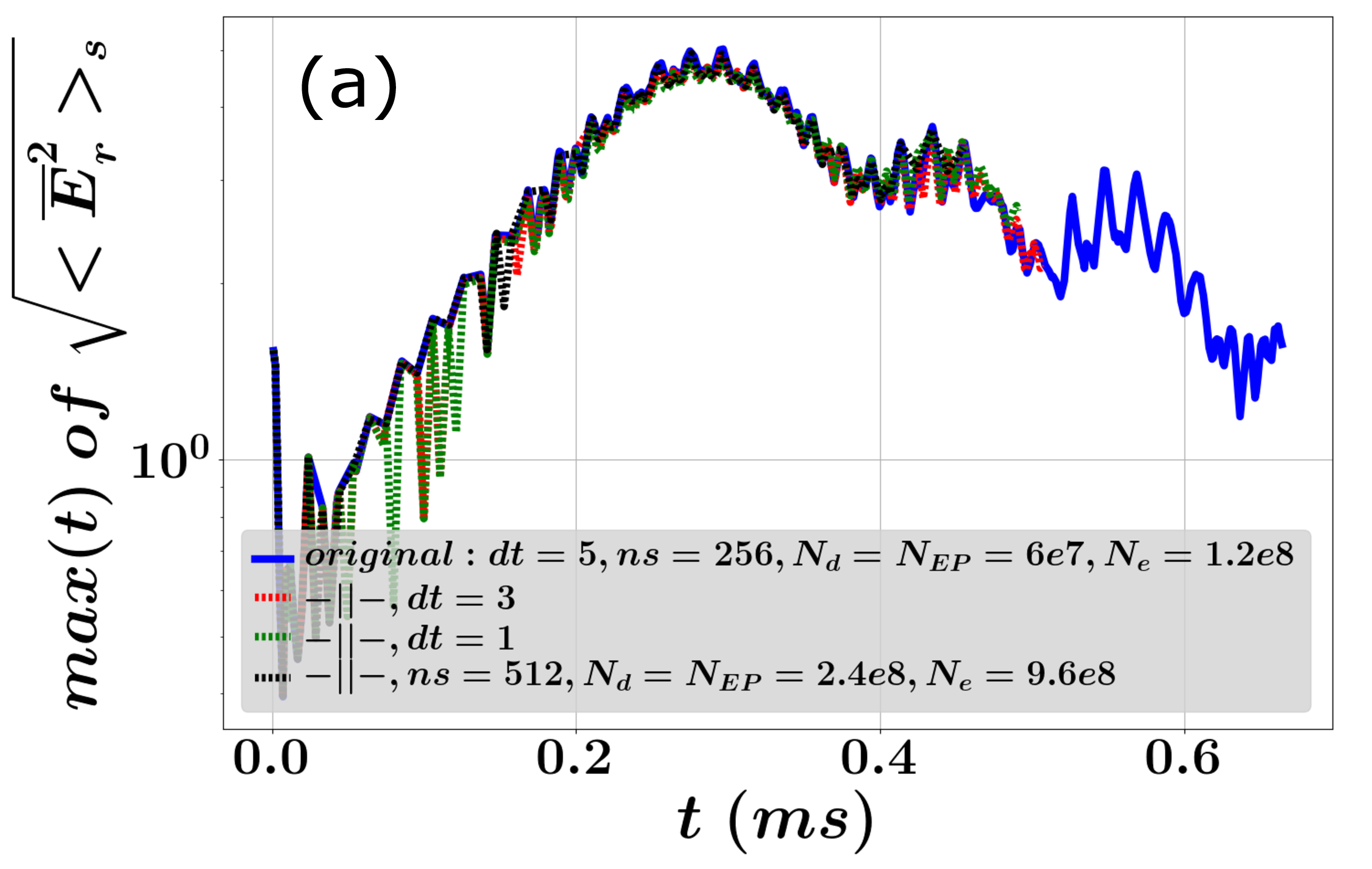}{fig:em-conv-a}
{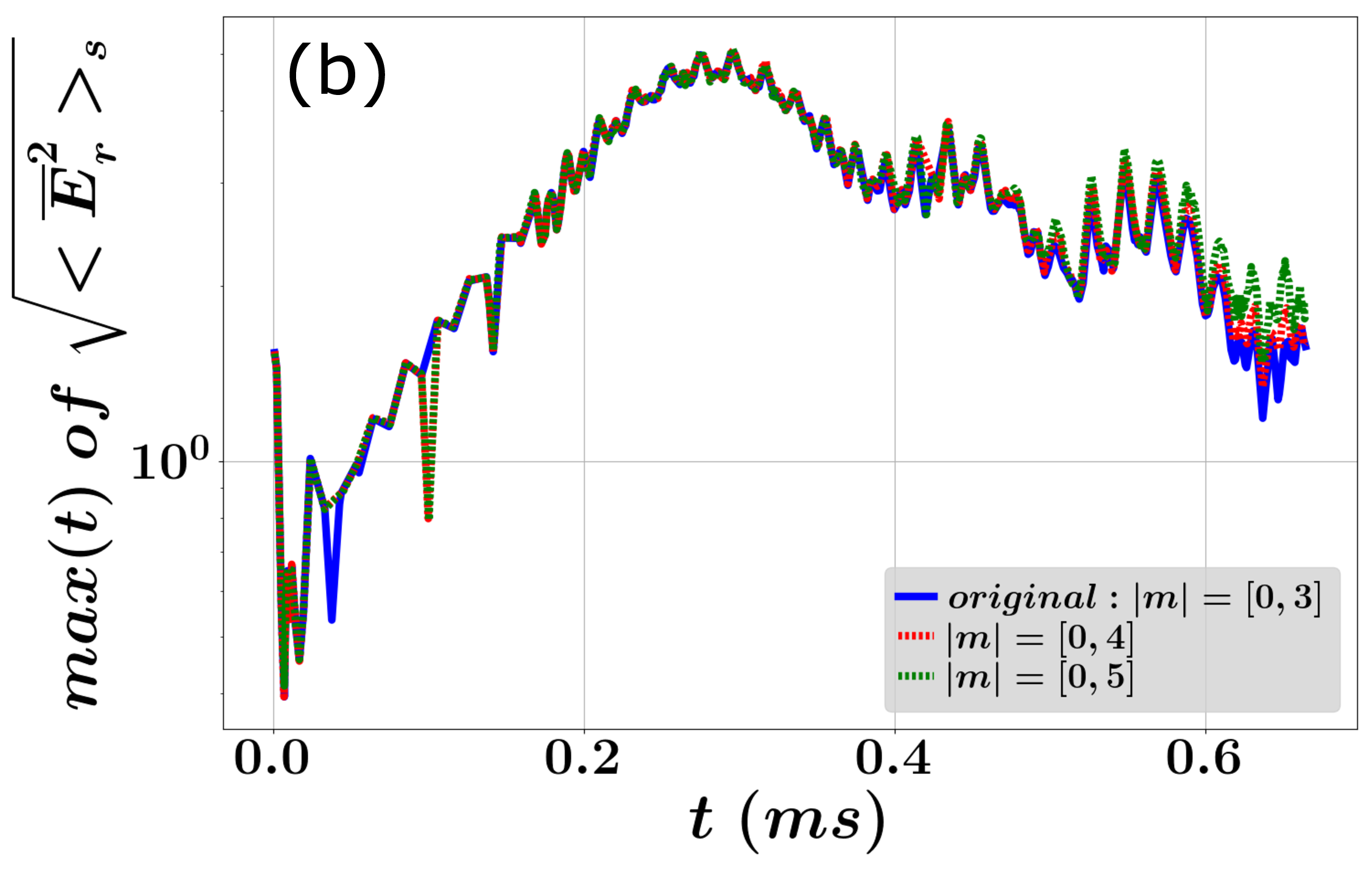}{fig:em-conv-b}
{Convergence tests for the nonlinear EM case with drift-kinetic electrons, described in Sec.~\ref{sec:kin}.}
{fig:em-conv}

Convergence tests of the nonlinear EM case with drift-kinetic electrons, described in Sec.~\ref{sec:kin}, are presented in 
Fig.~\ref{fig:em-conv}, where one can see that the EM simulation is converged.

\section{Detailed energy transfer analysis of EGAM dynamics}
\label{app:details-plasma-heating}
\begin{figure}[!t]
\subfloat{\includegraphics[width=0.50\textwidth]
	{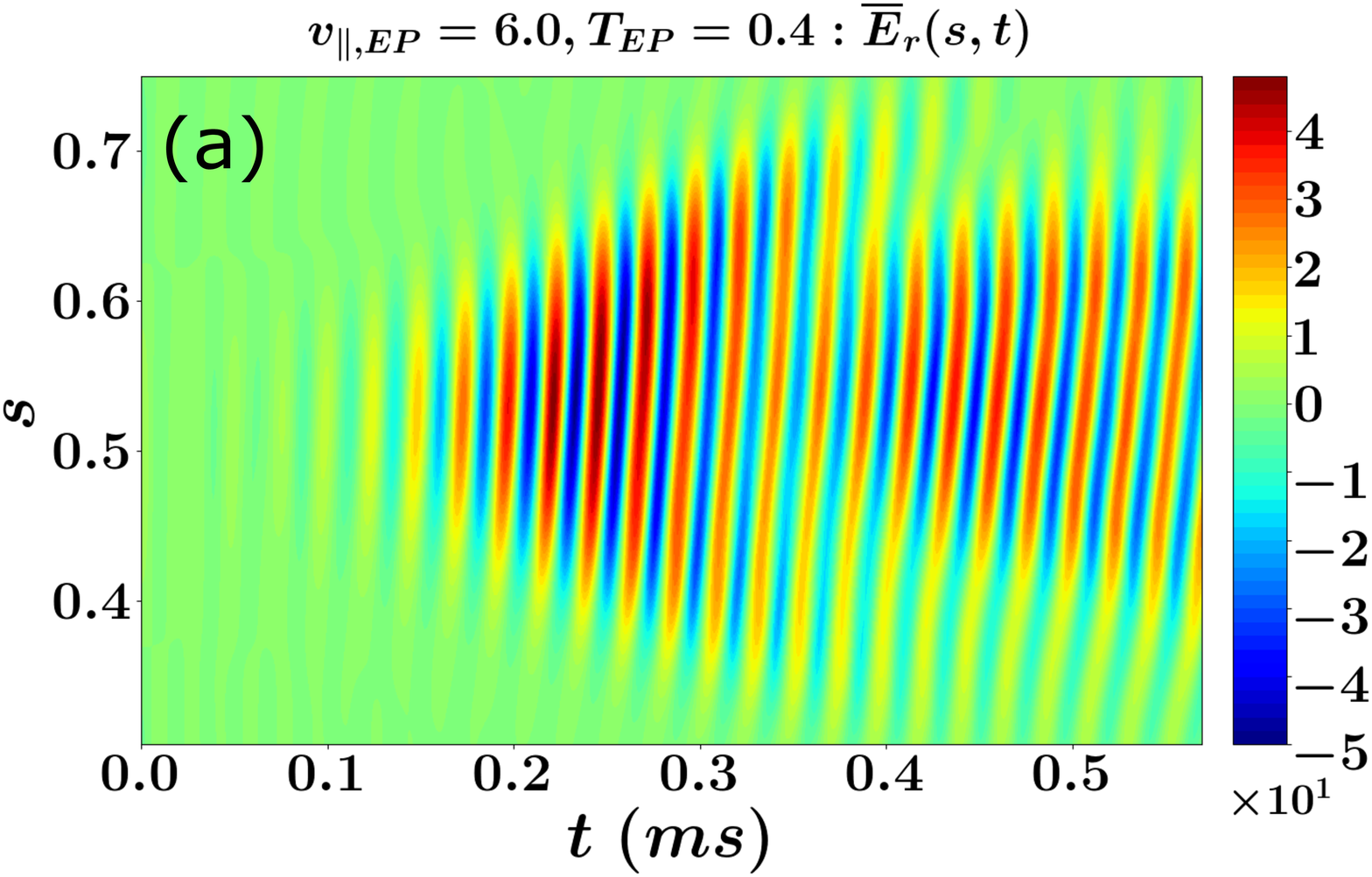}\label{fig:ES-HTA-a}}
\subfloat{\includegraphics[width=0.50\textwidth]
	{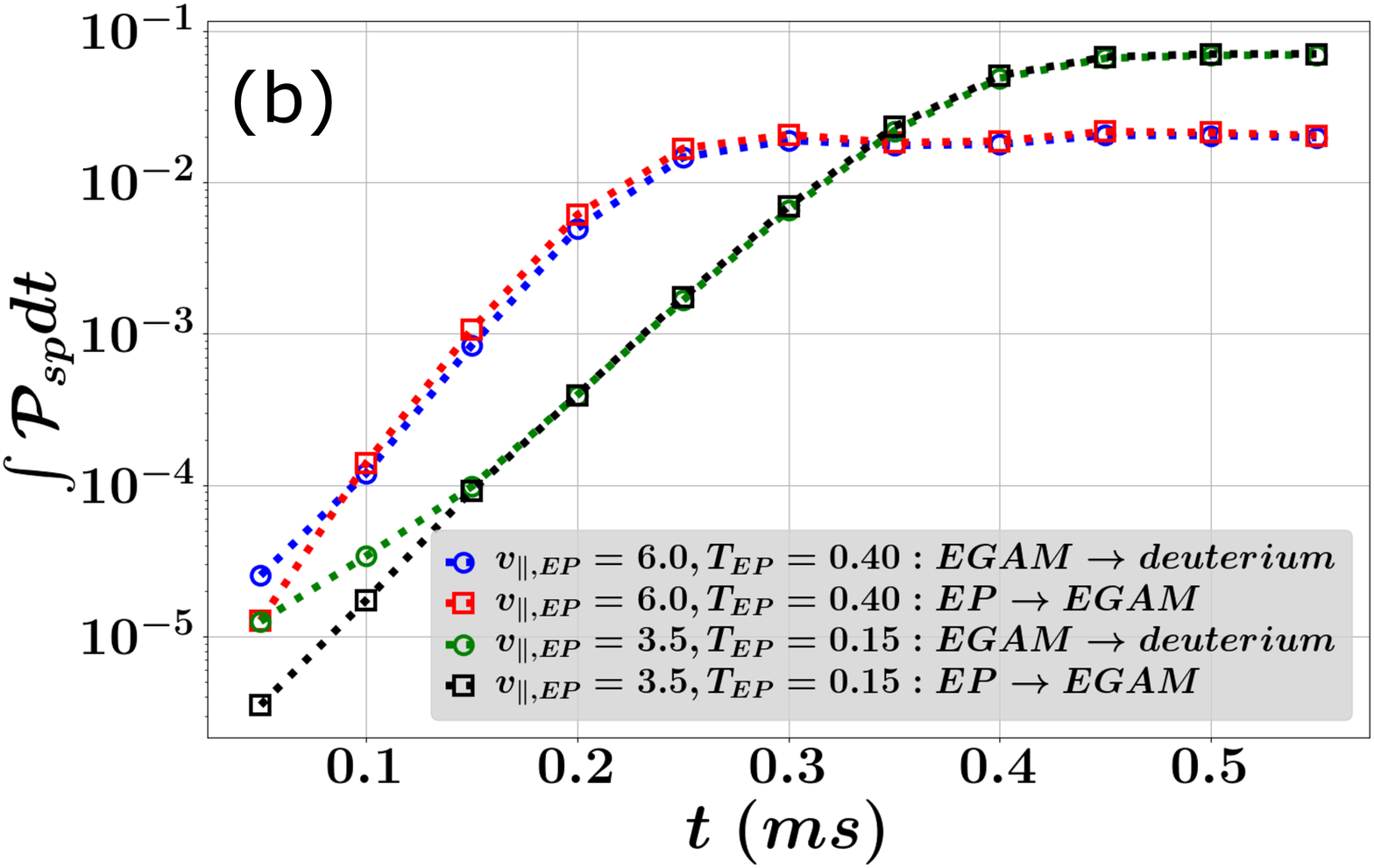}\label{fig:ES-HTA-b}}\\
\subfloat{\includegraphics[width=0.50\textwidth]
	{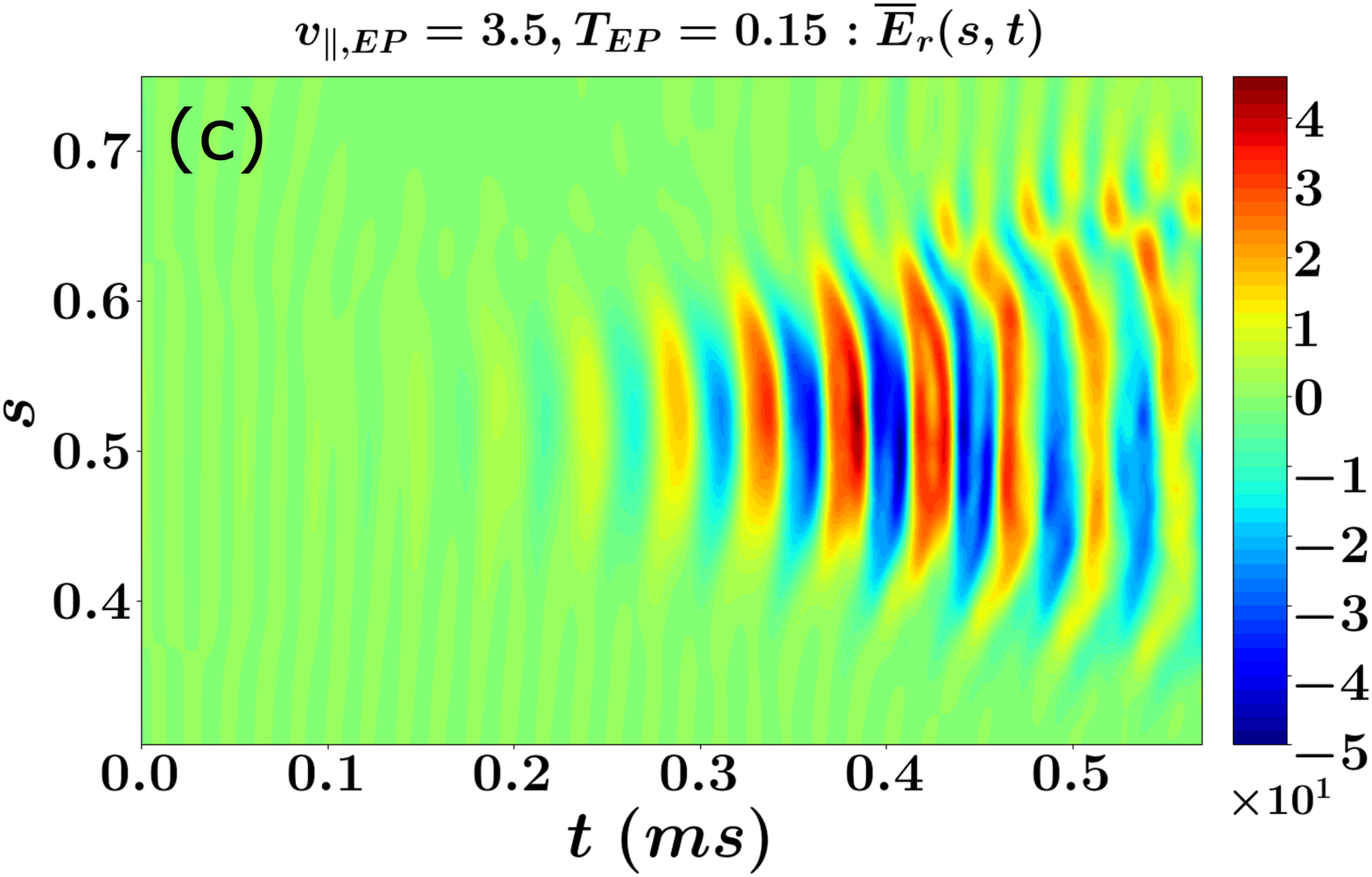}\label{fig:ES-HTA-c}}
\subfloat{\includegraphics[width=0.50\textwidth]
	{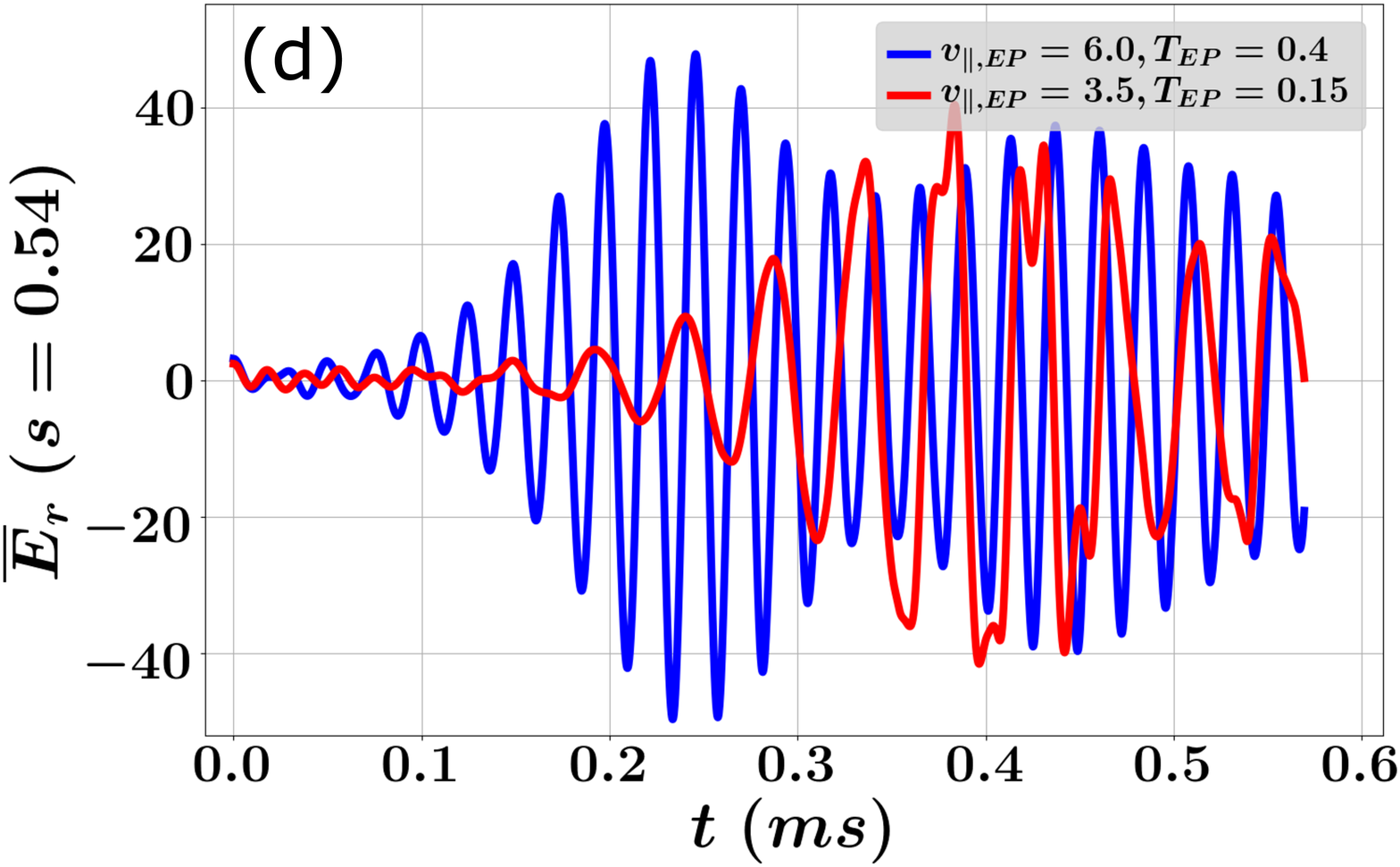}\label{fig:ES-HTA-d}}
\caption{EGAM radial structures are shown in Fig.~\ref{fig:ES-HTA-a} (case with $v_{\parallel, EP} = 6.0$) and Fig.~\ref{fig:ES-HTA-c}
(case with $v_{\parallel, EP} = 3.5$). 
Fig.~\ref{fig:ES-HTA-b}: energy exchange between the mode and ions.
Fig.~\ref{fig:ES-HTA-d}: time evolution of the mode at a radial point $s=0.54$. \label{fig:ES-HTA}}
\end{figure}

It has been mentioned in Sec.~\ref{sec:heating} that energetic particles of significantly different energies (parallel velocities) can drive EGAMs with similar saturation levels, but with a different efficiency in the energy transfer from the EPs to thermal plasma. 
Here, we are going to consider this process in detail on an example of two simulations.
In the first one, an energetic beam has $v_{\parallel, EP} = 3.5, T_{EP} = 0.15, n_{EP}/n_e = 0.09$, while in the second simulation, the EPs are represented by a $v_{\parallel, EP} = 6.0, T_{EP} = 0.40, n_{EP}/n_e = 0.01$ beam.
\begin{figure}[!t]
\subfloat{\includegraphics[width=0.50\textwidth]
	{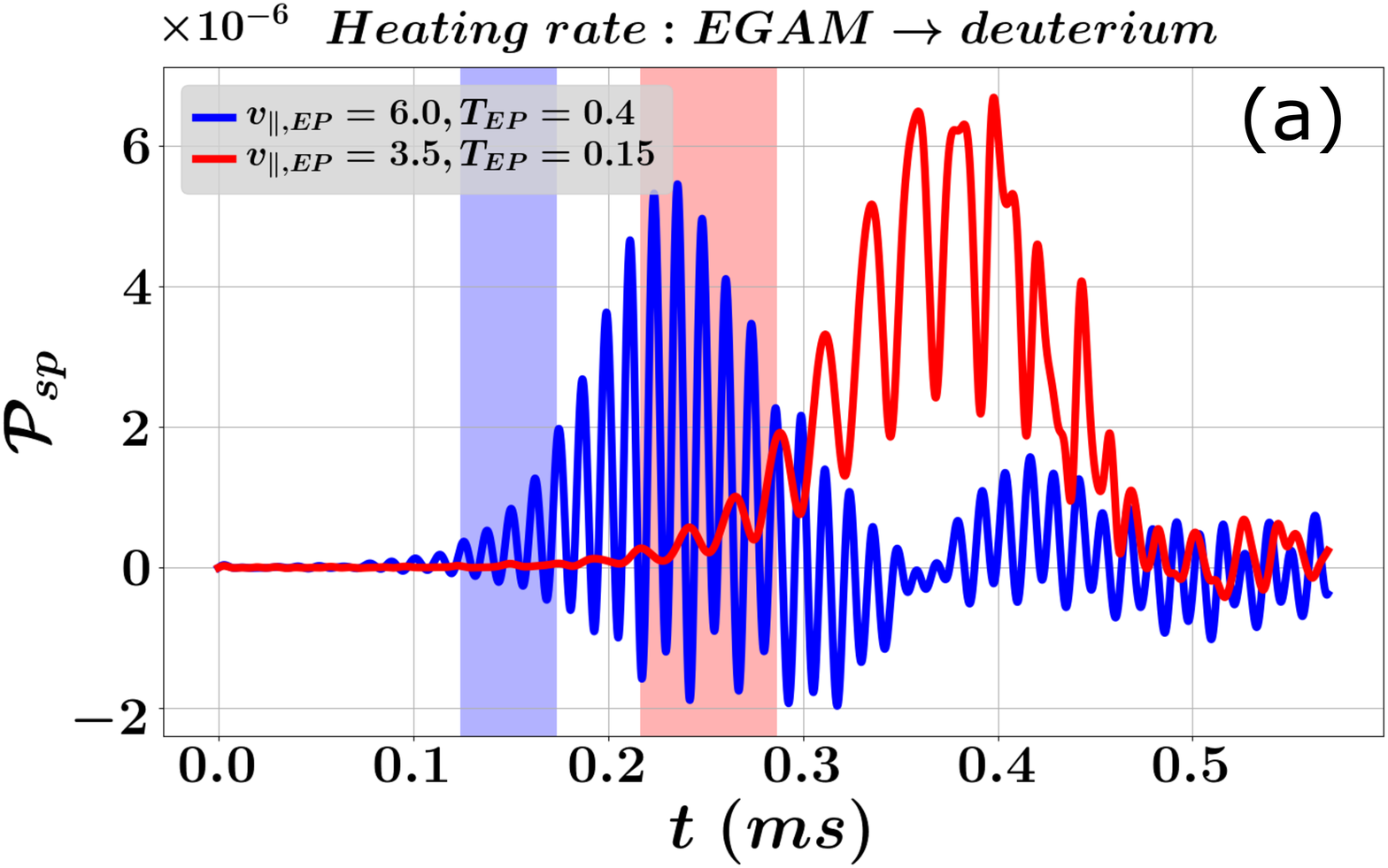}\label{fig:ES-HTA-res-a}}
\subfloat{\includegraphics[width=0.50\textwidth]
	{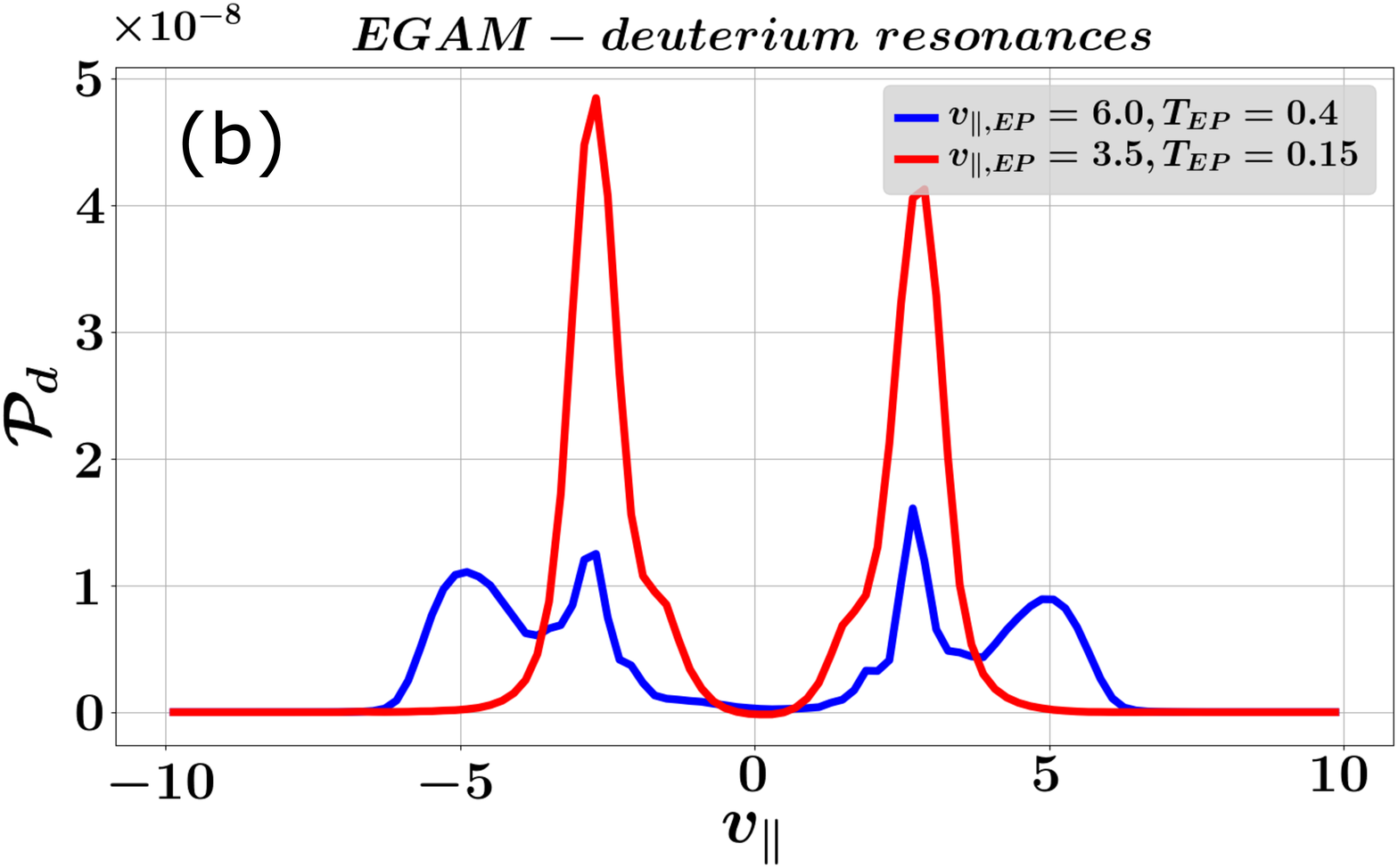}\label{fig:ES-HTA-res-b}}\\
\subfloat{\includegraphics[width=0.50\textwidth]
	{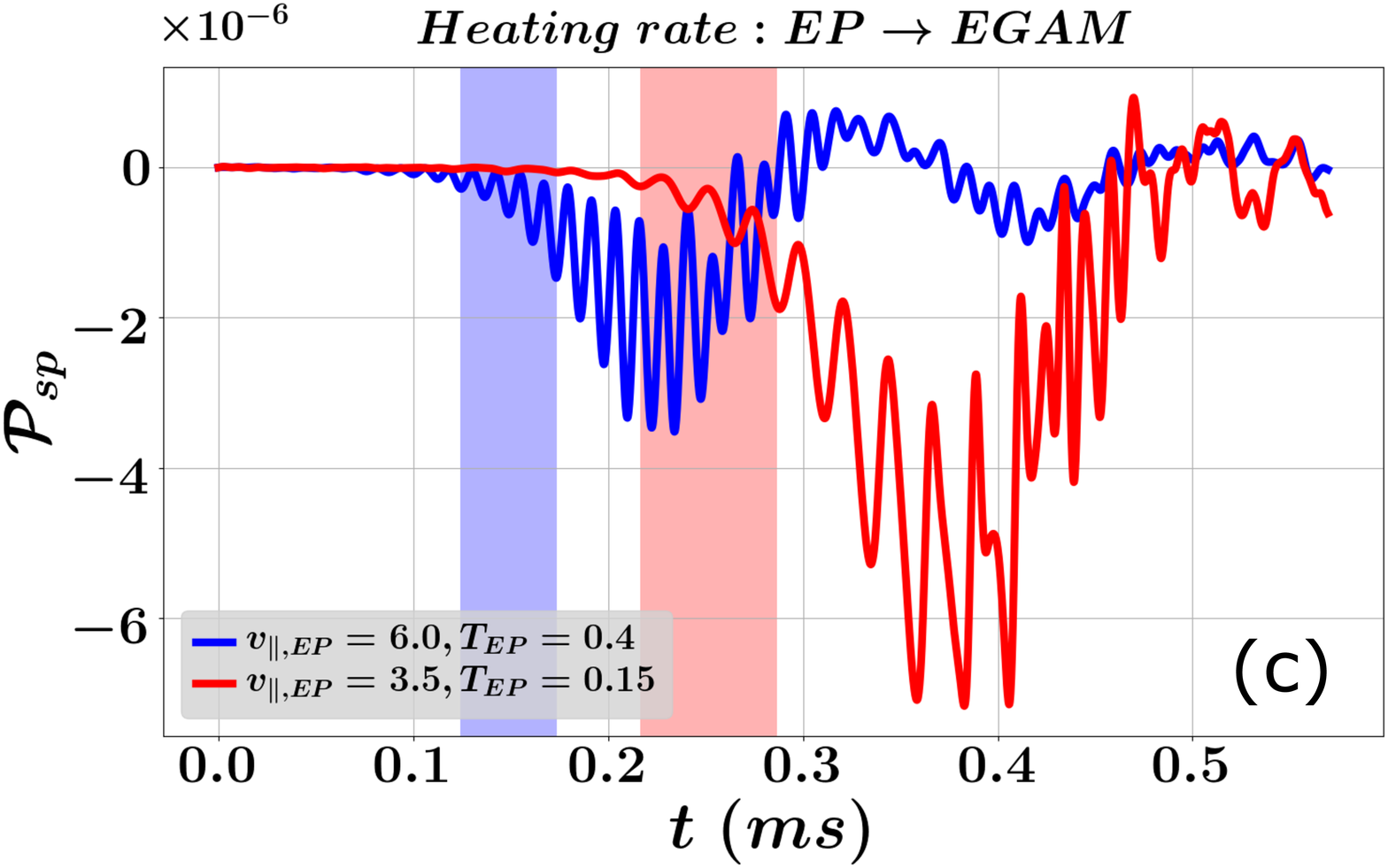}\label{fig:ES-HTA-res-c}}
\subfloat{\includegraphics[width=0.50\textwidth]
	{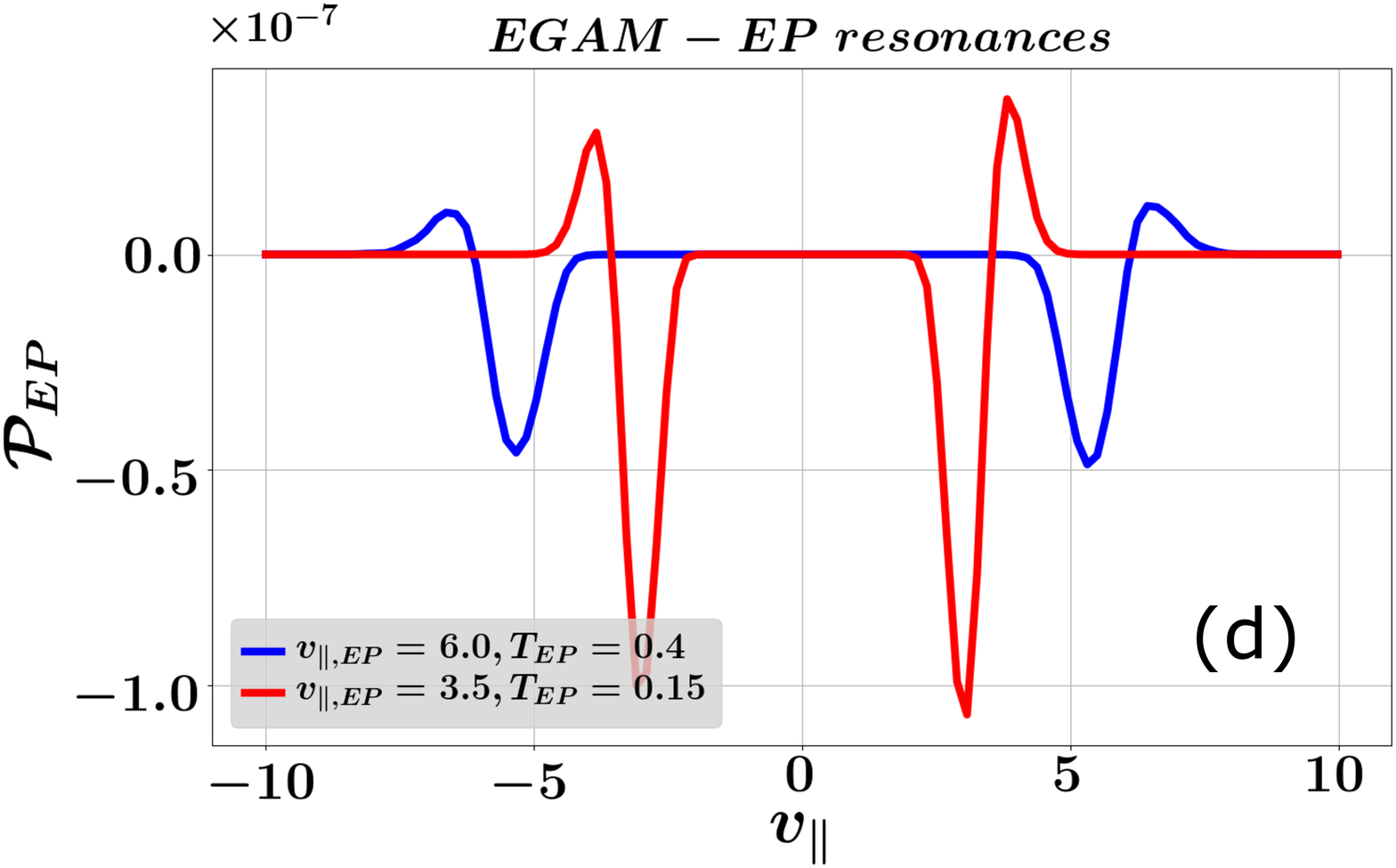}\label{fig:ES-HTA-res-d}}
\caption{Time evolution of energy transfer signals between an EGAM and thermal plasma (Fig.~\ref{fig:ES-HTA-res-a}), and between the mode and energetic particles (Fig.~\ref{fig:ES-HTA-res-c}). These signals are integrated in a real and velocity spaces.
Localisations of the EGAM-thermal deuterium resonances in case with $v_{\parallel,EP} = 6.0$ ($v_{\parallel,EP} = 3.5$) shown in Fig.~\ref{fig:ES-HTA-res-b} are obtained from the corresponding energy transfer signals integrated along a perpendicular velocity and time averaged in a blue (red) time intervals. 
The same procedure is applied to find the mode - EP resonances in Fig.~\ref{fig:ES-HTA-res-d}.
\label{fig:ES-HTA-res}}
\end{figure}
The time evolution of energy transfer signals between EGAM and thermal and energetic ions are shown in Fig.~\ref{fig:ES-HTA-res-a} and Fig.~\ref{fig:ES-HTA-res-c}, respectively.
In general, drive of an EGAM through higher order resonances, as it takes place in the case with $v_{\parallel, EP} = 3.5$, is less effective. 
To compensate that, the EP concentration has been increased for the $v_{\parallel, EP} = 3.5$ beam, while its temperature has been lowered to increase the velocity gradient of the EP distribution function at the localisation of the mode-particle resonance.
It leads to a higher EGAM drive by EPs and, as a result, to a higher amount of energy transferred from the mode to the thermal plasma (Fig.~\ref{fig:ES-HTA-b}).
However, the EGAM saturation levels in both cases with $v_{\parallel, EP} = 6.0$ and $v_{\parallel, EP} = 3.5$ are quite similar despite the difference in the EP concentration and temperature, as one can see from Fig.~\ref{fig:ES-HTA-d} or by comparing Fig.~\ref{fig:ES-HTA-a} and Fig.~\ref{fig:ES-HTA-c}.
It should be emphasized here that generally, rising the EP concentration without lowering the EP energy (velocity) leads to higher mode saturation levels.
\begin{figure}[!t]
\subfloat{
	\includegraphics[width=0.50\textwidth]
	{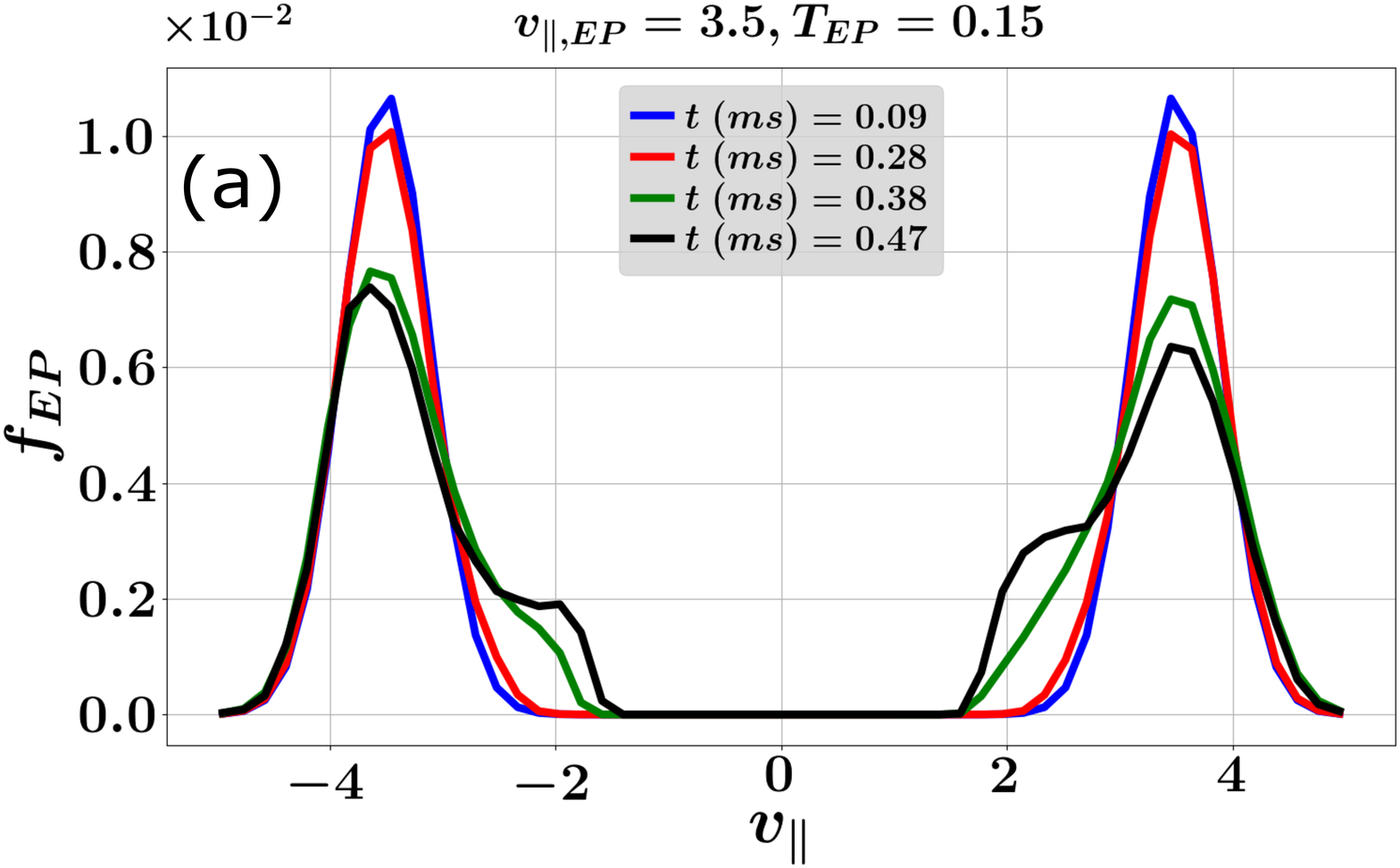}
	\label{fig:f-evolution-a}
}
\subfloat{
	\includegraphics[width=0.50\textwidth]
	{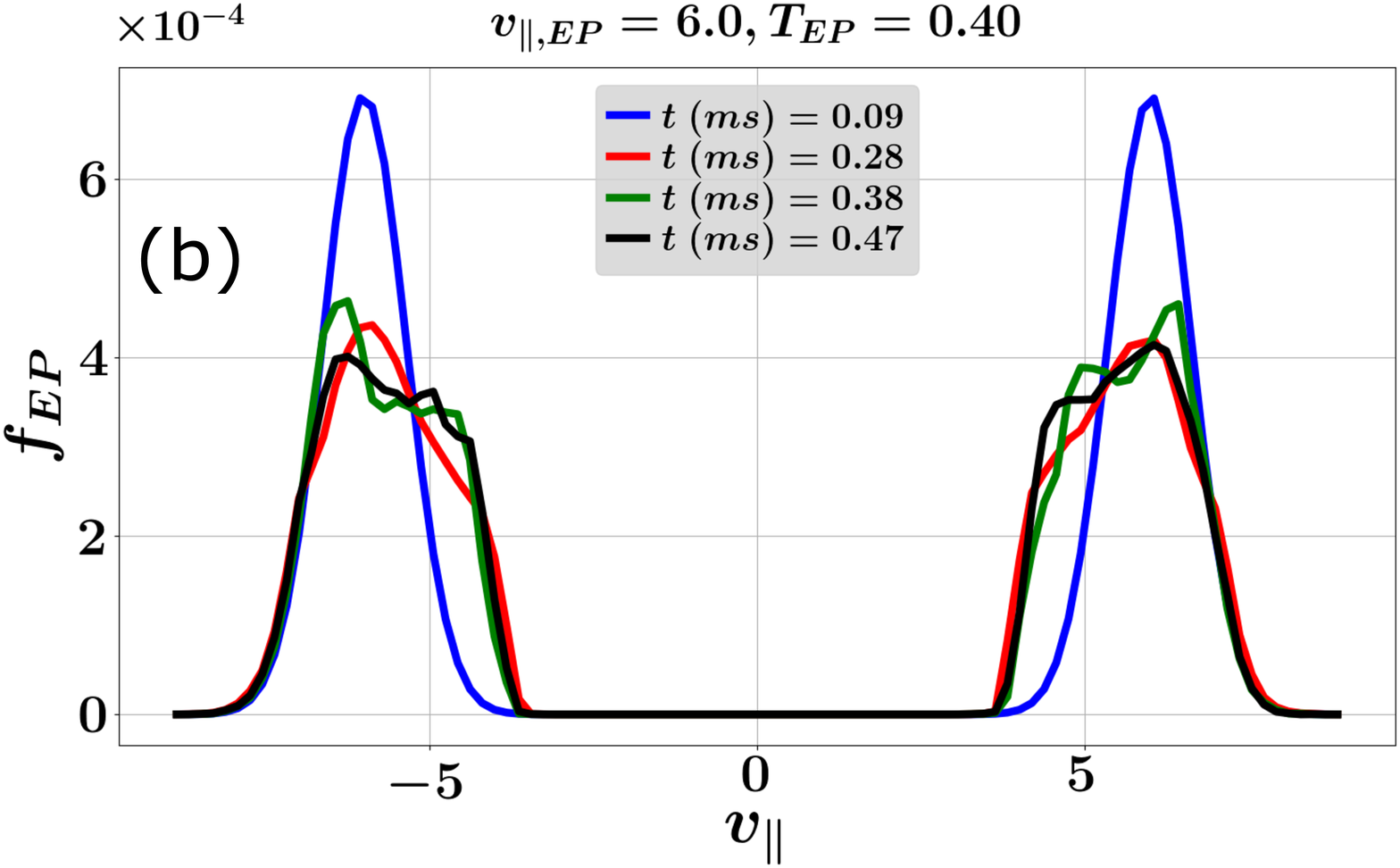}
	\label{fig:f-evolution-b}
}
\caption{Evolution in time of EP distribution functions for the case 
$v_{\parallel, EP} = 3.5, T_{EP} = 0.15$ 
(Fig.~\ref{fig:f-evolution-a}) and 
$v_{\parallel, EP} = 6.0, T_{EP} = 0.40$ 
(Fig.~\ref{fig:f-evolution-b}).
\label{fig:f-evolution}
}
\end{figure}
Apart from that, with $v_{\parallel, EP} = 3.5$, the mode grows slower, and the flattening of the EP distribution function develops slower as well (Fig.~\ref{fig:f-evolution}) than with $v_{\parallel, EP} = 6.0$.

\begin{figure}[!t]
\subfloat{
	\includegraphics[width=0.50\textwidth]
	{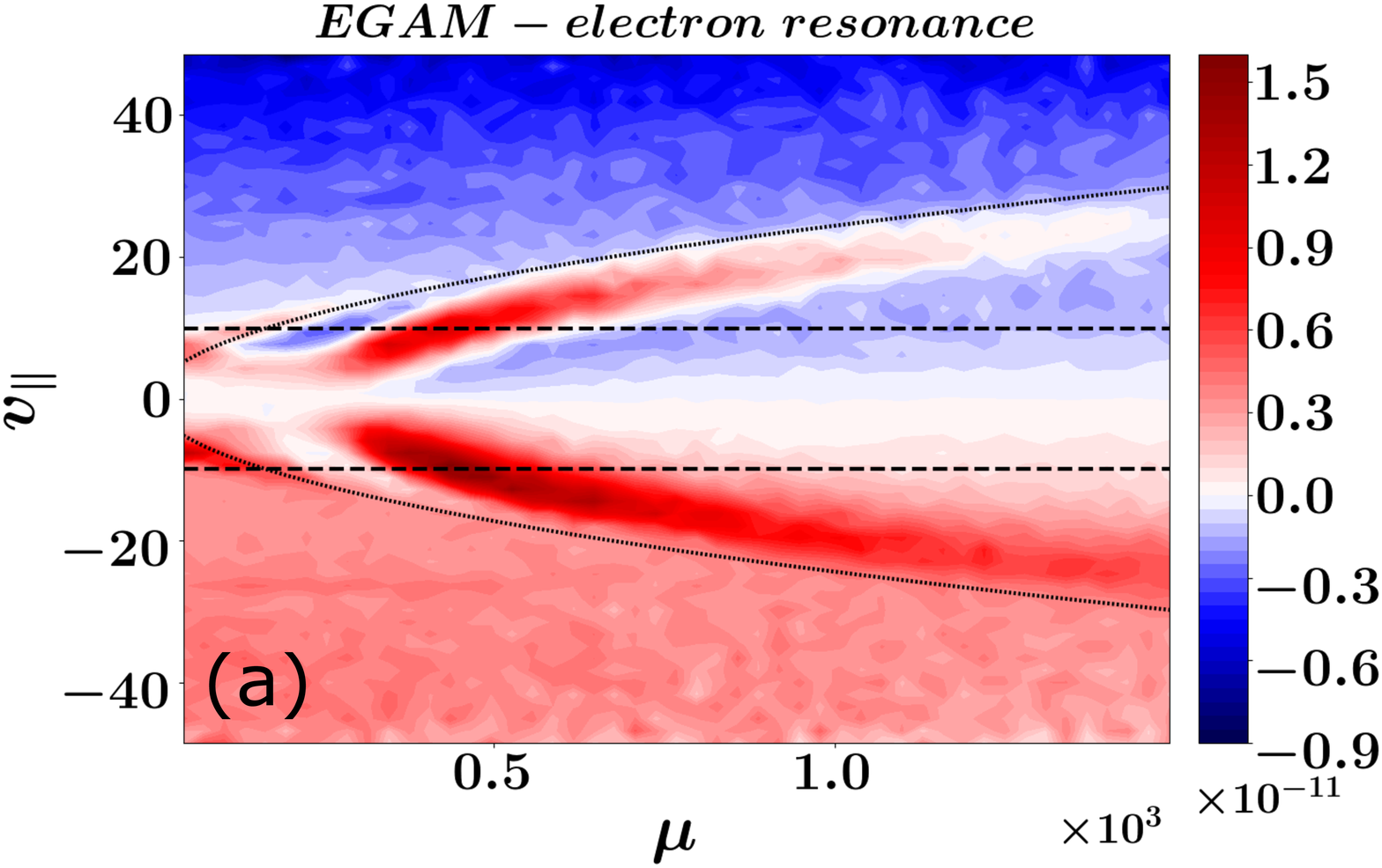}
	\label{fig:nl-em-mpr-a}
}
\subfloat{
	\includegraphics[width=0.50\textwidth]
	{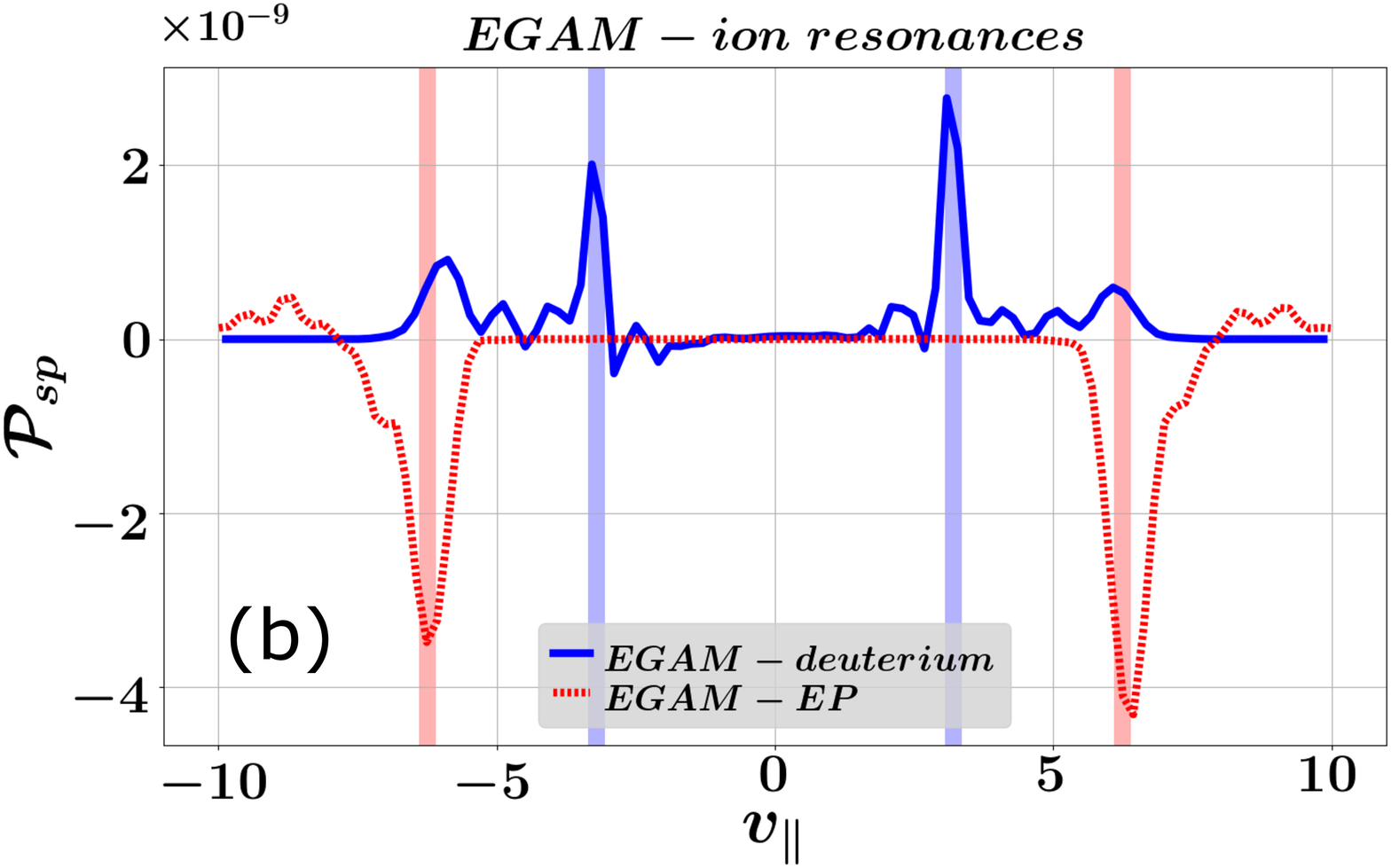}
	\label{fig:nl-em-mpr-b}
}
\caption{EGAM-electron resonance in a nonlinear EM case 
(Fig.~\ref{fig:nl-em-mpr-a}). The horizontal lines indicate an estimated position of the mode-particle $m = 1$ resonance (Eq.~\ref{eq:vresm}).
Fig.~\ref{fig:nl-em-mpr-b}: EGAM resonances with thermal deuterium and EPs. The vertical blue and red lines indicate analytical resonance positions for $m = 1$ and $m = 2$ respectively (Eq.~\ref{eq:vresm}). In this nonlinear EM simulation, the EGAM is localised around $s = 0.47$ having a linear frequency $\omega_{egam}[\omega_{ci}] \approx 3\cdot 10^{-3}$.
\label{fig:nl-em-mpr}
}
\end{figure}

The MPR diagnostic can also be applied to the nonlinear case with drift-kinetic electrons and $v_{EP} = 8.0$, described in Sec.~\ref{sec:kin}.
In Fig.~\ref{fig:nl-em-mpr-a}, the EGAM-electron resonances are localised near an estimated electron passing-trapped boundary (Eq.~\ref{eq:p-tr-boundary}), which is consistent to the linear computation shown in Fig.~\ref{fig:lin-AE-KE-b}.
Apart from that, as it has been observed in the ES cases discussed above, the mode here is driven through the $m = 1$ EGAM-EP resonance, while it transfers its energy to the thermal plasma mainly through the $m = 2$ resonance.

\end{appendices}

\bibliography{./AAPPS-bibl}

\end{document}